\documentstyle[11pt]{article}
\renewcommand{\thesection}{\arabic{section}.}
\renewcommand{\thesubsection}{\arabic{section}.\arabic{subsection}}

\renewcommand{\theequation}{\arabic{section}.\arabic{equation}}
\begin{document}
\baselineskip = 1.25\baselineskip
\begin{titlepage}
\begin{center}
{\large\bf MODAL EXPANSIONS AND ORTHOGONAL COMPLEMENTS\\[.2cm]
           IN THE THEORY OF COMPLEX MEDIA WAVEGUIDE EXCITATION\\[.2cm]
           BY EXTERNAL SOURCES FOR ISOTROPIC, ANISOTROPIC,\\[.2cm]
           AND BIANISOTROPIC MEDIA}\\[.7cm]

A. A. Barybin\\[.5cm]
{Electronics Department, Electrotechnical University,\\
                               St. Petersburg, 197376, Russia}\\[2cm]
\end{center}

\noindent
{\bf Abstract} -- A unified electrodynamic approach to the guided wave
excitation by external sources in the waveguiding structures with
bianisotropic media is developed. Effect of electric, magnetic, and
magneto-electric losses in such media manifests itself in the special form
of eigenmode orthogonality referred to as the {\it quasi-orthogonality
relation\/}. It reflects the existence of the {\it cross-power flow\/}
$P_{kl}$ and {\it loss\/} $Q_{kl}$ for any pair $(k,l)$ of modes which are
rigidly linked to each other by this relation. The quasi-orthogonality
relation remains true in the limiting case of lossless waveguides yielding
the customary relations of orthogonality and normalization for propagating
(active) modes and also their generalization for nonpropagating
(reactive)~modes.

It is shown that the eigenmode set for a waveguiding structure is complete
only outside the region of exciting sources. Inside this region the modal
expansions of fields are incomplete and must be supplemented with the
orthogonal {\it complementary fields\/} which extend the proper Hilbert
space spanned by waveguide eigenfunctions. Among exciting sources there
are the external bulk sources (currents, fields, and medium perturbations)
and the external surface currents. Besides, the ortho\-gonal complementary
fields generate the {\it effective surface currents\/} on boundaries of
the bulk exciting sources.

The problem of waveguide excitation by external sources is solved by means
of determining both the mode amplitudes for the modal field expansions and
the orthogonal complementary fields inside the source region. The equations
of mode excitation are derived on the basis of three approaches applying
the direct use of Maxwell's equations, the electrodynamic analogy with the
mathematical method of variation of constants, and the conjugate
reciprocity theorem.

\end{titlepage}

\newpage
\noindent
{\Large{\bf CONTENTS}}
\vspace{2\baselineskip}

\noindent
{\bf
1. INTRODUCTION\\[1mm]
2. GENERAL POWER-ENERGY RELATIONS OF ELECTRODYNAMICS FOR
                                           BIANISOTROPIC MEDIA \\[1mm]
2.1 {} Poynting's Theorem \\[1mm]
2.2 {} Mode Power Transmission and Dissipation \\[1mm]
3. ORTHOGONALITY AND QUASI-ORTHOGONALITY OF MODES IN LOSSLESS AND
                                                LOSSY WAVEGUIDES \\[1mm]
3.1 {} Quasi-orthogonality Relation for Lossy Waveguides \\[1mm]
3.2 {} Mode Orthogonality in Lossless Waveguides \\[1mm]
3.2.1 {} Orthogonality and Normalization Relations for Active Modes \\[1mm]
3.2.2 {} Orthogonality and Normalization Relations for
                                                     Reactive Modes\\ [1mm]
3.3 {} Time-average Stored Energy for Active Modes in
                                                Lossless Waveguides \\[1mm]
4. ORTHOGONAL COMPLEMENTS AND EFFECTIVE SURFACE CURRENTS
                                        INSIDE SOURCE REGION \\[1mm]
4.1 {} Bulk and Surface Exciting Sources \\[1mm]
4.2 {} Orthogonal Complementary Fields and Effective Surface
                                                       Currents \\[1mm]
5. EQUATIONS OF MODE EXCITATION \\[1mm]
5.1 {} Approach based on the Electrodynamic Method of Variation
                                                  of Constants \\[1mm]
5.2 {} Approach Based on the Reciprocity Theorem \\[1mm]
5.2.1 {} Derivation of the Conjugate Reciprocity Theorem \\[1mm]
5.2.2 {} Derivation of the Equations of Mode Excitation \\[1mm]
6. CONCLUSION \\[1mm]
APPENDIX \\[1mm]
REFERENCES \\[1mm]          }

\newpage
\section{INTRODUCTION}
\label{sec:1}

Guided-wave electrodynamics in the modern understanding deals with
the study of propagation, radiation, excitation, and interaction of
waves in a variety of waveguiding structures. The term "complex
media waveguide" applied in the title of the paper implies the medium
complexity of two types:\\
(i) the {\it physical\/} complexity associated with medium properties
diversified by the very  nature (gas and solid state plasmas
with drifting carriers;\, polarized and magnetized solids with
different properties: piezoelectric, electrooptic, acoustooptic,
magnetooptic, magnetoelastic;\, chiral, biisotropic, and
bianisotropic media);\\
(ii) the {\it geometrical\/} complexity due to using composite and
multilayered structures.

Electromagnetic theory has been developed up to its present state by
extensive works and efforts of a great number of researchers and
scientists. Besides pure scientific purposes, progress in classical
electrodynamics at all stages of its advancement was encouraged by
certain demands of technology.

At the first stage such a stimulating factor was related
to practical needs of then incipient radar and antenna engineering.
The consequent experience on electrodynamic properties of mostly
passive nondispersive media specified by phenomenological constants,
which had been gathered over a number of years, was accumulated
in many scientific publications. Among them we should
refer, for instance, to such famous and popular books
as~[1\,--\,5] which now constitute the theoretical foundation of
classical electrodynamics. Much attention was given to the study of
electromagnetic properties of gas plasma as a medium for wave
propagation. Later on the plasma wave aspects were extended to the
behavior of charge carriers in solids considered as a solid-state plasma.
At present the literature devoted to the electromagnetic properties of
plasmas is immense and the following books~[6\,--\,11] with
their bibliographies can give a good indication of the scope of plasma
electrodynamics.

Another direction of electrodynamic aspects was inspired by developing
the techno\-logy of microwave devices operating on wave principle. The
first to be developed were vacuum devices using the space charge,
cyclotron, and synchronous waves on an electron beam such as the
traveling-wave tube, backward-wave tube, and others~[12\,--\,14].
Later the similar idea to apply waves in solids for signal processing
gave rise to new lines of solid-state electronics. They are due to
applying the surface acoustic waves (SAW) in elastic
piezo-dielectrics~[15\,--\,17], the magnetostatic spin waves
(MSW) in magnetized ferrites~[18\,--\,23], and
the space charge waves (SCW) in semiconductors with negative
differential mobility of electrons~\cite{10,24,25}.
These waves refer to the quasistatic part of the electromagnetic
spectrum of waveguiding structures for which a relevant potential
field (electric for SAW and SCW or magnetic for MSW) dominates
over its curl counterpart. This fact caused some electrodynamic
formulations to be revised in order to separate such potential fields
and take into account the space-dispersive properties of these media
described by the proper equations of medium motion~\cite{15,22,24,25}.

For the last decades the macroscopic electrodynamics of waveguiding
structures has experienced two powerful stimulating actions. The first
is associated with needs of fiber and integrated optics and began
about twenty five years ago. A number of theoretical propositions in
electrodynamics were reformulated, as applied to optical waveguides,
and have been embodied in devices. The literature devoted to this
topic is enormous including the well-known books~[26\,--\,32].

Nowadays we observe the renewed interest in electrodynamic problems
caused by efforts to apply chiral, biisotropic, and bianisotropic media
for the control of electromagnetic radiation in waveguiding structures.
Phenomenon of optical activity in certain natural substances generated
by their handedness property (chirality) was already known last century.
The present revival of scientific and technological attention
to this problem is inspired by the modern progress of
material science and technology in synthesizing artificial composite
media. Such media possess unique properties to open new potential
possibilities in their utilizing in optics and at microwaves. This has
aroused a great wave of research followed by numerous publications,
among them there are the general books~[33\,--\,35] comprising
bianisotropic issues and the special books~[36\,--\,40] devoted entirely
to this~subject.        \\[.1cm]
\indent
Theoretical ground for many wave electrodynamics applications is the
modal expansion method. In the case of the eigenmode excitation
by external sources the question of completeness of the
eigenfunction basis chosen inside the source region is of crucial
importance in practice. Unfortunately, most authors solve this
question rather superficially assuming intuitively that the
set of eigenfunctions found as the general solution to the
boundary-value problem without sources is complete also inside
the source region. However, this is not the case in general.

From mathematical considerations given in Appendix\,A.1
it follows that the above statement is valid only for the desired
functions $\psi(x)$ tangential to the Hilbert space spanned by the
eigenfunction basis $\{\psi_k(x)\}$. Gene\-rally, for most functions
$f(x)$ their series expansion in terms of the base functions (convergent
in mean) is only a {\it projection\/} $\psi(x)$ of the function $f(x)$
on the Hilbert space. In addition, there may exist a nonzero function
$c(x)$ ortho\-gonal to this space, \,the so-called \,{\it orthogonal
complement\/}, \,which in general must be added to the projection
$\psi(x)$ in order for $f(x)$ to be considered as the complete
required function (see Eq.~(\ref{eq:A18}) and relevant relations in
Appendix\,A.2).
The above statement is fairly obvious for mathematicians but
unfortunately was fully ignored in developing the modern topics of
guided-wave electrodynamics by most authors, not counting Vainshtein
\cite{2} and Felsen and Marcuvitz~\cite{8}. Strange as it may seem,
when developing the excitation theory of optical waveguides, many authors
\cite{26,27,29,32,34} have correctly applied the modal expansions
to the transverse components of electromagnetic fields but entirely
dropped the orthogonal complements due to the longitudinal exciting
bulk currents. As will be shown, this causes the so-called effective
surface currents to be lost. Similar situation also holds for
electrodynamics of the waveguiding structures with chiral and
bianisotropic media [34,\,38\,--\,40] where the modal expansion method
is practically undeve\-loped and the problem of the orthogonal
complements and effective surface sources, worked out below, is
more complicated.   \\[.1cm]
\indent
The objective in writing this paper is to develop a unified
electrodynamic theory of wave\-guide excitation by external sources
(bulk and surface) applicable equally for any media and waveguiding
structures. Particular attention will be given to the study of the
unexpandable orthogonal complements to the eigenmode expansions which
should be expressed in terms of the given exciting currents as well
as the desired mode amplitudes of the modal expansions. To this end,
we begin with Sec.\,2 devoted to deriving the basic energy-power
relations of electrodynamics applied to the lossy bianisotropic media
including Poynting's theorem in the differential and integral forms
involving the {\it self-power\/} and {\it cross-power\/} quantities
(flows and losses) transmitted and dissipated by the eigenmodes of a
waveguiding structure. Sec.\,3 deals with a generalization of the known
orthogonality relation for the waveguides without losses to the so-called
{\it quasi-orthogonality\/} relation for lossy waveguides which describes,
as a special case, the orthogonality of the reactive (nonpropagating)
modes in lossless waveguides. In addition, an expression for the
time-average energy stored by the active (propagating) modes is proved.
Sec.\,4 is concerned with the consideration of external sources (currents,
fields, and medium perturbations) and electromagnetic fields inside the
source region. The complete representation of the fields, besides their
modal expansions, involves also the so-called {\it orthogonal
complementary fields\/} which necessarily generate the {\it effective
surface currents\/}. Sec.\,5 contains two different approaches
to the derivation of the equations of mode excitation by external
sources. The first approach applied only to the lossless waveguides is
based on an electrodynamic analogy with the known mathematical statements
such as the method of variation of cons\-tants and the relations of
functional analysis (see Appendix\,A). The second approach makes use of
the reciprocity theorem in the complex-conjugate form to obtain
the equations of mode excitation in the ge\-neral form valid for both
lossy and lossless wave\-guiding structures. Another alternative proof of
the excitation equations for lossless waveguides starting directly from
Maxwell's equations is adduced in Appendix\,B.

In this paper we restrict our consideration to the case of time-dispersive
media whose electrodynamic properties (isotropic, anisotropic,
bianisotropic) are characterized by the frequency-dependent constitutive
parameters consi\-dered as phenomenologically given. More complicated
case of space-dispersive media such as elastic piezo-dielectrics,
magnetized ferrites, nondege\-nerate plasmas with drifting charge carriers
whose electrodynamic description requires, besides Maxwell's equations,
employing the proper equation of medium motion will be the subject of
matter of the second part of the paper.  \\[.1cm]
\indent
In conclusion there are a few words concerning the notation applied:\\
(i) \,tensors of rank \,0 (scalars), \,1 (vectors), \,2 (dyadics),\,
and more than \,2 (tensors) are~denoted~as: {} $A$, {} $\bf A$, {}
$\bar{\bf A}$, {} and \,$\bar{\bar{\bf A}}$, {} respectively; \\
(ii) \,their products are denoted as: \,$AB$\, (for two scalars); {}
${\bf A}\cdot{\bf B}$,\, ${\bf A}\times{\bf B}$,\, and \,$\bf AB$\,
(for~scalar, vector, and dyad products of two vectors);
{} ${\bf AB}\cdot{\bf CD}= {\bf AD}({{\bf B}\cdot{\bf C}})$,\,
${\bar{\bf A}\cdot\bar{\bf B}} = A_{ij}B_{jk}$,\, and\,
${\bar{\bar{\bf A}}\cdot\bar{\bar{\bf B}}} = A_{ijk}B_{klm}$\,
(for scalar product of two vector dyads, dyadics, and tensors); {}
${\bf AB}:{\bf CD} = ({\bf A}\cdot{\bf D})({\bf B}\cdot{\bf C})$,\,
$\bar{\bf A}:\bar{\bf B}= A_{ij}B_{ji}$,\,
and~${\bar{\bar{\bf A}}\,:\,\bar{\bar{\bf B}}}~= A_{ijk}B_{kjl}$\,
(for~double scalar product of two vector dyads, dyadics,~and~tensors).

\section{GENERAL POWER-ENERGY RELATIONS OF ELECTRODYNAMICS
                           FOR BIANISOTROPIC MEDIA}
\label{sec:2}

\subsection{Poynting's Theorem}
\label{sec:2A}
\indent

In macroscopic electrodynamics, the electromagnetic properties of a
me\-dium are described by two field-intensity vectors, $\bf E$ (the
electric field) and $\bf H$ (the magnetic field), and two flux-density
vectors, $\bf D$ (the electric induction) and $\bf B$ (the magnetic
induction), which are related by means of Maxwell's equations (written
in the rationalized mks~system):
\begin{equation}
\mbox{\boldmath$\mbox{\boldmath$\nabla$}$}\!\times{\bf E}= -\,
{\partial{\bf B}\over\partial t} \,, \quad
\mbox{\boldmath$\mbox{\boldmath$\nabla$}$}\!\times{\bf H} =
{\partial{\bf D}\over\partial t}+ {\bf J} \,,\quad
\mbox{\boldmath$\mbox{\boldmath$\nabla$}$}\!\cdot{\bf D}= \rho\,, \quad
\mbox{\boldmath$\mbox{\boldmath$\nabla$}$}\!\cdot{\bf B}= 0 \,.
\label{eq:2.1}
\end{equation}

Mobile charge effects in the medium are specified by the charge and
current densities $\rho$ and $\bf J$, \,whereas the bound charges
arise as a result of polarization responses of the medium to
electromagnetic actions characterized by the electric and magnetic
polarization vectors ${\bf P}$ (the polarization vector) and
${\bf M}$ (the magnetization vector). These vectors yield the
corresponding conrtibutions to the electric and magnetic inductions:
\begin{equation}
{\bf D}\,=\,\epsilon_0{\bf E}\,+\,{\bf P} \qquad \mbox{and} \qquad
{\bf B}\,=\,\mu_0({\bf H}\,+\,{\bf M}) \,.
\label{eq:2.2}
\end{equation}

The conventional procedure applied to Eqs.~(\ref{eq:2.1}) reduces to
Poynting's theorem in the form involving the instantaneous values of
power-energy quantities:
\begin{equation}
{\partial w\over\partial t} \;+\;
\mbox{\boldmath$\mbox{\boldmath$\nabla$}$}\!\cdot{\bf S} \,=\,-\;
{\rm I}_{\bf J}\,-\,{\rm I}_{\bf P}\,-\,{\rm I}_{\bf M}
\label{eq:2.3}
\end{equation}
where \,$w= ({{\bf E}\cdot{\bf D}}+{{\bf H}\cdot{\bf B}})/2$\, is the
electromagnetic energy density and ${\bf S}= {\bf E}\times{\bf H}$ is
the electromagnetic energy flux density (Poynting's vector).
The terms on the right of Eq.~(\ref{eq:2.3})
\begin{equation}
{\rm I}_{\bf J} \,=\,{\bf J}\cdot{\bf E} \,,
\label{eq:2.4}
\end{equation}
\begin{equation}
{\rm I}_{\bf P} \,=\,{1\over2} \biggl( {\bf E}\cdot
{\partial{\bf P}\over\partial t} \,-\,
{\bf P}\cdot{\partial{\bf E}\over\partial t} \biggr) \,,
\label{eq:2.5}
\end{equation}
\begin{equation}
{\rm I}_{\bf M} \,=\,{1\over2} \biggl( {\bf H}\cdot
{\partial\mu_0{\bf M}\over\partial t} \,-\,
\mu_0{\bf M}\cdot{\partial{\bf H}\over\partial t} \biggr)
\label{eq:2.6} \\[.2cm]
\end{equation}
reflect specific properties of the medium under study and take into
account the instantaneous power of interaction between the electromagnetic
fields $(\bf E,\,\bf H)$ and the charges -- both mobile ones carrying
the current $\bf J$ and bound ones generating the polarization $\bf P$
and magnetization~$\bf M$.

In the literature the energy term ${\partial w/\partial t}$ is
conventionally identified with the sum ${\bf E}\cdot{\partial{\bf D}/
\partial t}+ {\bf H}\cdot{\partial{\bf B}/\partial t}$, which is true only
if \,${\bf D}=\bar{\!\mbox{\boldmath$\epsilon$}}\cdot{\bf E}$\, and
\,${\bf B}=\bar{\!\mbox{\boldmath$\mu$}}\cdot{\bf H}$ where the tensors
$\bar{\!\mbox{\boldmath$\epsilon$}}$ and $\bar{\!\mbox{\boldmath$\mu$}}$
are symmetric and time-independent. In this case only the first
interaction term~(\ref{eq:2.4}) is taken into account, whereas two
others~(\ref{eq:2.5}) and (\ref{eq:2.6}) are dropped without any
justification. As will be evident from our subsequent examination
including the second part of the paper, these terms play an important
role in the power-energy theorem.

For time-harmonic fields (with time dependence in the form of
${\rm exp}(i\omega t)$\,) one is usually interested in time-average
values of the power-energy quantities denoted~as~$\langle\ldots\rangle$.
In this case $\langle\partial w/\partial t\rangle= 0$ so that
Eq.~(\ref{eq:2.3}) takes the following form involving the time-average
values of quantities:
\begin{equation}
\mbox{\boldmath$\mbox{\boldmath$\nabla$}$}\!\cdot\langle\,{\bf S}\,\rangle\,=
\,-\,\langle\,{\rm I}_{\bf J}\rangle -
\langle\,{\rm I}_{\bf P}\rangle - \langle\,{\rm I}_{\bf M}\rangle \,.
\label{eq:2.7}
\end{equation}

Below we concentrate on bianisotropic media for which there is
no equation of motion. Their properties are usually described by
the constitutive equations establishing macroscopic local relations
among field vectors. It~should be emphasized that magneto-electric
effects (for instance, optical activity), by their microscopic nature,
are brought about by nonlocality of polarization response on
electromagnetic actions~\cite{35,38,39,44}. But their macroscopic
manifestations are usually similar to those of actual time-dispersive
media because for plane waves with the wave vector
${\bf k}= (\omega/c){\bf n}$ all the constitutive tensor parameters
of such media become solely frequency-dependent (see Ref.~\cite{44}).

There are a few forms of the constitutive relations for bianisotropic
media [33\,--\,38]. Among them we choose the following
form
\begin{eqnarray}
{\bf D} \!\!&=&\! \bar{\!\mbox{\boldmath$\epsilon$}}\cdot{\bf E}\;+\;\,
\bar{\!\mbox{\boldmath$\xi$}}\cdot{\bf H} \,,
\label{eq:2.8}\\
{\bf B} \!\!&=&\! \bar{\!\mbox{\boldmath$\zeta$}}\cdot{\bf E}\;+\;\,
\bar{\!\mbox{\boldmath$\mu$}}\cdot{\bf H} \,,
\label{eq:2.9}
\end{eqnarray}
as more convenient for our subsequent examination.

Four constitutive medium parameters \,$\bar{\!\mbox{\boldmath$\epsilon$}},
\,\;\bar{\!\mbox{\boldmath$\mu$}},\,\;\bar{\!\mbox{\boldmath$\xi$}}$\,
and \,$\bar{\!\mbox{\boldmath$\zeta$}}$\,
are considered as dyadic functions of frequency given phenomenologically.
They comprise all special cases of the physical media without
space dispersion:\\
(i) for the isotropic medium
\begin{equation}
\bar{\!\mbox{\boldmath$\epsilon$}}= \epsilon\,\bar{\bf I} \,, \qquad
\bar{\!\mbox{\boldmath$\mu$}}= \mu\,\bar{\bf I} \,,  \qquad
\bar{\!\mbox{\boldmath$\xi$}}=\,\bar{\!\mbox{\boldmath$\zeta$}}=\,0 \,;
\label{eq:2.10}
\end{equation}
(ii) for the double anisotropic medium
\begin{equation}
\bar{\!\mbox{\boldmath$\epsilon$}}\ne\epsilon\,\bar{\bf I} \,, \qquad
\bar{\!\mbox{\boldmath$\mu$}}\ne\mu\,\bar{\bf I} \,, \qquad
\bar{\!\mbox{\boldmath$\xi$}}=\,\bar{\!\mbox{\boldmath$\zeta$}}=\,0 \,;
\label{eq:2.11}
\end{equation}
(iii) for the chiral (biisotropic) medium
\begin{equation}
\bar{\!\mbox{\boldmath$\epsilon$}}= \epsilon\,\bar{\bf I}, \quad\;
\bar{\!\mbox{\boldmath$\mu$}}= \mu\,\bar{\bf I},  \quad\;
\bar{\!\mbox{\boldmath$\xi$}}=
(\chi- i\kappa)\sqrt{\epsilon_0\,\mu_0}\;\bar{\bf I}, \quad\;
\bar{\!\mbox{\boldmath$\zeta$}}=
(\chi+ i\kappa)\sqrt{\epsilon_0\,\mu_0}\;\bar{\bf I},
\label{eq:2.12}
\end{equation}
where $\chi$ and $\kappa$ are Tellegen's parameter of nonreciprocity
and Pasteur's parameter of chirality, respectively~\cite{38,41}.
It is known~[33\,--\,35] that for a bianisotropic medium without
losses the dyadics \,$\bar{\!\mbox{\boldmath$\epsilon$}}$\, and
\,$\bar{\!\mbox{\boldmath$\mu$}}$\, are hermitian (self-adjoint) while
the dyadics \,$\bar{\!\mbox{\boldmath$\xi$}}$
and \,$\bar{\!\mbox{\boldmath$\zeta$}}$\, are hermitian conjugate
(mutually adjoint), that is
\begin{equation}
\bar{\!\mbox{\boldmath$\epsilon$}}\,=\;
\bar{\!\mbox{\boldmath$\epsilon$}}\,^\dagger\,,\qquad\quad
\bar{\!\mbox{\boldmath$\mu$}}\,=\;
\bar{\!\mbox{\boldmath$\mu$}}\,^\dagger \,,  \qquad\quad
\bar{\!\mbox{\boldmath$\xi$}}\,=\;
\bar{\!\mbox{\boldmath$\zeta$}}\,^\dagger \,,
\label{eq:2.13}
\end{equation}
where superscript $^\dagger$ denotes transpose and complex conjugate
(hermitian conjugate). Relations~(\ref{eq:2.13}) imply that
in the general case of lossy media the antihermitian parts
\,$\bar{\!\mbox{\boldmath$\epsilon$}}\,^a\!=
(\,\bar{\!\mbox{\boldmath$\epsilon$}}-
\bar{\!\mbox{\boldmath$\epsilon$}}\,^\dagger)/2$,\,\,
$\bar{\!\mbox{\boldmath$\mu$}}\,^a\!=
(\,\bar{\!\mbox{\boldmath$\mu$}}-
\bar{\!\mbox{\boldmath$\mu$}}\,^\dagger)/2$\, and the difference\,
$(\,\bar{\!\mbox{\boldmath$\xi$}}-
\bar{\!\mbox{\boldmath$\zeta$}}\,^\dagger)$\,
are responsible for losses (dielectric, magnetic,
and magneto-electric, \,respectively).
If the medium has also the electric losses related to its conductive
properties and specified by the conductivity dyadic\,
$\bar{\!\mbox{\boldmath$\sigma$}}_c$,\, then in addition to
Eqs.~(\ref{eq:2.8}) and (\ref{eq:2.9}) there is another
constitutive relation
\begin{equation}
{\bf J}\,=\;\bar{\!\mbox{\boldmath$\sigma$}}_c\cdot{\bf E} \,.
\label{eq:2.14}
\end{equation}

Let us calculate the terms in the right-hand side of Eq.~(\ref{eq:2.7})
by using their definitions~(\ref{eq:2.4}) through (\ref{eq:2.6}) and
the constitutive relations~(\ref{eq:2.8}), (\ref{eq:2.9}),
and~(\ref{eq:2.14}):
\begin{equation} \!\!\!
\langle\,{\rm I}_{\bf J}\rangle \,\equiv\,
\langle\,{\bf J}\cdot{\bf E}\,\rangle \,=\,
{1\over2}\,{\rm Re}\,\{{\bf J}\cdot{\bf E}^*\} \,=\,
{1\over2}\,{\rm Re}\,\Bigl\{{\bf E}^*\cdot\,
\bar{\!\mbox{\boldmath$\sigma$}}_c\cdot{\bf E}\Bigr\} \,=\,
{1\over2}\;\bar{\!\mbox{\boldmath$\sigma$}}_c:{\bf E}{\bf E}^* ,
\label{eq:2.15}   \\[.2cm]
\end{equation}
\begin{equation}
\langle\,{\rm I}_{\bf P}\rangle \,\equiv\,
{1\over2}\,\biggl\langle{\bf E}\cdot{\partial{\bf P}\over\partial t} -
{\bf P}\cdot{\partial{\bf E}\over\partial t} \biggr\rangle \,=
\label{eq:2.16}
\end{equation}
\[
=\,{1\over2}\,\biggl\langle{\bf E}\cdot{\partial{\bf D}\over\partial t} -
{\bf D}\cdot{\partial{\bf E}\over\partial t} \biggr\rangle \,=\,
{1\over4}\,{\rm Re}\,\Bigl\{ {\bf E}^*\cdot(i\omega{\bf D})-
{\bf D}^*\cdot(i\omega{\bf E}) \Bigr\} \,=
\]
\[
=\,{1\over4}\,{\rm Re}\,\Bigl\{ i\omega
\Bigl( \,{\bf E}^*\cdot\,\bar{\!\mbox{\boldmath$\epsilon$}}\cdot{\bf E} +
{\bf E}^*\cdot\,\bar{\!\mbox{\boldmath$\xi$}}\cdot{\bf H} -
{\bf E}\cdot\,\bar{\!\mbox{\boldmath$\epsilon$}}\,^*\!\cdot{\bf E}^* -
{\bf E}\cdot\,\bar{\!\mbox{\boldmath$\xi$}}\,^*\!\cdot{\bf H}^*
\Bigr) \Bigr\} \,=
\]
\[
={1\over4} {\rm Re} \Bigl\{ i\omega
\Bigl[ {\bf E}^*\!\cdot(\,\bar{\!\mbox{\boldmath$\epsilon$}} -
\bar{\!\mbox{\boldmath$\epsilon$}}\,^\dagger)\cdot{\bf E} +
2{\bf E}^*\!\cdot\bar{\!\mbox{\boldmath$\xi$}}\cdot{\bf H}
\Bigr] \Bigr\} =
{1\over2} {\rm Re} \Bigl\{ i\omega
\Bigl( \bar{\!\mbox{\boldmath$\epsilon$}}\,^a\!:
{\bf E}{\bf E}^* \!+\,
\bar{\!\mbox{\boldmath$\xi$}}\!:
{\bf H}{\bf E}^* \Bigr) \Bigr\} ,    \\[.3cm]
\]
\begin{equation}
\langle\,{\rm I}_{\bf M}\rangle \,\equiv\,
{1\over2}\,\biggl\langle{\bf H}\cdot{\partial\mu_0{\bf M}\over\partial t}-
\mu_0{\bf M}\cdot{\partial{\bf H}\over\partial t} \biggr\rangle \,=
\label{eq:2.17}
\end{equation}
\[
=\,{1\over2}\,\biggl\langle{\bf H}\cdot{\partial{\bf B}\over\partial t} -
{\bf B}\cdot{\partial{\bf H}\over\partial t} \biggr\rangle \,=\,
{1\over4}\,{\rm Re}\,\Bigl\{ {\bf H}^*\cdot(i\omega{\bf B})-
{\bf B}^*\cdot(i\omega{\bf H}) \Bigr\} \,=
\]
\[
=\,{1\over4}\,{\rm Re}\,\Bigl\{ i\omega
\Bigl( {\bf H}^*\cdot\,\bar{\!\mbox{\boldmath$\mu$}}\cdot{\bf H} +
{\bf H}^*\cdot\,\bar{\!\mbox{\boldmath$\zeta$}}\cdot{\bf E} -
{\bf H}\cdot\,\bar{\!\mbox{\boldmath$\mu$}}\,^*\!\cdot{\bf H}^* -
{\bf H}\cdot\,\bar{\!\mbox{\boldmath$\zeta$}}\,^*\!\cdot{\bf E}^*
\Bigr) \Bigr\} \,=
\]
\[
={1\over4} {\rm Re} \Bigl\{ i\omega
\Bigl[ {\bf H}^*\!\cdot(\,\bar{\!\mbox{\boldmath$\mu$}} -
\bar{\!\mbox{\boldmath$\mu$}}\,^\dagger)\cdot{\bf H} -
2{\bf E}^*\!\cdot\,\bar{\!\mbox{\boldmath$\zeta$}}\,^\dagger
\!\cdot{\bf H} \Bigr] \Bigr\} =
{1\over2} {\rm Re} \Bigl\{ i\omega
\Bigl( \bar{\!\mbox{\boldmath$\mu$}}\,^a\!:{\bf H}{\bf H}^* -\,
\bar{\!\mbox{\boldmath$\zeta$}}\,^\dagger\!:{\bf H}{\bf E}^*
\Bigr) \Bigr\} .                \\[.2cm]
\]

Substitution of Eqs.~(\ref{eq:2.15}) -- (\ref{eq:2.17}) into
Eq.~(\ref{eq:2.7}) gives the time-average Poynting theorem in the
following form
\begin{equation}
\mbox{\boldmath$\nabla$}\!\cdot\langle\,{\bf S}\,\rangle \,+\,
\langle\,q\,\rangle\,=\, 0
\label{eq:2.18}
\end{equation}
involving the average Poynting vector
\begin{equation}
\langle\,{\bf S}\,\rangle\,=\,
{1\over2}\,{\rm Re}\,\{{\bf E}\times{\bf H}^*\}
\label{eq:2.19}
\end{equation}
and the average power loss density
\[
\langle\,q\,\rangle\,=\,\langle\,{\rm I}_{\bf J}\rangle\,+\,
\langle\,{\rm I}_{\bf P}\rangle\,+\,\langle\,{\rm I}_{\bf M}\rangle\,=
\]
\begin{equation}
=\,{1\over2}\;\bar{\!\mbox{\boldmath$\sigma$}}_e:{\bf E}{\bf E}^* \,+\,
{1\over2}\;\bar{\!\mbox{\boldmath$\sigma$}}_m:{\bf H}{\bf H}^* \,+\,
{1\over2}\,{\rm Re}\,\{\,\bar{\!\mbox{\boldmath$\sigma$}}_{me}:
{\bf H}{\bf E}^* \}
\label{eq:2.20}        \\[.2cm]
\end{equation}
where we have introduced the total tensor of {\it electric
conductivity\/}
\begin{equation}
\bar{\!\mbox{\boldmath$\sigma$}}_e\,=\,
\bar{\!\mbox{\boldmath$\sigma$}}_c +\,
\bar{\!\mbox{\boldmath$\sigma$}}_d\,=\,
\bar{\!\mbox{\boldmath$\sigma$}}_c\,+\,
i\omega\,\bar{\!\mbox{\boldmath$\epsilon$}}\,^a \,\equiv\,
\bar{\!\mbox{\boldmath$\sigma$}}_c\,+\,
i\omega\,{\;\bar{\!\mbox{\boldmath$\epsilon$}}-\,
\bar{\!\mbox{\boldmath$\epsilon$}}\,^\dagger\over 2}
\label{eq:2.21}
\end{equation}
associated with conductor $(\,\bar{\!\mbox{\boldmath$\sigma$}}_c)$
and dielectric $(\,\bar{\!\mbox{\boldmath$\sigma$}}_d =
i\omega\,\bar{\!\mbox{\boldmath$\epsilon$}}\,^a)$
losses of a medium, \,the tensor of {\it magnetic conductivity\/}
\begin{equation}
\bar{\!\mbox{\boldmath$\sigma$}}_m\,=\,
i\omega\,\bar{\!\mbox{\boldmath$\mu$}}\,^a\,\equiv\,
i\omega\,{\;\bar{\!\mbox{\boldmath$\mu$}}-\,
\bar{\!\mbox{\boldmath$\mu$}}\,^\dagger\over 2}
\label{eq:2.22}
\end{equation}
associated with magnetic losses of a medium, \,and the tensor of
{\it magneto-electric conductivity\/}
\begin{equation}
\bar{\!\mbox{\boldmath$\sigma$}}_{me}\,=\,
i\omega(\,\bar{\!\mbox{\boldmath$\xi$}}-\,
\bar{\!\mbox{\boldmath$\zeta$}}\,^\dagger) \,\equiv\,
i\omega \Bigl[ (\,\bar{\!\mbox{\boldmath$\xi$}}\,^a +\,
\bar{\!\mbox{\boldmath$\zeta$}}\,^a)\,+\,
(\,\bar{\!\mbox{\boldmath$\xi$}}\,^h -\,
\bar{\!\mbox{\boldmath$\zeta$}}\,^h) \Bigr]
\label{eq:2.23}\\[.2cm]
\end{equation}
consisting of both antihermitian (with superscript $a$) and hermitian
(with superscript~$h$) parts of the cross susceptibilities
\,$\bar{\!\mbox{\boldmath$\xi$}}$ and $\bar{\!\mbox{\boldmath$\zeta$}}$.
Unlike $\bar{\!\mbox{\boldmath$\sigma$}}_{me}$, the dyadics
\,$\bar{\!\mbox{\boldmath$\sigma$}}_e\!=
\bar{\!\mbox{\boldmath$\sigma$}}_c+i\omega\,
\bar{\!\mbox{\boldmath$\epsilon$}}\,^a$ and
\,$\bar{\!\mbox{\boldmath$\sigma$}}_m\!=
i\omega\,\bar{\!\mbox{\boldmath$\mu$}}\,^a$ are
hermitian so that they produce the real (positive) definite quadratic
forms in Eq.~(\ref{eq:2.20}).

\subsection{Mode Power Transmission and Dissipation}
\label{sec:2B}
\indent

In order to obtain expressions for the power carried by modes along a
waveguiding structure involving complex (anisotropic and bianisotropic)
media and to find the dissipation of mode power it is necessary to
go from the time-average Poynting theorem in differential form
(\ref{eq:2.18}) to its integ\-ral form. For this purpose let us integrate
Eq.~(\ref{eq:2.18}) over the composite (multilayered) cross section
$S= \sum S_i$ formed from a few medium parts $S_i$ with interface
contours $L_i$ by using the two-dimensional divergence theorem
(e.\,g.,~see~Ref.~\cite{5}, p.\,150)
\begin{equation}
\int_{S_i} \mbox{\boldmath$\nabla$}\!\cdot{\bf A}\,dS=
{\partial\over\partial z} \int_{S_i} {\bf z}_0\cdot{\bf A}\,dS +
\oint_{L_i} {\bf n}_0\cdot{\bf A}\,dl
\label{eq:2.24}
\end{equation}
where ${\bf A}$ is the arbitrary field vector and ${\bf n}_0$ is the
{\it outward\/} unit vector normal to the contour $L_i$ and
perpendicular to the longitudinal unit vector ${\bf z}_0$.

Application of the integral relation~(\ref{eq:2.24}) to\,
$\mbox{\boldmath$\nabla$}\!\cdot\langle\,{\bf S}\,\rangle$\, gives
\begin{equation} \!\!\!
\int_S \mbox{\boldmath$\nabla$}\!\cdot\langle\,{\bf S}\,\rangle\,dS =
{\partial\over\partial z}
\int_S {\bf z}_0\cdot\langle\,{\bf S}\,\rangle\,dS -
\sum_i \oint_{L_i} \Bigl[ {\bf n}_i^{+}\!\cdot\langle\,{\bf S}^{+}\rangle+
{\bf n}_i^{-}\!\cdot\langle\,{\bf S}^{-}\rangle \Bigr]\,dl
\label{eq:2.25}
\end{equation}
where $\langle\,{\bf S}^\pm\rangle$ means values of the time-average
Poynting vector taken at points of contour $L_i$ lying on its
different sides marked by the {\it inward\/} (for either adjacent medium)
unit vectors {} ${\bf n}_i^\pm$. The parts of interfaces between two
adjacent nonconducting media do not contribute to the line integrals
in Eq.~(\ref{eq:2.25}) owing to continuity in tangential components of
the electric and magnetic fields. The only contribution may
appear from the parts of $L_i$ due to conducting surfaces on
which there is the known boundary condition~\cite{2,5,38}
\begin{equation}
{\bf E}_\tau = \bar{\bf Z}_s\cdot({\bf H}_\tau\times{\bf n}_i)
\label{eq:2.26}
\end{equation}
where ${\bf E}_\tau$ and ${\bf H}_\tau$ are the electric and magnetic
fields tangential to the surface and $\bar{\bf Z}_s$ is the surface
impedance tensor. For the special case of the isotropic metallic surface
with the conductivity $\sigma$ and the skin depth \,$\delta=
\sqrt{2/\omega\mu_0\sigma}$\,  we~have~\cite{2,5}
\begin{equation}
\bar{\bf Z}_s= (1+i){\cal R}_s\,\bar{\bf I} \qquad \mbox{where} \qquad
{\cal R}_s= 1/\sigma\delta= \sqrt{\omega\mu_0/2\sigma} \,.
\label{eq:2.27}
\end{equation}

In this case the integrand of the line integral in Eq.~(\ref{eq:2.25})
yields the surface loss power density $\langle\,q^\prime\,\rangle$ in
addition to the bulk loss power density $\langle\,q\,\rangle$ entering
into Poynting's theorem~(\ref{eq:2.18}).

The result of integrating Eq.~(\ref{eq:2.18}) over the cross
section $S$ of a waveguide and apply\-ing Eqs.~(\ref{eq:2.19}),
(\ref{eq:2.20}), and (\ref{eq:2.25}) -- (\ref{eq:2.27}) gives
Poynting's theorem in the integral~form
\begin{equation}
{dP\over dz}\,+\,Q\,=\,0
\label{eq:2.28}
\end{equation}
where the total real power carried by electromagnetic fields in the
direction of increasing coordinate $z$ is equal to
\begin{equation}
P= \int_S \langle\,{\bf S}\,\rangle\cdot{\bf z}_0\,dS =
{1\over2}\,{\rm Re} \int_S ({\bf E}\times{\bf H}^*)\cdot{\bf z}_0\,dS
\label{eq:2.29}
\end{equation}
and the total power loss per unit length caused by the bulk losses
$Q^{(b)}$ (obtained by integrating $\langle\,q\,\rangle$ over
$S= \sum S_i$) and the surface (skin) losses $Q^{(s)}$ (obtained by
integrating $\langle\,q^\prime\,\rangle$ along $L=\sum L_i$) is equal to
\[
Q\,=\, Q^{(b)}\,+\,Q^{(s)} = \int_S \langle\,q\,\rangle\,dS \,+
\int_L \langle\,q^\prime\,\rangle\,dl \,=
\]
\[
=\,{1\over2}\int_S(\,\bar{\!\mbox{\boldmath$\sigma$}}_e:
{\bf E}{\bf E}^*)\,dS+\,
{1\over2} \int_S(\,\bar{\!\mbox{\boldmath$\sigma$}}_m :
{\bf H}{\bf H}^*)\,dS\,+        \\[.15cm]
\]
\begin{equation}
+\;{1\over2}\,{\rm Re} \int_S(\,\bar{\!\mbox{\boldmath$\sigma$}}_{me}:
{\bf H}{\bf E}^*)\,dS +\,
{1\over2} \int_L {\cal R}_s\,({\bf H}_\tau\cdot{\bf H}_\tau^*)\,dl  \,.
\label{eq:2.30} \\[.2cm]
\end{equation}

Power relation~(\ref{eq:2.28}) is valid only for the source-free
region of a wave\-guiding structure whose electromagnetic fields
can be expanded in terms of its eigenmodes (cf.~Eqs.~(\ref{eq:A29}))
\begin{eqnarray}
{\bf E}({\bf r}_t,z)  \!\!&=&\!\!
\sum_k A_k\,\hat{\bf E}_k({\bf r}_t)\,{\rm e}^{-\,\gamma_kz} =\,
\sum_k a_k(z)\,\hat{\bf E}_k({\bf r}_t) \,,
\label{eq:2.31}\\
{\bf H}({\bf r}_t,z)  \!\!\!&=&\!\!\!
\sum_k A_k\,\hat{\bf H}_k({\bf r}_t)\,{\rm e}^{-\,\gamma_kz} =\,
\sum_k a_k(z)\,\hat{\bf H}_k({\bf r}_t) \,.
\label{eq:2.32}
\end{eqnarray}

Every $k$th mode is specified by the propagation constant
$\gamma_k\!= \alpha_k+ i\beta_k$\, and the eigenfunctions $\{\hat{\bf E}_k,
\,\hat{\bf H}_k\}$\, (where the hat sign over field vectors
implies their dependence only on transverse coordinates ${\bf r}_t$,\,
see Appendix\,A.\,2), which are regarded as known quantities found from
solving the appropriate boundary-value problem. The amplitudes $A_k$ are
determined by the exciting sources and called the
{\it excitation amplitudes\/}. Inside the source region they depend
on $z$\, as a result of source actions but for the source-free
region $A_k(z)= const.$,\, as in the case of Eqs.~(\ref{eq:2.31})
and (\ref{eq:2.32}). It is often convenient instead of $A_k(z)$ to
introduce the {\it mode~amplitudes\/}
\begin{equation}
a_k(z)= A_k(z)\,{\rm e}^{-\,\gamma_kz}
\label{eq:2.33}
\end{equation}
which take into account the total $z$-dependence related both to the
mode propagation $(\exp(-\gamma_kz))$ and to the exciting sources
$(A_k(z))$,\, if any.     \\[.1cm]
\indent
Let us employ the modal expansions~(\ref{eq:2.31}) and (\ref{eq:2.32})
to calculate the power flow $P(z)$ and the power loss $Q(z)$ given by
Eqs.~(\ref{eq:2.29}) and (\ref{eq:2.30}) for the source-free region.
The final result of calculations is the following:
\[
P(z)= {1\over4} \int_S ({\bf E}^*\times{\bf H}+
{\bf E}\times{\bf H}^*)\cdot{\bf z}_0\,dS\,=
\]
\begin{equation}
=\,{1\over4}\sum_k\sum_l\,N_{kl}\,a_k^*(z) a_l(z)
\equiv\sum_k\sum_l P_{kl}(z)  \,,
\label{eq:2.34}   \\[.2cm]
\end{equation}
\[
Q(z)= {1\over2}\int_S
({\bf E}^*\cdot\,\bar{\!\mbox{\boldmath$\sigma$}}_e\cdot{\bf E})\,dS+\,
{1\over2}\int_S
({\bf H}^*\cdot\,\bar{\!\mbox{\boldmath$\sigma$}}_m\cdot{\bf H})\,dS\,+
\]
\[
+\,{1\over4} \int_S
({\bf E}^*\cdot\,\bar{\!\mbox{\boldmath$\sigma$}}_{me}\cdot{\bf H} +
{\bf H}^*\cdot\,\bar{\!\mbox{\boldmath$\sigma$}}_{me}^{\,\dagger}
\cdot{\bf E})\,dS
+\,{1\over2} \int_L {\cal R}_s\,({\bf H}_\tau^*\cdot{\bf H}_\tau)\,dl\,=
\]
\begin{equation}
=\,{1\over4}\sum_k\sum_l\,M_{kl}\,a_k^*(z) a_l(z)
\equiv\sum_k\sum_l Q_{kl}(z)
\label{eq:2.35} \\[.2cm]
\end{equation}
where we have introduced the normalizing coefficients
\begin{equation}
N_{kl}= \int_S (\hat{\bf E}_k^*\times\hat{\bf H}_l +
\hat{\bf E}_l\times\hat{\bf H}_k^*)\cdot{\bf z}_0\,dS
\label{eq:2.36}
\end{equation}
and the dissipative coefficients
\[
M_{kl}= 2\int_S (\hat{\bf E}_k^*\cdot\,\bar{\!\mbox{\boldmath$\sigma$}}_e
\cdot\hat{\bf E}_l)\,dS+\,
2\int_S(\hat{\bf H}_k^*\cdot\,\bar{\!\mbox{\boldmath$\sigma$}}_m
\cdot\hat{\bf H}_l)\,dS\,+
\]
\begin{equation}
+ \int_S (\hat{\bf E}_k^*\cdot\,\bar{\!\mbox{\boldmath$\sigma$}}_{me}
\cdot\hat{\bf H}_l +
\hat{\bf H}_k^*\cdot\,\bar{\!\mbox{\boldmath$\sigma$}}_{me}^{\,\dagger}
\cdot\hat{\bf E}_l)\,dS
+\,2\int_L {\cal R}_s\,(\hat{\bf H}_{\tau,k}^*
\cdot\hat{\bf H}_{\tau,l})\,dl
\label{eq:2.37} \\[.2cm]
\end{equation}
constructed of the cross-section eigenfield vectors (marked with
the hat sign above them,\, see Eq.~(\ref{eq:3.5})).

From Eqs.~(\ref{eq:2.36}) and (\ref{eq:2.37}) it follows that
the matrices $\{N_{kl}\}$ and $\{M_{kl}\}$ are hermitian, that is
\begin{equation}
N_{kl}= N_{lk}^*  \qquad \mbox{and} \qquad M_{kl}= M_{lk}^* \,,
\label{eq:2.38}
\end{equation}
and have dimensions of watts and watts per meter, respectively, because
the amplitudes $A_k$ and $a_k$ are dimensionless.

The quantities $P_k(z)$ and $Q_k(z)$ appearing in Eqs.~(\ref{eq:2.34})
and (\ref{eq:2.35}) for $l= k$ in the following form
\begin{equation}
P_k(z)\equiv P_{kk}(z)= {1\over4}\,N_{kk}\,a_k^*(z)a_k(z)=
{1\over4}\,N_k\,|A_k|^2\,{\rm e}^{-2\alpha_kz}     \;\,
\label{eq:2.39}            \\[.1cm]
\end{equation}
\begin{equation}
Q_k(z)\!\equiv\! Q_{kk}(z)= {1\over4}\,M_{kk}\,a_k^*(z)a_k(z)=
{1\over4}\,M_k\,|A_k|^2\,{\rm e}^{-2\alpha_kz}
\label{eq:2.40}            \\[.2cm]
\end{equation}
are the real {\it self powers\/} transmitted and dissipated
at point $z$ by the $k$th mode which was excited at point $z= 0$ with
amplitude $A_k$.

Similarly, the quantities $P_{kl}(z)$ and $Q_{kl}(z)$ for $l\ne k$
equal to
\begin{equation}
P_{kl}(z)= {1\over4}\,N_{kl}\,a_k^*(z)a_l(z)=
{1\over4}\,N_{kl}\,A_k^*A_l\,{\rm e}^{-\,(\gamma_k^* \,+\,\gamma_l)z}
\label{eq:2.41}       \\[.1cm]
\end{equation}
\begin{equation}
Q_{kl}(z)= {1\over4}\,M_{kl}\,a_k^*(z)a_l(z)=
{1\over4}\,M_{kl}\,A_k^*A_l\,{\rm e}^{-\,(\gamma_k^* \,+\,\gamma_l)z}
\label{eq:2.42}        \\[.15cm]
\end{equation}
can be interpreted as the complex {\it cross powers\/} transmitted and
dissipated at point $z$ jointly by the $k$th and $l$th modes which
were excited at point $z= 0$ with amplitudes $A_k$ and $A_l$. Owing to
(\ref{eq:2.38}), the quantities defined by Eqs.~(\ref{eq:2.41}) and
(\ref{eq:2.42}) are also hermitian:
\begin{equation}
P_{kl}(z)= P_{lk}^*(z)  \qquad \mbox{and} \qquad
Q_{kl}(z)= Q_{lk}^*(z) \,.
\label{eq:2.43}
\end{equation}

From Eqs.~(\ref{eq:2.41}) through (\ref{eq:2.43}) it follows that in
a lossy waveguiding structure every pair of modes always transmits and
dissipates the real {\it combined cross powers\/}
\begin{equation}
P_{kl}^c(z)\equiv P_{kl}(z)+ P_{lk}(z)= 2\,{\rm Re}\,P_{kl}(z) =
{1\over2}\,{\rm Re}\,\{N_{kl}\,a_k^*(z)a_l(z)\}  \,,
\label{eq:2.44}
\end{equation}
\begin{equation}
Q_{kl}^c(z)\equiv Q_{kl}(z)+ Q_{lk}(z)= 2\,{\rm Re}\,Q_{kl}(z) =
{1\over2}\,{\rm Re}\,\{M_{kl}\,a_k^*(z)a_l(z)\} .  \!\!\!
\label{eq:2.45} \\[.2cm]
\end{equation}

Therefore, the double sums in Eqs.~(\ref{eq:2.34}) and (\ref{eq:2.35})
yield the real (time-average) {\it total powers\/} transmitted and
dissipated by all modes in a lossy waveguide:
\[
P(z)= \sum_k\sum_l P_{kl}(z)= \sum_k P_k(z) +
\sum_k\sum_{l\ne k} P_{kl}^c(z)\,=
\]
\begin{equation}
=\,{1\over4} \sum_k N_k\,|a_k(z)|^2 +\,
{1\over2}\,{\rm Re}\sum_k\sum_{l>k} N_{kl}\,a_k^*(z)a_l(z)  \,,
\label{eq:2.46}        \\[.2cm]
\end{equation}
\[
Q(z)= \sum_k\sum_l Q_{kl}(z)= \sum_k Q_k(z) +
\sum_k\sum_{l\ne k} Q_{kl}^c(z)\,=
\]
\begin{equation}
=\,{1\over4} \sum_k M_k\,|a_k(z)|^2 +\,
{1\over2}\,{\rm Re}\sum_k\sum_{l>k} M_{kl}\,a_k^*(z)a_l(z) \,,
\label{eq:2.47} \\[.2cm]
\end{equation}
where {} $N_k\equiv N_{kk}$ and {} $M_k\equiv M_{kk}$.  \\[.1cm]
\indent      In the next section we shall derive a relation named the
{\it quasi-orthogonality\/} relation to link the cross powers $P_{kl}$
and $Q_{kl}$ for every pair of modes in a lossy waveguide or for every
pair of the so-called {\it twin-conjugate\/} modes in a lossless
waveguide.

\setcounter{equation}{0}
\section{ORTHOGONALITY\, AND\, QUASI-ORTHOGONALITY\, OF MODES
                           IN LOSSLESS AND LOSSY WAVEGUIDES}
\label{sec:3}

\subsection{Quasi-orthogonality Relation for Lossy Waveguides}
\label{sec:3A}
\indent

Let us begin our examination with the general case of the composite
(multilayered) wave\-guiding structure containing bianisotropic media
with bulk (electric, magnetic, magneto-electric) losses and surface
(skin) losses. Consider the $k$th and $l$th modes propagating in the
source-free region of the waveguide which obey the curl Maxwell
equations~(\ref{eq:2.1}) rewritten by using the constitutive
relations~(\ref{eq:2.8}), (\ref{eq:2.9}), and (\ref{eq:2.14}) in the
following form
\begin{eqnarray}
\mbox{\boldmath$\nabla$}\times{\bf E}_{k(l)} \!&=&\!
-\,i\omega\,\bar{\!\mbox{\boldmath$\mu$}}\cdot{\bf H}_{k(l)}\,-\,
i\omega\,\bar{\!\mbox{\boldmath$\zeta$}}\cdot{\bf E}_{k(l)}\;,
\label{eq:3.1}\\    [.25cm]
\mbox{\boldmath$\nabla$}\times{\bf H}_{k(l)} \!&=&\!
(\,\bar{\!\mbox{\boldmath$\sigma$}}_c+
i\omega\,\bar{\!\mbox{\boldmath$\epsilon$}}\,)\cdot{\bf E}_{k(l)}\,+\,
i\omega\,\bar{\!\mbox{\boldmath$\xi$}}\cdot{\bf H}_{k(l)} \,.
\label{eq:3.2}
\end{eqnarray}
A conventional procedure applied to Eqs.~(\ref{eq:3.1}) and
(\ref{eq:3.2}) gives
\[
\mbox{\boldmath$\nabla$}\cdot({\bf E}_k^*\times{\bf H}_l +
{\bf E}_l\times{\bf H}_k^*)\,=
\]
\begin{equation}
=\,-\,2\,{\bf E}_k^*\cdot\,\bar{\!\mbox{\boldmath$\sigma$}}_e
\cdot{\bf E}_l -
2\,{\bf H}_k^*\cdot\,\bar{\!\mbox{\boldmath$\sigma$}}_m
\cdot{\bf H}_l -
(\,{\bf E}_k^*\cdot\,\bar{\!\mbox{\boldmath$\sigma$}}_{me}
\cdot{\bf H}_l +
\,{\bf H}_k^*\cdot\,\bar{\!\mbox{\boldmath$\sigma$}}_{me}^{\,\dagger}
\cdot{\bf E}_l)
\label{eq:3.3}        \\[.2cm]
\end{equation}
where we have used formulas~(\ref{eq:2.21}) -- (\ref{eq:2.23}). \\[.1cm]
\indent
Application of the two-dimensional divergence theorem~(\ref{eq:2.24})
to the left-hand side of Eq.~(\ref{eq:3.3}), by analogy with formula
(\ref{eq:2.25}) and by using the boundary condition~(\ref{eq:2.26}),
results in the following expression
\[
{\partial\over\partial z} \int_S ({\bf E}_k^*\times{\bf H}_l +
{\bf E}_l\times{\bf H}_k^*)\cdot{\bf z}_0\,dS\,=   \\[-0.1cm]
\]
\[
=\,-\,2\int_S ({\bf E}_k^*\cdot\,\bar{\!\mbox{\boldmath$\sigma$}}_e
\cdot{\bf E}_l)\,dS-\,
2\int_S({\bf H}_k^*\cdot\,\bar{\!\mbox{\boldmath$\sigma$}}_m
\cdot{\bf H}_l)\,dS\,-
\]
\begin{equation}
- \int_S ({\bf E}_k^*\cdot\,\bar{\!\mbox{\boldmath$\sigma$}}_{me}
\cdot{\bf H}_l +
{\bf H}_k^*\cdot\,\bar{\!\mbox{\boldmath$\sigma$}}_{me}^{\,\dagger}
\cdot{\bf E}_l)\,dS -\,
2\int_L {\cal R}_s\,({\bf H}_{\tau,k}^*
\cdot{\bf H}_{\tau,l})\,dl \,.
\label{eq:3.4} \\[.2cm]
\end{equation}

Representation of the fields for the $k$th and $l$th modes in the form
\begin{equation}
{\bf E}_{k(l)}({\bf r}_t,z) = \hat{\bf E}_{k(l)}({\bf r}_t)\,
{\rm e}^{-\,\gamma_{k(l)}z},      \;\quad
{\bf H}_{k(l)}({\bf r}_t,z) = \hat{\bf H}_{k(l)}({\bf r}_t)\,
{\rm e}^{-\,\gamma_{k(l)}z}
\label{eq:3.5}
\end{equation}
and their substitution into Eq.~(\ref{eq:3.4}) give, by comparing with
formulas~(\ref{eq:2.36}) and (\ref{eq:2.37}) for $N_{kl}$ and $M_{kl}$,
the desired relation
\begin{equation}
(\gamma_k^* +\gamma_l)\,N_{kl}=\,M_{kl}
\label{eq:3.6}
\end{equation}
referred to as the {\it quasi-orthogonality relation\/}. It will
play the same role in deriving the excitation equations (see
Eq.~(\ref{eq:5.46})) as the ordinary orthogonality~relation.

Expressions~(\ref{eq:2.41}) and (\ref{eq:2.42}) relate the normalizing
and dissipative coefficients $N_{kl}$ and $M_{kl}$ to the cross-power
flow $P_{kl}$ and the cross-power loss $Q_{kl}$, respectively,
transmitted and dissipated jointly by the $k$th and $l$th modes. The use
of these expressions allows us to rewrite the quasi-orthogonality
relation~(\ref{eq:3.6}) in the~power~form
\begin{equation}
(\gamma_k^* +\gamma_l)\,P_{kl} =\,Q_{kl} \,.
\label{eq:3.6a}
\end{equation}

This formulation provides the following power interpretation of the mode
quasi-ortho\-gonality:\, outside the source region every pair of modes,
independently of other modes, transmits the complex cross-power flow
$P_{kl}$ rigidly coupled to the complex cross-power loss $Q_{kl}$ by
the factor $(\gamma_k^* +\gamma_l)$ consisting of the mode propagation
constants $\gamma_{k(l)}= \alpha_{k(l)}+ i\beta_{k(l)}\,,$\, and in doing
so the combined cross powers $P_{kl}^c= P_{kl}+ P_{lk}$ and $Q_{kl}^c=
Q_{kl}+ Q_{lk}$ always remain real. The quasi-orthogonality relation
(\ref{eq:3.6a}) means that outside the source region Poynting's
theorem~(\ref{eq:2.28}) takes place for any one of mode pairs~$(k,l)$:
\begin{equation}
{dP_{kl}\over dz}\,+\,Q_{kl}\,=\,0 \,.
\label{eq:3.7}
\end{equation}

Besides, every single mode has the real self-power flow $P_k\equiv P_{kk}$
and self-power loss $Q_k\equiv Q_{kk}$ in the form of Eqs.~(\ref{eq:2.39})
and Eq.~(\ref{eq:2.40}). These self powers are coupled to each other by
the same relations~(\ref{eq:3.6a}) and (\ref{eq:3.7}) which for \,$l= k$\,
yield the following expression for the attenuation constant:
\begin{equation}
\alpha_k =\,{Q_k\over2\,P_k} =\,{M_k\over2\,N_k} \,.
\label{eq:3.8}
\end{equation}

Hence, there occurs the following pattern of mode power transfer in the
lossy wave\-guiding structures. Every $k$th mode propagates from the source
region with the fixed value of amplitude $A_k$ (the loss attenuation is
taken into account by the amplitude constant $\alpha_k$ appearing in
$\gamma_k$) which was excited by the sources. Outside them the mode, being
a linearly independent solution to the boundary-value problem, does not
interact with other modes owing to their linear independence. The $k$th
mode transfers the self power $P_k$ on its own and the cross powers
$P_{kl}$ in conjunction with the other $l$th modes which were also
excited inside the source region and outside retain constant their
excitation amplitudes $A_l$ as well as the $k$th mode.

\subsection{Mode Orthogonality in Lossless Waveguides}
\label{sec:3B}
\indent

The orthogonality relation for a lossless waveguiding structure is
obtained from the general relation~(\ref{eq:3.6}) as the special
case of $M_{kl}= 0$ and has the following form
\begin{equation}
(\gamma_k^* +\gamma_l)\,N_{kl}=\,0 \,.
\label{eq:3.9}
\end{equation}

In spite of the absence of dissipation, in the eigenmode spectrum
of the lossless waveguide, besides propagating modes with
$\alpha_k\equiv 0$ and $\gamma_k= i\beta_k$, there are also modes
having complex values of the propagation constant $\gamma_k= \alpha_k+
i\beta_k$ with $\alpha_k\ne 0$. These modes exist in the cutoff regime
of propagation and their attenuation is of reactive (nondissipative)
character associated with the storage of reactive power. As a token of
this, it seems reasonable to refer to such modes as the
{\it reactive\/} modes to distinguish between them and the
{\it active\/} (propagating) modes carrying an active~(real)~power.

In the literature the reactive (in our terminology) modes are variously
termed the complex, cutoff, nonpropagating, and evanescent modes.
The last term is usually assigned only to cutoff modes with pure decay
($\alpha_k\ne 0$) and without phase delay ($\beta_k= 0$). The latter
feature of evanescent modes makes appropriate for them also the term
"nonpropagating" because there is no phase propagation. But for
the complex modes with $\alpha_k\ne 0$ and $\beta_k\ne 0$ their reactive
decay as $\exp(-\,\alpha_kz)$ is accompanied by the phase variation in
accordance with the wave factor $\exp[i(\omega t- \beta_kz)]$. For this
reason it is more preferable to refer to the complex modes as reactive
modes rather than nonpropagating ones. However, we shall apply both
terms, the {\it reactive\/} and {\it nonpropagating\/} modes, as well as
their antitheses, the {\it active\/} and {\it propagating\/} modes,
to reflect the fact that the former do not transfer any self power,
whereas the latter carry it.

Let us show that in any lossless waveguide, independently of its
structure and media used, every reactive mode with number $k$ has its
own {\it twin\/} mode with number $\tilde k$ (marked by tilde) so that
their propagation constants are related~by~the~equality
\begin{equation}
\gamma_{\tilde k}= -\,\gamma_k^*   \qquad \mbox{or} \qquad
\alpha_{\tilde k}= -\,\alpha_k \;, \quad
\beta_{\tilde k}= \beta_k \,.
\label{eq:3.10}
\end{equation}

Such mode twins with pair of numbers $(k,\,\tilde k)$ that satisfy
the relation (\ref{eq:3.10}) will be referred to as the
{\it twin-conjugate\/} modes. As is seen from Eq.~(\ref{eq:3.10}),
these modes have the same phase velocity ($\beta_k =\beta_{\tilde k}$)
but decay in opposite directions $(\alpha_k = -\,\alpha_{\tilde k})$.

The existence of twin-conjugate modes possessing the property
expressed by Eq.~(\ref{eq:3.10}) can be justified by means of the
following reasoning. In our treatment of complex amplitude technique,
we have chosen the wave factor in the form {}
$\exp[i(\omega t -{\bf k}\cdot{\bf r})]$ where $k_z\!\equiv\!-\,i\,\gamma$.
However, there is another alternative form \,$\exp[-i(\omega t-
\tilde{\bf k}\cdot{\bf r})]$ with $\tilde{k}_z\!\equiv\!-\,i\,\tilde\gamma$
which differs from the first form in opposite sign of imaginary unity
and having tilde above the wave vector. It is clearly evident that
the alternative case can be obtained from our solution by applying
complex conjugation, then
\begin{equation}
\tilde{k}_z\equiv -\,i\tilde\gamma\,=\,k_z^*\equiv\,i\gamma^*
\qquad \mbox{whence} \qquad  \tilde\gamma= -\,\gamma^* .
\label{eq:3.11}
\end{equation}

Equalities for gammas in Eqs.~(\ref{eq:3.10}) and (\ref{eq:3.11})
are fully coincident not counting different positions of tilde (the
former marking the mode number~$\tilde k$ in subscripts will be used
later on). This result substantiates the existence of twin-conjugate
modes for which \,$\gamma_k\!=-\,\gamma_{\tilde k}^*$\, or \,$k_{z,k}\!=
k_{z,\tilde k}^*$. In other words, any dispersion equation obtained as
a result of solving the boundary-value problem for lossless systems
has the complex roots with complex-conjugate values of the longitudinal
wavenumber $k_z$ which appear in pairs. Such a pair of complex roots
corresponds to the twin-conjugate modes.   \\[.1cm]
\indent
Sign of the amplitude constant $\alpha_k$ can be used as the basis for
classification of the reactive (nonpropagating) modes under two types,
{\it forward\/} and {\it backward\/}, as is usually done for the active
(propagating) modes but on the basis of a sign of the group velocity
\,$v_{gr,k}=[d\beta_k(\omega)/d\omega]^{-1}$. In reference to the source
region location between $z= 0$ and $z= L$, all the modes (active and
reactive) can be classified into two categories:\\
(i) the {\it forward\/} modes marked by subscript $k=+\,n\!>\!0$
(active with $v_{gr,+n}\!>\!0$ or reactive with $\alpha_{+n}\!>\!0$)
which, being excited inside the source region, leave it (without or with
reactive damping) across the right boundary and exist outside at $z> L$;\\
(ii) the {\it backward\/} modes marked by subscript $k=-\,n\!<\!0$
(active with $v_{gr,-n}\!<\!0$ or reactive with $\alpha_{-n}\!<\!0$)
which, being excited inside the source region, leave it (without or with
reactive damping) across the left boundary and exist outside at $z< 0$.

\subsubsection{Orthogonality and Normalization Relations
                      for Active Modes}
\label{sec:3B1}
\indent

The active (propagating) modes exist in the pass band of lossless
waveguides where they have zero amplitude attenuation ($\alpha_{k(l)}= 0$),
so that their propagation constants $\gamma_{k(l)}= i\beta_{k(l)}$
are pure imaginary. In this case the orthogonality relation~(\ref{eq:3.9})
rewritten in the form
\begin{equation}
(\beta_k- \beta_l)\,N_{kl}=\,0
\label{eq:3.12}
\end{equation}
along with expression~(\ref{eq:2.36}) for $N_{kl}$ furnishes two
alternatives:
\begin{equation}
N_{kl}\equiv \int_S (\hat{\bf E}_k^*\times\hat{\bf H}_l +
\hat{\bf E}_l\times\hat{\bf H}_k^*)\cdot{\bf z}_0\,dS = 0
\qquad \mbox{for} \qquad    l\ne k
\label{eq:3.13}
\end{equation}
or
\begin{equation}
N_k\equiv N_{kk}= 2\,{\rm Re} \int_S
(\hat{\bf E}_k^*\times\hat{\bf H}_k)\cdot{\bf z}_0\,dS \ne\,0
\qquad \mbox{for} \qquad    l= k  \,.
\label{eq:3.14}
\end{equation}

Expression (\ref{eq:3.13}) is the orthogonality relation between the
different pro\-pagating modes for which $\beta_k- \beta_l\ne 0$,
whereas formula~(\ref{eq:3.14}) defines the {\it norm\/}
$N_k\equiv N_{kk}$ of the $k$th mode. It should be noted that
Eq.~(\ref{eq:3.13}) does not necessarily hold for different but
degenerate modes with $\beta_k= \beta_l$. In this case one can employ
the conventional technique commonly used for usual
waveguides~\cite{2,3} to ensure the orthogonality among degenerate
modes by constructing from them such linear combinations that
constitute a new orthogonal subset for which relation~(\ref{eq:3.13})
is applicable. For this reason we shall no longer turn special
attention to degenerate modes.

From Eqs.~(\ref{eq:2.41}) and (\ref{eq:3.13}) it follows that two
different propagating modes (with numbers $k\ne l$) have zero
cross-power flow $(P_{kl}= 0)$, i.\,e., they are orthogonal in power
sense. Any mode carries along a waveguide only the self power $P_k$
defined by formula~(\ref{eq:2.39}), which gives the following power
interpretation for the norm of active modes:\, $N_k$ is equal
to $4P_k^{\rm o}$ where $P_k^{\rm o}$ means the time-average
power carried in the posi\-tive $z$-direction by the $k$th mode with
unit amplitude ($|A_k|= 1$). In some instances it may be
more convenient to normalize the mode amplitude to unit power
($|P_k^{\rm o}|= 1$ watt). Then $|N_k|= 4$ watts and according to
Eq.~(\ref{eq:2.39})
\begin{equation}
P_k= \pm\,|a_k|^2= \pm\,|A_k|^2
\label{eq:3.15}
\end{equation}
where subscripts should be read as $k\!= \pm\,n$,\, with upper and
lower signs corresponding to the forward and backward modes for which,
respectively, $N_{+n}\!=\!4$ watts and $N_{-n}\!=\!-\,4$ watts. In~the
special case of a reciprocal waveguide wherein for every forward mode
there is a backward one with the same law of dispersion, their norms
are related to each other by the equality
\begin{equation}
N_{+n}= -\,N_{-n}  \,.
\label{eq:3.16}
\end{equation}

In conclusion, let us write the relation of orthonormalization for the
active (propagating) modes in the following form
\begin{equation}
N_{kl}= N_{kk}\,\delta_{kl}\equiv \left\{
\begin{array}{cl} 0 & \qquad\mbox{for}\qquad  l\ne k \;,\\
\!\!N_{kk}\equiv N_k    & \qquad\mbox{for}\qquad  l = k   \;,
\end{array}  \right.
\label{eq:3.17}
\end{equation}
where the normalizing coefficient $N_{kl}$ and the norm $N_k$ are
given by Eqs.~(\ref{eq:3.13}) and (\ref{eq:3.14}), respectively.

\subsubsection{Orthogonality and Normalization Relations
                      for Reactive Modes}
\label{sec:3B2}
\indent

The reactive modes of a lossless waveguide are cutoff modes whose
propa\-gation constants \,$\gamma_{k(l)}=\alpha_{k(l)}+ i\beta_{k(l)}$\,
are generally complex-valued or parti\-cularly real-valued for the
evanescent modes with $\beta_{k(l)}= 0$. So the general relation of
orthogonality~(\ref{eq:3.9})~holds for them and by using
Eq.~(\ref{eq:3.10}) for the twin-conjugate modes gives two alternatives:
\begin{equation}
N_{kl}\equiv \int_S (\hat{\bf E}_k^*\times\hat{\bf H}_l +
\hat{\bf E}_l\times\hat{\bf H}_k^*)\cdot{\bf z}_0\,dS = 0
\qquad \mbox{for} \qquad      l\ne {\tilde k}
\label{eq:3.18}
\end{equation}
or
\begin{equation}
N_k\equiv N_{k\tilde k}=
\int_S (\hat{\bf E}_k^*\times\hat{\bf H}_{\tilde k} +
\hat{\bf E}_{\tilde k}\times\hat{\bf H}_k^*)\cdot{\bf z}_0\,dS \ne 0
\qquad \mbox{for} \qquad      l= {\tilde k}  \,.
\label{eq:3.19}
\end{equation}

Expression (\ref{eq:3.18}) fulfils a role of the orthogonality relation
for the reactive modes. As is seen from here, every reactive $k$th mode
is orthogonal to all the $l$th modes (reactive with $\alpha_l\ne 0$ and
active with $\alpha_l= 0$) for which $\gamma_l +\gamma_k^*=\gamma_l -
\gamma_{\tilde k}\ne 0$ and $N_{kl}= 0$, including itself since
$\gamma_k +\gamma_k^* = 2\alpha_k\ne 0$ and $N_{kk}= 0$. The only mode
nonorthogonal to the given $k$th mode is its own twin with number
$\tilde k$ for which $\gamma_{\tilde k} +\gamma_k^* = 0$.
Formula~(\ref{eq:3.19}) defines the {\it norm} $N_k\equiv N_{k\tilde k}$
\,for the reactive $k$th mode which is constructed of the fields of
twin-conjudate modes $(k,\,\tilde k)$.

From Eqs.~(\ref{eq:2.39}), (\ref{eq:2.41}), and (\ref{eq:3.18}) it
follows that every reactive mode has no both the self power
($P_{kk}=0$) and the cross powers with the other modes $(P_{kl}= 0)$
for which $\gamma_l\!\ne\!-\,\gamma_k^*$,\, i.\,e., these modes are
orthogonal in power sense.

Each mode forming a twin-conjugate pair, being nonorthogonal to its
twin, has its own norm defined by Eq.~(\ref{eq:3.19}) as
\begin{equation}
N_k\equiv N_{k\tilde k}    \qquad \mbox{or} \qquad
N_{\tilde k}\equiv N_{{\tilde k}k} \,,
\label{eq:3.20}
\end{equation}
whence, according to the general property of hermitian symmetry for
the normalizing coefficients expressed by equality~(\ref{eq:2.38}),
it follows that
\begin{equation}
N_k=\,N_{\tilde k}^* \,,
\label{eq:3.21}
\end{equation}
i.\,e., the reactive twin-conjugate modes have the complex-conjugate norms.

Although the reactive mode has no self power $(P_{kk}\equiv 0)$, the
twin-conjugate modes in pair carry the real combined cross power
(cf.~Eq.~(\ref{eq:2.44}))
\begin{equation}
P_{k\tilde k}^c\equiv P_{k\tilde k} + P_{{\tilde k}k}=
2\,{\rm Re}\,P_{k\tilde k} =
{1\over2}\,{\rm Re}\,\{N_k\,A_k^*A_{\tilde k}\} \equiv
{1\over2}\,{\rm Re}\,\{N_{\tilde k}\,A_{\tilde k}^*A_k\}
\label{eq:3.22}
\end{equation}
where subscripts should be read as $k\!=\!+\,n$ and $\tilde k\!=\!-\,n$.
This is a consequence of relations~(\ref{eq:3.10}) for the twin-conjugate
modes and the definition of forward and backward reactive modes:
if the $k$th mode is a forward one with $N_k= N_{+n}
\equiv N_{+n,-n}$, then the $\tilde k$th mode is a backward one with
$N_{\tilde k}= N_{-n}\equiv N_{-n,+n}= N_{+n}^*$.

As evident from Eq.~(\ref{eq:3.22}), to transfer the real power by
reactive modes it is necessary that both constituents of a twin-conjugate
pair should have nonzero amplitudes ($A_k$ and $A_{\tilde k}$) and to be
in such a phase relationship that their combined cross power
$P_{k\tilde k}^c$ would be other than zero. Similar situation usually
takes place in the regular waveguide of finite length bounded by two
irregularities and excited at frequences below its cutoff
frequency~[1--3]. Reflections from these irregularities
can form inside this length two evanescent (cutoff) modes with
numbers $k=+\,n$ (forward mode) and $\tilde k=-\,n$ (backward mode)
constituting the twin-conjugate pair for which $\beta_{\pm n}\!= 0$ and
$\gamma_{+n}\!\equiv\alpha_{+n}\!=-\,\alpha_{-n}\!\equiv-\,\gamma_{-n}^*$.
It is easy to see that the norms for the forward and backward evanescent
modes are pure imaginary-valued and related to each other by the general
relation~(\ref{eq:3.21}). Superposition of fields for the two evanescent
modes with opposite decay sense furnishes nonzero real cross-power flow
along a short length of the cutoff waveguide.   \\[.1cm]
\indent
In conclusion, let us write the relation of orthonormalization for the
reactive (nonpropagating) modes in the following form
\begin{equation}
N_{kl}= N_{k{\tilde k}}\,\delta_{{\tilde k}l}\equiv \left\{
\begin{array}{cl}     0 & \qquad\mbox{for}\qquad  l\ne{\tilde k}\;,\\
\!\!N_{k\tilde k}\equiv N_k & \qquad\mbox{for}\qquad  l = {\tilde k}\;,
\end{array}  \right.
\label{eq:3.23}
\end{equation}
where the normalizing coefficient $N_{kl}$ and the norm $N_k$ are
given by Eqs.~(\ref{eq:3.18}) and (\ref{eq:3.19}), respectively. From
comparison of Eqs.~(\ref{eq:3.17}) and (\ref{eq:3.23}) it is seen that the
latter relation is of general form because it comprises the former one
for the active modes as a special case obtained by replacing subscript
$\tilde k$ with $k$ so that, in particular, the norm $N_k\equiv N_{kk}$
takes the form given by Eq.~(\ref{eq:3.14}).

It is pertinent to note that all the above expressions for the norms and
the relations of orthogonality and orthonormalization can contain the total
field vectors in place of their cross section parts related to each other
by Eqs.~(\ref{eq:3.5}), i.\,e.,\, the hat sign over the field vectors can be
dropped. This is obvious for the active modes and follows from the
equality $\gamma_k +\gamma_{\tilde k}^*= 0$ for the reactive
twin-conjugate modes.

If in a waveguiding structure there are both the active (propagating) and
reactive (nonpropagating) modes, the total power flow~(\ref{eq:2.46})
carried by them, in accordance with the aforesaid, is given by the
following expression
\[
P=\,P_{act}\,+\,P_{react}\,=
\!\!\!\sum_{\scriptstyle k\atop\scriptstyle(active)}\!\!\! P_k\;+
\!\!\!\!\mathop{{\sum}'}_{\scriptstyle k\atop\scriptstyle(reactive)}
\!\!\!\!\! P_{k\tilde k}^c\,=
\]
\begin{equation}
=\,{1\over4}\!\!\!\sum_{\scriptstyle k\atop\scriptstyle(active)}
\!\!\! N_k\,|a_k(z)|^2\;+\,\;{1\over2}\;{\rm Re}
\!\!\!\!\!\mathop{{\sum}'}_{\scriptstyle k\atop\scriptstyle(reactive)}
\!\!\!\!\! N_k\,a_k^*(z)a_{\tilde k}(z)
\label{eq:3.24}
\end{equation}
where prime on the sum sign means summation of the twin-conjugate
modes in pairs rather than that of the single reactive modes.

\subsection{Time-average Stored Energy for Active Modes in
                                            Lossless Waveguides}
\label{sec:3C}
\indent

The time-average Poynting theorem written in the form of Eq.~(\ref{eq:2.18})
for time-harmonic fields does not contain a stored energy density.
In order to find it one usually applies variational technique (e.\,g., see
Ref.~\cite{29}). To this end, it is necessary to obtain a relation
between variations of the electromagnetic fields $(\delta{\bf E}_k,\,
\delta{\bf H}_k)$ for the $k$th mode and perturbations of the frequency
and medium parameters $(\,\delta(\omega\,
\bar{\!\mbox{\boldmath$\epsilon$}}\,),\;
\delta(\omega\,\bar{\!\mbox{\boldmath$\mu$}}),\;
\delta(\omega\,\bar{\!\mbox{\boldmath$\xi$}}),\;
\delta(\omega\,\bar{\!\mbox{\boldmath$\zeta$}}))$
which bring about these variations.

The $k$th mode is governed by Maxwell's equations (\ref{eq:3.1}) and
(\ref{eq:3.2}) with \,$\bar{\!\mbox{\boldmath$\sigma$}}_c= 0$ for a
lossless medium whose other parameters satisfy the
requirements~(\ref{eq:2.13}). By taking variations in these
equations we obtain
\begin{equation}
\mbox{\boldmath$\nabla$}\!\times\delta{\bf E}_k \,=
-\,i\omega\,\bar{\!\mbox{\boldmath$\mu$}}\cdot\delta{\bf H}_k
-\,i\omega\,\bar{\!\mbox{\boldmath$\zeta$}}\cdot\delta{\bf E}_k
-\,i\delta(\omega\,\bar{\!\mbox{\boldmath$\mu$}})\cdot{\bf H}_k
-\,i\delta(\omega\,\bar{\!\mbox{\boldmath$\zeta$}})\cdot{\bf E}_k \,,
\label{eq:3.25}
\end{equation}
\begin{equation}
\mbox{\boldmath$\nabla$}\!\times\delta{\bf H}_k \,=
\;\;i\omega\,\bar{\!\mbox{\boldmath$\epsilon$}}\cdot\delta{\bf E}_k \,+\,
i\omega\,\bar{\!\mbox{\boldmath$\xi$}}\cdot\delta{\bf H}_k \,+\,
i\delta(\omega\,\bar{\!\mbox{\boldmath$\epsilon$}}\,)\cdot{\bf E}_k \,+\,
i\delta(\omega\,\bar{\!\mbox{\boldmath$\xi$}})\cdot{\bf H}_k \,.
\label{eq:3.26} \\[.2cm]
\end{equation}

A conventional procedure applied to Eqs.~(\ref{eq:3.25}) and
(\ref{eq:3.26}) reduces to the following relation
\[
\mbox{\boldmath$\nabla$}\cdot({\bf E}_k^*\times\delta{\bf H}_k +
\delta{\bf E}_k\times{\bf H}_k^*) \,=
-\,i\omega\Bigl[ \, {\bf E}_k^*\cdot(\,\bar{\!\mbox{\boldmath$\epsilon$}}-
\bar{\!\mbox{\boldmath$\epsilon$}}\,^\dagger)\cdot\delta{\bf E}_k\,+
\]
\[
+\;{\bf H}_k^*\cdot(\,\bar{\!\mbox{\boldmath$\mu$}} -
\bar{\!\mbox{\boldmath$\mu$}}\,^\dagger)\cdot\delta{\bf H}_k +
{\bf E}_k^*\cdot(\,\bar{\!\mbox{\boldmath$\xi$}} -
\bar{\!\mbox{\boldmath$\zeta$}}\,^\dagger)\cdot\delta{\bf H}_k +
{\bf H}_k^*\cdot(\,\bar{\!\mbox{\boldmath$\zeta$}} -
\bar{\!\mbox{\boldmath$\xi$}}\,^\dagger)\cdot
\delta{\bf E}_k\, \Bigr] \,-   \\[.15cm]
\]
\[
-\;i \Bigl[ \,
{\bf E}_k^*\cdot\delta(\omega\,
\bar{\!\mbox{\boldmath$\epsilon$}}\,)\cdot{\bf E}_k +
{\bf H}_k^*\cdot\delta(\omega\,
\bar{\!\mbox{\boldmath$\mu$}})\cdot{\bf H}_k +
{\bf E}_k^*\cdot\delta(\omega\,
\bar{\!\mbox{\boldmath$\xi$}})\cdot{\bf H}_k +
{\bf H}_k^*\cdot\delta(\omega\,
\bar{\!\mbox{\boldmath$\zeta$}})\cdot{\bf E}_k \, \Bigr]   \\[.2cm]
\]
where the terms inside the first square brackets vanish because of
relations~(\ref{eq:2.13}) for lossless media so that
\[
\mbox{\boldmath$\nabla$}\cdot({\bf E}_k^*\times\delta{\bf H}_k +
\delta{\bf E}_k\times{\bf H}_k^*)\,=
\]
\begin{equation}
= -\,i \Bigl[\,
\delta(\omega\,\bar{\!\mbox{\boldmath$\epsilon$}}\,):
{\bf E}_k{\bf E}_k^* +
\delta(\omega\,\bar{\!\mbox{\boldmath$\mu$}}):
{\bf H}_k{\bf H}_k^* +
2\,{\rm Re}\,\{ \delta(\omega\,\bar{\!\mbox{\boldmath$\xi$}}):
{\bf H}_k{\bf E}_k^* \,\Bigr]\,.
\label{eq:3.27} \\[.2cm]
\end{equation}

Electromagnetic fields of a propagating mode and their variations can
be written on the basis of Eq.~(\ref{eq:3.5}) as
\begin{eqnarray}
{\bf E}_k \,=\, \hat{\bf E}_k\,{\rm e}^{-i\,\beta_kz}\,, &\qquad&
\delta{\bf E}_k = (\delta\hat{\bf E}_k- i\delta\beta_kz\,\hat{\bf E}_k)
\,{\rm e}^{-i\,\beta_kz}\,,
\label{eq:3.28} \\  [.25cm]
{\bf H}_k = \hat{\bf H}_k\,{\rm e}^{-i\,\beta_kz}\,, &\qquad&
\delta{\bf H}_k = (\delta\hat{\bf H}_k- i\delta\beta_kz\,\hat{\bf H}_k)
\,{\rm e}^{-i\,\beta_kz} .
\label{eq:3.29}
\end{eqnarray}

When substituting Eqs.~(\ref{eq:3.28}) and (\ref{eq:3.29}) into
Eq.~(\ref{eq:3.27}) and applying the integral relation~(\ref{eq:2.25})
where the line integral vanishes owing to continuity in tangential
components of the fields, the integration over the cross section~$S$
yields
\[
\delta\beta_k\;{1\over2}{\rm Re}\!
\int_S ({\bf E}_k\times{\bf H}_k^*)\cdot{\bf z}_0\,dS\,=
\]
\begin{equation}
=\,{1\over4}\int_S \Bigl[\,
\delta(\omega\,\bar{\!\mbox{\boldmath$\epsilon$}}\,):
\hat{\bf E}_k\hat{\bf E}_k^* +
\delta(\omega\,\bar{\!\mbox{\boldmath$\mu$}}):
\hat{\bf H}_k\hat{\bf H}_k^* +
2\,{\rm Re}\{ \delta(\omega\,\bar{\!\mbox{\boldmath$\xi$}}):
\hat{\bf H}_k\hat{\bf E}_k^*\} \,\Bigr] dS .
\label{eq:3.30} \\[.2cm]
\end{equation}

The left-hand side of Eq.~(\ref{eq:3.30}) involves the time-average
power flow $P_k$ as a multiplier of $\delta\beta_k$. By using the
known relation
\begin{equation}
P_k= v_{gr,k}\,W_k  \qquad \mbox{where} \qquad
v_{gr,k}= \biggl( {\partial\beta_k(\omega)\over\partial\omega} \biggr)^{-1}
\label{eq:3.31}
\end{equation}
is the group velocity, we obtain from Eq.~(\ref{eq:3.30}) the desired
expression for the time-average energy stored per unit length of a
waveguide (dropping the mode index~$k$):
\[
W = {1\over4} \int_S \biggl(
{\bf E}^* \cdot
{\partial(\omega\,\bar{\!\mbox{\boldmath$\epsilon$}}\,)
\over\partial\,\omega}
\cdot{\bf E} \,+\, {\bf H}^*\cdot
{\partial(\omega\,\bar{\!\mbox{\boldmath$\mu$}}\,)\over\partial\,\omega}
\cdot{\bf H} \biggr)\,dS \,+
\]
\begin{equation}
+\,{1\over2}\,{\rm Re} \int_S \biggl(
{\bf E}^* \cdot
{\partial(\omega\,\bar{\!\mbox{\boldmath$\xi$}}\,)\over\partial\,\omega}
\cdot{\bf H} \biggr)\,dS \equiv \int_S w\,dS  \,.
\label{eq:3.32} \\[.2cm]
\end{equation}

In accordance with expression (\ref{eq:3.32}), the time-average stored
energy density for a lossless bianisotropic medium is given by the
following formula
\begin{equation}
w = {1\over4}\,\Bigl( {\bf E}^* \;\;{\bf H}^* \Bigr) \cdot
\left( \begin{array}{cc}
\partial(\omega\,\bar{\!\mbox{\boldmath$\epsilon$}}\,)/\partial\,\omega
\;\;&\;\;
\partial(\omega\,\bar{\!\mbox{\boldmath$\xi$}}\,)/\partial\,\omega \\[.3cm]
\partial(\omega\,\bar{\!\mbox{\boldmath$\zeta$}}\,)/\partial\,\omega
\;\;&\;\;
\partial(\omega\,\bar{\!\mbox{\boldmath$\mu$}}\,)/\partial\,\omega
\end{array}  \right) \cdot
\left( \begin{array}{c}
{\bf E} \\ {\bf H}
\end{array} \right)
\label{eq:3.33} \\[.1cm]
\end{equation}
which is a generalization of the usual expression for lossless
anisotropic media~\cite{7,10,22}.

\setcounter{equation}{0}
\section{ORTHOGONAL\, COMPLEMENTS\, AND\, EFFECTIVE\, SURFACE\,
                 CURRENTS\, INSIDE\, SOURCE\, REGION}
\label{sec:4}

\subsection{Bulk and Surface Exciting Sources}
\label{sec:4A}
\indent

Up to the present, the external sources exciting the composite
(multilayered) wave\-guiding structures involving isotropic, anisotropic,
and bianisotropic media have been dropped. From this point onward,
the special atten\-tion will be given to investigating the behavior of
modes inside the source region. In doing so, we assume that all the
eigenfields $\{{\bf E}_k,{\bf H}_k\}$
in the form of Eq.~(\ref{eq:3.5}), including their eigenfunctions of
cross-section coordinates $\{\hat{\bf E}_k,\hat{\bf H}_k\}$ (marked
by hat over them) and their eigenvalues of propagation constants
$\gamma_k=\alpha_k+ i\beta_k$, are known from solving the corresponding
boundary-value problem. As shown in Appendix\,A, these
eigenfields constitute an infinite countable set of the vector
functions quadratically integrable on the cross section $S$
of a waveguiding structure. This set can be taken as a
basis of the proper Hilbert space to expand the required fields
${\bf E}$ and ${\bf H}$ not only outside sources, as was done by
Eqs.~(\ref{eq:2.31}) and (\ref{eq:2.32}), but also inside the region
of external sources. In general, this eigenvector basis is not
complete inside the source region since it cannot take into account
entirely the potential fields of the sources. This requires to supplement
the {\it modal expansions\/} ${\bf E}_a$ and ${\bf H}_a$ with unknown
modal amplitudes $A_k(z)$ by the {\it orthogonal complements\/}
${\bf E}_b$ and ${\bf H}_b$ (see Eqs.~(\ref{eq:A27}) and (\ref{eq:A28})).
Hence, the desired issues to be obtained inside the source region are both
the longitudinal dependence of modal amplitudes and the orthogonal
complements to the modal expansions.   \\[.1cm]
\indent
In the most general case there exist three physical reasons to excite
the waveguiding structure under examination:\\
(a)\, the external currents -- electric \,${\bf J}_{ext}^e$ and magnetic
\,${\bf J}_{ext}^m$ ,\\
(b)\, the external fields -- electric \,${\bf E}_{ext}$ and magnetic
\,${\bf H}_{ext}$ ,\\
(c)\, the external perturbations of bianisotropic medium parameters\,
$\Delta\,\bar{\!\mbox{\boldmath$\epsilon$}}$,\,
$\Delta\,\bar{\!\mbox{\boldmath$\mu$}}$,\,
$\Delta\,\bar{\!\mbox{\boldmath$\xi$}}$,\,
and~$\Delta\,\bar{\!\mbox{\boldmath$\zeta$}}$.

Owing to these medium perturbations, the total electric $({\bf E}+
{\bf E}_{ext})$ and magnetic $({\bf H}+{\bf H}_{ext})$ fields create
the excess electric $\Delta{\bf D}$ and magnetic $\Delta{\bf B}$
inductions linked by the constitutive relations
(\ref{eq:2.8}) and (\ref{eq:2.9}), that is
\begin{eqnarray}
\Delta{\bf D} \!&=&\! \Delta\,\bar{\!\mbox{\boldmath$\epsilon$}}\cdot
({\bf E}+ {\bf E}_{ext})\,+\;
\Delta\,\bar{\!\mbox{\boldmath$\xi$}}\cdot({\bf H}+ {\bf H}_{ext}) \;,
\label{eq:4.1}\\  [.2cm]
\Delta{\bf B} \!&=&\! \Delta\,\bar{\!\mbox{\boldmath$\zeta$}}\cdot
({\bf E}+ {\bf E}_{ext})\,+\;
\Delta\,\bar{\!\mbox{\boldmath$\mu$}}\cdot({\bf H}+ {\bf H}_{ext}) \;.
\label{eq:4.2}
\end{eqnarray}

These excess inductions bring about the induced displacement currents --
electric \,${\bf J}_{ind}^e= i\omega\Delta{\bf D}$\, and magnetic
\,${\bf J}_{ind}^m= i\omega\Delta{\bf B}$\, which, being added to
the external currents \,${\bf J}_{ext}^e$\, and \,${\bf J}_{ext}^m$,\,
yield the {\it bulk exciting currents\/}
\begin{equation}
{\bf J}_b^e= {\bf J}_{ext}^e+ {\bf J}_{ind}^e=
{\bf J}_{ext}^e+ i\omega\Delta{\bf D} \,, \quad
{\bf J}_b^m= {\bf J}_{ext}^m+ {\bf J}_{ind}^m=
{\bf J}_{ext}^m+ i\omega\Delta{\bf B}
\label{eq:4.3}
\end{equation}
entering into the curl Maxwell equations~(\ref{eq:2.1}) in the
following form
\begin{eqnarray}
\mbox{\boldmath$\nabla$}\times{\bf E}\,= -\,
i\omega{\bf B} \!&-&\! {\bf J}_b^m \,,
\label{eq:4.4}\\  [.15cm]
\mbox{\boldmath$\nabla$}\times{\bf H}\,=\;\;
i\omega{\bf D} \!&+&\! {\bf J}_b^e \;\,.
\label{eq:4.5}
\end{eqnarray}

The conduction current ${\bf J}\,$ of a conductive medium defined by
Eq.~(\ref{eq:2.14}) is now assumed to be incorporated with the electric
displacement current $i\omega{\bf D}$, whereas the induction vectors
${\bf D}$ and ${\bf B}$ are taken, as before, to be related to the
intrinsic electromagnetic fields ${\bf E}$ and ${\bf H}$ inside the
medium in question by the same constitutive relations~(\ref{eq:2.8})
and (\ref{eq:2.9}). So the permittivity tensor
$\bar{\!\mbox{\boldmath$\epsilon$}}$ is now regarded as a sum
$(\,\bar{\!\mbox{\boldmath$\epsilon$}}\,+\;
\bar{\!\mbox{\boldmath$\sigma$}}_c/i\omega)$ whose
antihermitian part defines the total tensor of electric conducttivity
\,$\bar{\!\mbox{\boldmath$\sigma$}}_e=
\bar{\!\mbox{\boldmath$\sigma$}}_c+\,\bar{\!\mbox{\boldmath$\sigma$}}_d$
\,given by Eq.~(\ref{eq:2.21}).   \\[.1cm]
\indent
Besides the bulk exciting currents  ${\bf J}_b^e$ and ${\bf J}_b^m$,
there may exist the {\it surface exciting currents\/}  ${\bf J}_s^e$ and
${\bf J}_s^m$  which give discontinuities of the appropriate tangential
components of fields at points of the surface whereon these sources are
located, written in the form of the following boundary conditions:
\begin{equation} \;\;
{\bf n}_s^+ \times\,{\bf E}^+ +\,{\bf n}_s^-\times{\bf E}^- =
-\,{\bf J}_s^m ,
\label{eq:4.6}
\end{equation}
\begin{equation}
{\bf n}_s^+ \times\,{\bf H}^+ +\, {\bf n}_s^-\times{\bf H}^- =\,
\;{\bf J}_s^e \;.
\label{eq:4.7}
\end{equation}

Here the field vectors with superscripts $^\pm$ mean their values taken
at points of the source location contour $L_s$ lying on its different
sides marked by the {\it inward\/} (for either adjacent medium) unit
vectors \,${\bf n}_s^\pm$.

\subsection{Orthogonal Complementary Fields and Effective
                                            Surface Currents}
\label{sec:4B}
\indent

The general electrodynamic eigenmode treatment (see Appendix\,A.2)
ba\-sed on the well-known mathematical formulations
(see Appendix\,A.1) yields the complete representation of the
desired field vector ${\bf F}$ inside sources as a sum of the the modal
expansion $\bf\Psi$ giving a {\it projection\/} of ${\bf F}$ onto the
Hilbert space and the complement ${\bf C}$ {\it orthogonal\/} to the
Hilbert space (see Eq.~(\ref{eq:A25})). Thus, the
electromagnetic fields inside the source region have the complete
representation given by Eqs.~(\ref{eq:A27}) and (\ref{eq:A28}), namely
(cf.~Eqs. (\ref{eq:2.31}) and~(\ref{eq:2.32})):
\begin{eqnarray}
{\bf E}({\bf r}_t, z)=
{\bf E}_a({\bf r}_t, z)\,+\,{\bf E}_b({\bf r}_t, z) \!\!&=&\!\!
\sum_k a_k(z)\,\hat{\bf E}_k({\bf r}_t) + {\bf E}_b({\bf r}_t, z) \,,
\label{eq:4.8}\\    [.1cm]
{\bf H}({\bf r}_t, z)=
{\bf H}_a({\bf r}_t, z)+{\bf H}_b({\bf r}_t, z) \!\!&=&\!\!
\sum_k a_k(z)\,\hat{\bf H}_k({\bf r}_t) + {\bf H}_b({\bf r}_t, z) \,,
\label{eq:4.9}
\end{eqnarray}
where  ${\bf E}_b({\bf r}_t, z)$ and ${\bf H}_b({\bf r}_t, z)$ are
the required orthogonal complements. The mode amp\-litude $a_k(z)=
A_k(z)\exp(-\,\gamma_k z)$ allows for the total dependence on $z$
due to both the unperturbed propagation of the $k$th mode with
a constant $\gamma_k$ and the perturbed amp\-litude $A_k(z)$ as a result
of source actions. Using Eqs.~(\ref{eq:2.8}) and (\ref{eq:2.9}) gives
the similar expressions for the induction vectors
\begin{eqnarray} \!\!\!\!\!
{\bf D}({\bf r}_t, z)\!=
{\bf D}_a({\bf r}_t, z)+{\bf D}_b({\bf r}_t, z) \!\!&=&\!\!
\sum_k a_k(z)\,\hat{\bf D}_k({\bf r}_t) + {\bf D}_b({\bf r}_t, z) ,
\label{eq:4.10}\\        [.1cm]
{\bf B}({\bf r}_t, z)=
{\bf B}_a({\bf r}_t, z)+{\bf B}_b({\bf r}_t, z) \!\!&=&\!\!
\sum_k a_k(z)\,\hat{\bf B}_k({\bf r}_t) + {\bf B}_b({\bf r}_t, z) .
\label{eq:4.11}
\end{eqnarray}

In order to find the orthogonal complements let us substitute
Eqs.~(\ref{eq:4.8}) through (\ref{eq:4.11}) into Maxwell's equations
(\ref{eq:4.4}) and (\ref{eq:4.5}) with taking into account the fact
that the eigenfields~(\ref{eq:3.5}) satisfy the homogeneous (with no
sources) Maxwell equations~(\ref{eq:3.1}) and (\ref{eq:3.2}). Some
transformations yield
\begin{equation}
\sum_k {dA_k\over dz}\,({\bf z}_0\times{\bf E}_k)\,=\,
-\,\mbox{\boldmath$\nabla$}\times{\bf E}_b\,-\,
i\omega{\bf B}_b\,-\,{\bf J}_b^m \;,
\label{eq:4.12}   \\[-0.15cm]
\end{equation}
\begin{equation}
\sum_k {dA_k\over dz}\,({\bf z}_0\times{\bf H}_k)\,=\,
-\,\mbox{\boldmath$\nabla$}\times{\bf H}_b\,+\,
i\omega{\bf D}_b\,+\,{\bf J}_b^e \;.
\label{eq:4.13}
\end{equation}

The left-hand side of Eqs.~(\ref{eq:4.12}) and (\ref{eq:4.13}) has only
transverse components. From here it necessarily follows that there must
exist nonzero ortho\-gonal complementary fields. Otherwise (when ${\bf E}_b=
{\bf H}_b= {\bf D}_b= {\bf B}_b= 0$) these equations become physically
contradictory because then they require the longitudinal components of
the arbitrary bulk currents \,${\bf J}_b^e$ and \,${\bf J}_b^m$ to be
always equal to zero, which is of course not the case.

Hence, the required orthogonal complementary fields should be chosen
so as to make the longitudinal component of the right-hand part of
Eqs.~(\ref{eq:4.12}) and (\ref{eq:4.13}) vanish,~that~is
\begin{equation}
\mbox{\boldmath$\nabla$}\cdot({\bf z}_0\times{\bf E}_b)\,-\,{\bf z}_0\cdot
(i\omega{\bf B}_b + {\bf J}_b^m)= 0 \,,
\label{eq:4.14}
\end{equation}
\begin{equation}
\mbox{\boldmath$\nabla$}\cdot({\bf z}_0\times{\bf H}_b)\,+\,{\bf z}_0\cdot
(i\omega{\bf D}_b + {\bf J}_b^e\,)= 0 \,,
\label{eq:4.15}      \\[.1cm]
\end{equation}
where the identity ${\bf z}_0\cdot(\mbox{\boldmath$\nabla$}\times{\bf a})=
-\,\mbox{\boldmath$\nabla$}\cdot({\bf z}_0\times{\bf a})$ has been used.

Since the field parts ${\bf E}_b$ and ${\bf H}_b$ form the orthogonal
complement to the Hilbert space spanned by the base eigenvectors
$\{{\bf E}_k,{\bf H}_k\}$, they must be orthogonal to the fields of
any eigenmode in power sense given by the relation similar to
Eqs.~(\ref{eq:3.13}) and (\ref{eq:3.18}) (cf.~Eq.~(\ref{eq:A26})):
\[
\int_S ({\bf E}_k^*\times{\bf H}_b +\,
{\bf E}_b\times{\bf H}_k^*)\cdot{\bf z}_0\,dS \equiv
\]
\begin{equation}
\equiv \int_S \Bigl[\,({\bf z}_0\times{\bf E}_b)\cdot{\bf H}_k^* -\,
({\bf z}_0\times{\bf H}_b)\cdot{\bf E}_k^* \,\Bigr] dS \,=\,0 \;.
\label{eq:4.16} \\[.2cm]
\end{equation}

In virtue of arbitrary choice of the $k$th eigenmode taken from the
base set, zero equality in Eq.~(\ref{eq:4.16}) can occur
if and only if
\begin{eqnarray}
{\bf z}_0\times{\bf E}_b\,=\,0   \qquad &\mbox{or}& \qquad
{\bf E}_b=\,{\bf z}_0\,E_b \;,
\label{eq:4.17}\\    [.2cm]
{\bf z}_0\times{\bf H}_b\,=\,0   \qquad &\mbox{or}& \qquad
{\bf H}_b=\,{\bf z}_0\,H_b  \;,
\label{eq:4.18}
\end{eqnarray}
i.\,e., both complementary fields are longitudinal. In order for their
magnitude to be found, it is necessary to insert Eqs.~(\ref{eq:2.8}),
(\ref{eq:2.9}), (\ref{eq:4.17}), and (\ref{eq:4.18}) into
Eqs.~(\ref{eq:4.14}) and (\ref{eq:4.15}), then
\begin{equation}
{\bf z}_0\cdot{\bf D}_b\,\equiv\,
\epsilon_{zz}\,E_b\,+\,\xi_{zz}\,H_b\,= -\,{1\over i\omega}\,J_{bz}^e\;,
\label{eq:4.19}
\end{equation}
\begin{equation}
{\bf z}_0\cdot{\bf B}_b\,\equiv\,
\zeta_{zz}\,E_b\,+\,\mu_{zz}\,H_b\,= -\,{1\over i\omega}\,J_{bz}^m \;.
\label{eq:4.20} \\[.2cm]
\end{equation}

From here it finally follows that
\begin{equation}
E_b =\,-\,{J_{bz}^e - \nu^m J_{bz}^m\over
i\omega\epsilon_{zz}\,(1- \nu^e\nu^m)}  \;,
\label{eq:4.21}
\end{equation}
\begin{equation}
H_b =\,-\,{J_{bz}^m - \nu^e J_{bz}^e\over
i\omega\mu_{zz}\,(1- \nu^e\nu^m)} \;,
\label{eq:4.22}
\end{equation}
where we have denoted
\begin{equation}
\nu^e =\,{\zeta_{zz}\over\epsilon_{zz}}\,=\,
{\bar{\!\mbox{\boldmath$\zeta$}}:{\bf z}_0{\bf z}_0\over
\bar{\!\mbox{\boldmath$\epsilon$}}:{\bf z}_0{\bf z}_0} \;, \qquad
\nu^m =\,{\xi_{zz}\over\mu_{zz}}\,=\,
{\bar{\!\mbox{\boldmath$\xi$}}:{\bf z}_0{\bf z}_0\over
\bar{\!\mbox{\boldmath$\mu$}}:{\bf z}_0{\bf z}_0} \;.
\label{eq:4.23}
\end{equation}

Hence, both the orthogonal complementary fields are longitudinal
and produced by the longitudinal components of the bulk exciting
currents.      \\[.1cm]
\indent
Existence of the complementary fields ${\bf E}_b$ and ${\bf H}_b$
immediately reduces to appearance of the so-called {\it effective
surface currents\/} ${\bf J}_{s,ef}^e$ and ${\bf J}_{s,ef}^m$.\,
Consider the bulk source region having the cross section $S_b$
with a boundary contour $L_b$ and write the complete electric field
inside and outside this area:
\begin{equation}
{\bf E}({\bf r}_t,z) = \left\{
\begin{array}{lcl}
\sum_k A_k(z)\,{\bf E}_k({\bf r}_t,z)+ {\bf E}_b({\bf r}_t,z)
&\equiv& {\bf E}^-  \;\;\mbox{-- inside}\;\;\; S_b \,,  \\  [.3cm]
\sum_k A_k(z)\,{\bf E}_k({\bf r}_t,z)
&\equiv& {\bf E}^+  \;\;\mbox{-- outside}\; S_b \,.
\end{array}           \right.
\label{eq:4.24}
\end{equation}
Analogous expressions can be written for the magnetic field
${\bf H}({\bf r}_t,z)$.

The eigenfields~${\bf E}_k$ and ${\bf H}_k$, being obtained for the
situation without sources, are generally continuous at points of the
line $L_b$. Then the tangential components of the complete fields
${\bf E}$ and ${\bf H}$ prove discontinuous:
\begin{eqnarray}
{\bf n}_b^+\times\,{\bf E}^+ \!\!\!\!&+&\!\! {\bf n}_b^-\times\,{\bf E}^- =
-\,{\bf n}_b\times{\bf E}_b  \,,
\label{eq:4.25} \\   [.2cm]
{\bf n}_b^+\times{\bf H}^+   \!\!\!\!&+&\!\! {\bf n}_b^-\times{\bf H}^- =
-\,{\bf n}_b\times{\bf H}_b \,,
\label{eq:4.26}
\end{eqnarray}
where ${\bf n}_b ={\bf n}_b^+ = -\,{\bf n}_b^-$\, is the
{\it outward\/} unit vector normal to both the line $L_b$ and the
longitudinal unit vector ${\bf z}_0$. The comparison of these relations
with the boundary conditions~(\ref{eq:4.6}) and (\ref{eq:4.7}) yields
the desired effective surface currents
\begin{equation}
{\bf J}_{s,ef}^e = -\,{\bf n}_b\times{\bf H}_b\,\Big|_{L_b} =
-\,{\mbox{\boldmath$\tau$}}\;{J_{bz}^m(L_b) -\,\nu^e J_{bz}^e(L_b)\over
i\omega\mu_{zz}\,(1- \nu^e\nu^m)} \;,
\label{eq:4.27}  \\[.2cm]
\end{equation}
\begin{equation}
{\bf J}_{s,ef}^m =\;\;{\bf n}_b\times{\bf E}_b\,\Big|_{L_b}\,=\,
\;\;{\mbox{\boldmath$\tau$}}\;{J_{bz}^e(L_b) -\,\nu^m J_{bz}^m(L_b)\over
i\omega\epsilon_{zz}\,(1- \nu^e\nu^m)} \;,
\label{eq:4.28} \\[.2cm]
\end{equation}
where $J_{bz}^e(L_b)$ and $J_{bz}^m(L_b)$ mean the longitudinal components
of the bulk currents taken at points lying on the boundary $L_b$ of
their existence area $S_b$ and ${\mbox{\boldmath$\tau$}}=
{\bf z}_0\times{\bf n}_b$\,is the unit vector tangential to
the contour $L_b$.

The general expressions~(\ref{eq:4.21}) and (\ref{eq:4.22}) for the
complementary fields ${\bf E}_b$ and ${\bf H}_b$ and the general expressions
(\ref{eq:4.27}) and (\ref{eq:4.28}) for the effective surface currents
${\bf J}_{s,ef}^e$ and ${\bf J}_{s,ef}^m$ take the following simplified
form in special~cases~of:\\[.15cm]
\indent
(i)\, the isotropic medium with parameters~(\ref{eq:2.10})
{} ($\nu^e=\nu^m= 0$)
\begin{eqnarray}
{\bf E}_b= -\,{{\bf z}_0\over i\omega\epsilon}\,J_{bz}^e
\qquad &\mbox{and}& \qquad
{\bf H}_b= -\,{{\bf z}_0\over i\omega\mu}\,J_{bz}^m \;,
\label{eq:4.29} \\     [.2cm]
{\bf J}_{s,ef}^m=\,{\mbox{\boldmath$\tau$}\over
i\omega\epsilon}\,J_{bz}^e(L_b)
\qquad &\mbox{and}& \qquad
{\bf J}_{s,ef}^e= -\,{\mbox{\boldmath$\tau$}\over
i\omega\mu}\,J_{bz}^m(L_b) \,;
\label{eq:4.30}
\end{eqnarray}

(ii) the anisotropic medium with parameters~(\ref{eq:2.11})
{} ($\nu^e=\nu^m= 0$)
\begin{eqnarray}
{\bf E}_b= -\,{{\bf z}_0\over i\omega\epsilon_{zz}}\,J_{bz}^e
\qquad &\mbox{and}& \qquad
{\bf H}_b= -\,{{\bf z}_0\over i\omega\mu_{zz}}\,J_{bz}^m \;,
\label{eq:4.31} \\     [.2cm]
{\bf J}_{s,ef}^m=\,{\mbox{\boldmath$\tau$}\over
i\omega\epsilon_{zz}}\,J_{bz}^e(L_b)
\qquad &\mbox{and}& \qquad
{\bf J}_{s,ef}^e= -\,{\mbox{\boldmath$\tau$}\over
i\omega\mu_{zz}}\,J_{bz}^m(L_b) \,.
\label{eq:4.32}
\end{eqnarray}

Formulas~(\ref{eq:4.29}) and (\ref{eq:4.31}) are in agreement with those
obtained first by Vainshtein~\cite{2} and Felsen and Marcuvitz~\cite{8},
respectively. As for
the effective surface currents~(\ref{eq:4.30}) and (\ref{eq:4.32}),
Vainshtein did not consider them at all but the excitation integrals in
the theory of Felsen and Marcuvitz allow for them implicitly, which will
be shown later (see~Sec.\,5.1).

Therefore, in the absence of medium bianisotropy the bulk currents,
{\it electric\/}~${\bf J}_b^e$ and {\it magnetic\/} ${\bf J}_b^m,$\,
generate the effective surface currents, respectively, {\it magnetic\/}
${\bf J}_{s,ef}^m$ and {\it electric\/}~${\bf J}_{s,ef}^e$.
As is seen from Eqs.~(\ref{eq:4.27}) and (\ref{eq:4.28}), the
bianisotropic properties of a medium intermix the contributions from
the bulk currents into the effective surface currents owing to the
longitudinal components $\xi_{zz}$ and $\zeta_{zz}$.

The newly obtained effective surface currents ${\bf J}_{s,ef}^e$ and
${\bf J}_{s,ef}^m$, as well as the actual surface currents ${\bf J}_s^e$
and ${\bf J}_s^m$ entering into the boundary conditions~(\ref{eq:4.6})
and (\ref{eq:4.7}), make contributions to the excitation
amplitudes $A_k$ along with the bulk currents ${\bf J}_b^e$
and~${\bf J}_b^m$.   \\[.1cm]
\indent
The next step should be done toward deriving the differential
equations to find the functions $A_k(z)$ inside the region of
bulk and surface sources. For this purpose we shall apply three
independent approaches set forth in the next section and Appendix\,B.

\setcounter{equation}{0}
\section{EQUATIONS OF MODE EXCITATION}
\label{sec:5}

\subsection{Approach based on the Electrodynamic Method of
                                    Variation of Constants}
\label{sec:5A}
\indent

Mathematical method of variation of constants is applied to solve an
inhomogeneous differential equation (with driving terms) by representing
its general solution in the form of a superposition of the known linearly
independent solutions of the proper homogeneous equation with coefficients
which are no longer considered constant and assumed to be the desired
functions of an independent variable~\cite{45}. Electrodynamic analog of
the mathematical method of variation of constants is built by
representing the fields ${\bf E}({\bf r}_t,z)$ and ${\bf H}({\bf r}_t,z)$
inside the source region in the form of the expansions
(\ref{eq:4.8}) and (\ref{eq:4.9}) in terms of eigenfunctions of the
proper homogeneous boundary-value problem (without sources) whose
amplitude coefficients $A_k(z)$ are the desired functions of $z$ rather
than constants, as they are outside sources.

According to the conventional mathematical technique, the method
of variation of constants is to give differential equations for the
mode amplitudes in the following~form
\begin{equation}
{d A_k(z)\over dz}\,=\, f_k(z)\;,  \qquad  k= 1,2,\ldots
\label{eq:5.1}
\end{equation}
where the functions $f_k(z)$ take into account the longitudinal
distribution of exciting sources (bulk and surface). Integration of
Eq.~(\ref{eq:5.1}) yields the required dependence
\begin{equation}
A_k(z)\,=\,A_k^{\rm o}\,+\int f_k(z)\,dz\,\equiv\,
A_k^{\rm o} \,+\,\Delta A_k(z) \;.
\label{eq:5.2}
\end{equation}

The integration constant $A_k^{\rm o}$ should be determined from a
boundary condition given at one of two boundaries ($z= 0$ \,or\, $z= L$)
of the source region depending on the type of modes for a lossless
waveguiding structure:\\
(i)\, for the {\it forward\/} modes (active and reactive, $\;k= +\,n$)
supplied at the~left~input
\begin{equation}
A_{+n}(0)\,\ne\,0  \qquad \mbox{or} \qquad  A_{+n}(0)\,=\,0 \;,
\label{eq:5.3}
\end{equation}
(ii) for the {\it backward\/} modes (active and reactive, $\;k= -\,n$)
supplied at the right~input
\begin{equation}
A_{-n}(L)\,\ne\,0  \qquad \mbox{or} \qquad  A_{-n}(L)\,=\,0 \;.
\label{eq:5.4}
\end{equation}

Substitution of Eq.~(\ref{eq:5.2}) into Eqs.~(\ref{eq:4.8}) and
(\ref{eq:4.9}) allows us to represent the complete solution for the
electromagnetic fields inside the source region as the sum of the
general solution $({\bf E}_{gen},\,{\bf H}_{gen})$ to the homogeneous
boundary-value problem (without exciting sources) involving the constant
amplitude coefficients $A_k^{\rm o}$ and the particular solution
$({\bf E}_{par},\,{\bf H}_{par})$ to the proper inhomogeneous problem
(with exciting sources), \,namely:
\[
{\bf E}({\bf r}_t, z)\,=\,{\bf E}_{gen} + {\bf E}_{par} \equiv \\[-0.1cm]
\]
\[
\equiv\sum_k A_k^{\rm o}\,{\bf E}_k({\bf r}_t,z) \,+\,
\biggl[\, \sum_k\Delta A_k(z)\,{\bf E}_k({\bf r}_t,z) +
{\bf E}_b({\bf r}_t, z) \,\biggr] \,,       \\[.1cm]
\]
\[
{\bf H}({\bf r}_t, z)\,=\,{\bf H}_{gen} + {\bf H}_{par} \equiv
\]
\[
\equiv\sum_k A_k^{\rm o}\,{\bf H}_k({\bf r}_t,z) \,+\,
\biggl[\, \sum_k\Delta A_k(z)\,{\bf H}_k({\bf r}_t,z) +
{\bf H}_b({\bf r}_t, z) \,\biggr] \,.           \\[.2cm]
\]

Thus, the general technique of solving the electrodynamic problem of
waveguide excitation by external sources based on the method of
variation of constants gives rise to the representation of the desired
electromagnetic fields as the sum of the general and particular solutions
adopted in the theory of linear differential equations. The next task
is to obtain a specific form for the excitation equation like
Eq.~(\ref{eq:5.1}).     \\[.1cm]
\indent
To derive the equation of mode excitation let us vector-multiply both
sides of Eqs.~(\ref{eq:4.8}) and (\ref{eq:4.9}) by \,$\hat{\bf H}_l^*$
and \,$-\hat{\bf E}_l^*$,\, respectively,\, and add them. Then after
scalar-multiplying the result of summation by ${\bf z}_0$ and
integrating over the cross section $S$ of a waveguide we obtain
\[
\int_S (\hat{\bf E}_l^*\times{\bf H} + {\bf E}\times
\hat{\bf H}_l^*)\cdot {\bf z}_0\,dS \,=\, \sum_k a_k
\int_S (\hat{\bf E}_l^*\times\hat{\bf H}_k +\hat{\bf E}_k\times
\hat{\bf H}_l^*)\cdot {\bf z}_0\,dS\;+
\]
\begin{equation}
+\,\int_S (\hat{\bf E}_l^*\times{\bf H}_b+
{\bf E}_b\times\hat{\bf H}_l^*) \cdot {\bf z}_0\,dS \;.
\label{eq:5.5} \\[.2cm]
\end{equation}

The last integral in the right-hand side of Eq.~(\ref{eq:5.5}) vanishes
because of the ortho\-gonality relation~(\ref{eq:4.16}) or (\ref{eq:A26}).
According to the orthonormalization relations~(\ref{eq:3.17}) and
(\ref{eq:3.23}), the integral under the sign of summation is equal to
$N_l\delta_{lk}$ for the active (propagating) modes and to
$N_l\delta_{{\tilde l}k}$ for the reactive (nonpropagating) modes.
Hence, from Eq.~(\ref{eq:5.5}) we obtain
(cf.~Eq.~(\ref{eq:A31})):   \\[.1cm]
\indent
(i)\, for the active modes (with replacing subscripts $l\!\to\!k$)
\[
a_k= {1\over N_k}
\int_S (\hat{\bf E}_k^*\times{\bf H}+
{\bf E}\times\hat{\bf H}_k^*)\cdot{\bf z}_0\,dS\,\equiv\,
A_k\,{\rm e}^{-i\,\beta_kz}
\]
\begin{equation} \!\!\!\!\!\!\!\!\!\!\!\!\!       \mbox{or}\qquad
A_k= {1\over N_k} \int_S ({\bf E}_k^*\times{\bf H}+
{\bf E}\times{\bf H}_k^*)\cdot{\bf z}_0\,dS
\label{eq:5.6} \\[.2cm]
\end{equation}
where the norm $N_k$ is defined by formula~(\ref{eq:3.14})\,,   \\[.1cm]
\indent
(ii) for the reactive modes (with replacing subscripts
$l\!\to\!{\tilde k}$ and ${\tilde l}\!\to\!k$)
\[
a_k= {1\over N_{\tilde k}}
\int_S (\hat{\bf E}_{\tilde k}^*\times{\bf H}+
{\bf E}\times\hat{\bf H}_{\tilde k}^*)\cdot{\bf z}_0\,dS\,\equiv\,
A_k\,{\rm e}^{-\,\gamma_kz}
\]
\begin{equation} \!\!\!\!\!\!\!\!\!\!\!\!       \mbox{or}\qquad
A_k= {1\over N_{\tilde k}}
\int_S ({\bf E}_{\tilde k}^*\times{\bf H}+
{\bf E}\times{\bf H}_{\tilde k}^*)\cdot{\bf z}_0\,dS
\label{eq:5.7}
\end{equation}
where subscript $\tilde k$ corresponds to the mode which together with
the $k$th mode constitute the twin-conjugate pair and have the
propagation constant \,$\gamma_{\tilde k}\!=\!-\,\gamma_k^*$\, and the
norm \,$N_{\tilde k}\!=\! N_k^*$\, defined by formula~(\ref{eq:3.19}).
From comparison of Eqs.~(\ref{eq:5.6}) and (\ref{eq:5.7}) it is seen
that the latter expression can be considered as the general form valid
not only for the reactive modes but also for the active modes with
replacing \,$\tilde k$\, by \,$k$.

It is pertinent to note that expression~(\ref{eq:5.6}) for \,$a_k$\,
is in agreement with Eq.~(\ref{eq:A31}) obtained by minimizing the
mean-square difference $D_n$ (defined by Eq.~(\ref{eq:A33})) between
the mode series expansion $\bf\Psi$ and the partial sum ${\bf S}_n$ of
the $n$th order (given by Eq.~(\ref{eq:A32})) to provide
convergence in mean for the modal expansion.

Formulas (\ref{eq:5.6}) and (\ref{eq:5.7}) give a rule to find the
mode excitation amplitude if the electromagnetic fields are known.
However, this is usually not the case because the exciting currents
(bulk and surface) are assumed to be given rather than the fields.
In order to go from the fields to the currents, let us differentiate
the general relation~(\ref{eq:5.7}) with respect to $z$:
\[
N_{\tilde k}\,{d A_k\over dz}\,=\,{\partial\over\partial z}
\int_S ({\bf E}_{\tilde k}^*\times{\bf H}+
{\bf E}\times{\bf H}_{\tilde k}^*)\cdot{\bf z}_0\,dS\,=   \\[-0.1cm]
\]
\[
= \int_S \mbox{\boldmath$\nabla$}\cdot({\bf E}_{\tilde k}^*\times{\bf H}+
{\bf E}\times{\bf H}_{\tilde k}^*)\,dS\;+
\]
\begin{equation}
+\,\sum_i \oint_{L_i} \Bigl[\,
{\bf n}_i^+\cdot ({\bf E}_{\tilde k}^*\times{\bf H} +
{\bf E}\times{\bf H}_{\tilde k}^*)^+ \,+\,
{\bf n}_i^-\cdot ({\bf E}_{\tilde k}^*\times{\bf H} +
{\bf E}\times{\bf H}_{\tilde k}^*)^- \Bigr] dl
\label{eq:5.8}
\end{equation}
where the last equality is written by using the relation similar
to~Eq.~(\ref{eq:2.25}).

The complete fields ${\bf E}$ and ${\bf H}$ inside the source region
satisfy the inhomogeneous Maxwell equations~(\ref{eq:4.4}) and
(\ref{eq:4.5}), whereas the fields ${\bf E}_{\tilde k}^*$ and
${\bf H}_{\tilde k}^*$\, of the $\tilde k$th mode obey the following
homogeneous equations
\begin{eqnarray}
\mbox{\boldmath$\nabla$}\times{\bf E}_{\tilde k}^*\!&=&\!\,\;
i\omega{\bf B}_{\tilde k}^* \,,
\label{eq:5.9}\\      [.15cm]
\mbox{\boldmath$\nabla$}\times{\bf H}_{\tilde k}^*\!&=&\!\!\! -\,
i\omega{\bf D}_{\tilde k}^* \,.
\label{eq:5.10}
\end{eqnarray}

By using the constitutive relations~(\ref{eq:2.8}), (\ref{eq:2.9})
and Eqs.~(\ref{eq:4.4}), (\ref{eq:4.5}), (\ref{eq:5.9}), and
(\ref{eq:5.10}) it is easy to prove that
\begin{equation}
\mbox{\boldmath$\nabla$}\cdot({\bf E}_{\tilde k}^*\times{\bf H} +
{\bf E}\times{\bf H}_{\tilde k}^*) =
-\,({\bf J}_b^e\cdot{\bf E}_{\tilde k}^* +
{\bf J}_b^m\cdot{\bf H}_{\tilde k}^*)\,-
\label{eq:5.11}
\end{equation}
\[
-\,i\omega \Bigl[\,
(\,\bar{\!\mbox{\boldmath$\epsilon$}}-
\bar{\!\mbox{\boldmath$\epsilon$}}\,^\dagger):
{\bf E}{\bf E}_{\tilde k}^* +
(\,\bar{\!\mbox{\boldmath$\mu$}}-
\bar{\!\mbox{\boldmath$\mu$}}\,^\dagger):
{\bf H}{\bf H}_{\tilde k}^* +
(\,\bar{\!\mbox{\boldmath$\xi$}}-
\bar{\!\mbox{\boldmath$\zeta$}}\,^\dagger):
{\bf H}{\bf E}_{\tilde k}^* +
(\,\bar{\!\mbox{\boldmath$\zeta$}}-
\bar{\!\mbox{\boldmath$\xi$}}\,^\dagger):
{\bf E}{\bf H}_{\tilde k}^* \,\Bigr]
\]
where the square bracket equals zero for a lossless medium owing
to Eq.~(\ref{eq:2.13}).

The contour integrals in the right-hand side of Eq.~(\ref{eq:5.8})
include two contributions:\\
(i) from the {\it actual surface currents\/} \,${\bf J}_s^e$\, and
\,${\bf J}_s^m$\, which are located on a contour $L_s$ and meet the
boundary conditions (\ref{eq:4.6}) and (\ref{eq:4.7}),\\
(ii) from the {\it effective surface currents\/} ${\bf J}_{s,ef}^e$
and ${\bf J}_{s,ef}^m$ given by Eqs.~(\ref{eq:4.27}) and
(\ref{eq:4.28}) which are located on a contour $L_b$ bounding the
bulk current area $S_b$ and meet the boundary conditions~(\ref{eq:4.25})
and (\ref{eq:4.26}). \\[.2cm]
\indent
On the strength of the aforesaid we can write
\[
\sum_i \oint_{L_i} \Bigl[\,
{\bf n}_i^+\cdot ({\bf E}_{\tilde k}^*\times{\bf H} +
{\bf E}\times{\bf H}_{\tilde k}^*)^+ \,+\,
{\bf n}_i^-\cdot ({\bf E}_{\tilde k}^*\times{\bf H} +
{\bf E}\times{\bf H}_{\tilde k}^*)^- \Bigr] dl =             \\[-0.3cm]
\]
\[
= \int_{L_s} \Bigl[\,
{\bf n}_s^+\cdot ({\bf E}_{\tilde k}^*\times{\bf H} +
{\bf E}\times{\bf H}_{\tilde k}^*)^+ \,+\,
{\bf n}_s^-\cdot ({\bf E}_{\tilde k}^*\times{\bf H} +
{\bf E}\times{\bf H}_{\tilde k}^*)^- \Bigr] dl \,+
\]
\[
+ \int_{L_b} \Bigl[\,
{\bf n}_b^+\cdot ({\bf E}_{\tilde k}^*\times{\bf H} +
{\bf E}\times{\bf H}_{\tilde k}^*)^+ \,+\,
{\bf n}_b^-\cdot ({\bf E}_{\tilde k}^*\times{\bf H} +
{\bf E}\times{\bf H}_{\tilde k}^*)^- \Bigr] dl =
\]
\begin{equation}
=\,-\int_{L_s} ({\bf J}_s^e\cdot{\bf E}_{\tilde k}^* +
{\bf J}_s^m\cdot{\bf H}_{\tilde k}^*)\,dl \,-
\int_{L_b} ({\bf J}_{s,ef}^e\cdot{\bf E}_{\tilde k}^* +
{\bf J}_{s,ef}^m\cdot{\bf H}_{\tilde k}^*)\,dl \,.
\label{eq:5.12} \\[.2cm]
\end{equation}

Eqs.~(\ref{eq:5.8}), (\ref{eq:5.11}), and (\ref{eq:5.12}) finally give
the desired equations             
written as      \\[.2cm]
(i) for the excitation amplitudes $A_k(z),\;\;k= \pm\,n$:
\[
{dA_k\over dz} \,=\,-\,{1\over N_{\tilde k}} \int_{S_b}
({\bf J}_b^e\cdot{\bf E}_{\tilde k}^* +
{\bf J}_b^m\cdot{\bf H}_{\tilde k}^*)\,dS \;-
\]
\begin{equation}
-\;{1\over N_{\tilde k}} \int_{L_s}
({\bf J}_s^e\cdot{\bf E}_{\tilde k}^* +
{\bf J}_s^m\cdot{\bf H}_{\tilde k}^*)\,dl \,-\,
{1\over N_{\tilde k}} \int_{L_b}
({\bf J}_{s,ef}^e\cdot{\bf E}_{\tilde k}^* +
{\bf J}_{s,ef}^m\cdot{\bf H}_{\tilde k}^*)\,dl \,;
\label{eq:5.13} \\[.2cm]
\end{equation}
(ii) for the mode amplitudes $a_k(z)\!=\!
A_k(z)\exp(-\,\gamma_k z),\;\;k= \pm\,n$:
\[
{da_k\over dz}+\,\gamma_k\,a_k\,=\,-\,{1\over N_{\tilde k}} \int_{S_b}
({\bf J}_b^e\cdot\hat{\bf E}_{\tilde k}^* +
{\bf J}_b^m\cdot\hat{\bf H}_{\tilde k}^*)\,dS \;-
\]
\begin{equation}
-\;{1\over N_{\tilde k}} \int_{L_s}
({\bf J}_s^e\cdot\hat{\bf E}_{\tilde k}^* +
{\bf J}_s^m\cdot\hat{\bf H}_{\tilde k}^*)\,dl \,-\,
{1\over N_{\tilde k}} \int_{L_b}
({\bf J}_{s,ef}^e\cdot\hat{\bf E}_{\tilde k}^* +
{\bf J}_{s,ef}^m\cdot\hat{\bf H}_{\tilde k}^*)\,dl \,.
\label{eq:5.14} \\[.2cm]
\end{equation}

The excitation equations~(\ref{eq:5.13}) and (\ref{eq:5.14}) written
for the amplitudes of reactive modes hold true also for an active
mode if one assumes ${\tilde k}\!=\!k$ and $\gamma_k\!=\!i\beta_k$.
The excitation integrals in the right-hand side of these equations
represent the complex power of interaction between the external
currents (bulk and surface) and the eigenfields of the $k$th mode
(for active ones) or those of its twin-conjugate $\tilde k$th mode
(for reactive ones).      \\[.1cm]
\indent
As distinct from the theory developed, Vainshtein~\cite{2} fully
excluded from consideration the reactive (nonpropagating) modes and
the effective surface currents and restricted his analysis only to
the reciprocal waveguides with isotropic media. In this case every
forward-propagating mode $(k\!=\!+\,n)$ has a backward counterpart
$(k\!=\!-\,n)$ of the same type so that their common norm is defined
by Vainshtein as
\[
N_n= \int_S (\hat{\bf E}_{+n}\times\hat{\bf H}_{-n} -
\hat{\bf E}_{-n}\times\hat{\bf H}_{+n})\cdot{\bf z}_0\,dS \,.
\]

Unlike the definition~(\ref{eq:3.14}), Vainshtein's norm has no power
sense and does not allow a generalization to nonreciprocal waveguides
to be made.      \\[.1cm]
\indent
Theory of Felsen and Marcuvitz~\cite{8}, unlike Vainshtein's theory,
takes into consideration anisotropic (not bianisotropic) media but
also does not allow for the reactive modes. The excitation integral in
their equation similar to our Eq.~(\ref{eq:5.14}) has a visually
different form which does not involve the effective surface currents
explicitly. In order for their implicit existence to be displayed, let
us convert our excitation integral containing the bulk currents.

To this end, it is necessary to transform the products of longitudinal
components such as $J_{bz}^e\,E_{kz}^*$ and $J_{bz}^m\,H_{kz}^*$ (where
\,$k\!=\!{\tilde k}$\, for reactives modes). The use of the
constitutive relations
${\bf D}= \bar{\!\mbox{\boldmath$\epsilon$}}\cdot{\bf E}$\, and\,
${\bf B}= \bar{\!\mbox{\boldmath$\mu$}}\cdot{\bf H}$ for a
double-anisotropic medium in Eqs.~(\ref{eq:5.9}) and (\ref{eq:5.10})
written for the $k$th mode gives the longitudinal projections of these
equations:
\begin{eqnarray}
\!\!\!\!\!\!\!\!\!\!(\mbox{\boldmath$\nabla$}\times{\bf E}_k^*)_z\equiv
(\mbox{\boldmath$\nabla$}_t\times{\bf E}_{kt}^*)\cdot{\bf z}_0
\!\!&=&\!\! i\omega\,
(\mu_{zx}^* H_{kx}^* \!+\mu_{zy}^* H_{ky}^* \!+\mu_{zz}^* H_{kz}^*) \,,
\label{eq:5.15}\\     [.2cm]
\!\!\!\!\!\!\!\!\!\!(\mbox{\boldmath$\nabla$}\times{\bf H}_k^*)_z\!\equiv\!
(\mbox{\boldmath$\nabla$}_t\times{\bf H}_{kt}^*)\cdot{\bf z}_0
\!\!&=&\!\!\! -\,i\omega\,
(\epsilon_{zx}^* E_{kx}^* \!+ \epsilon_{zy}^* E_{ky}^*\!+
\epsilon_{zz}^* E_{kz}^*) \,.
\label{eq:5.16}
\end{eqnarray}

Taking into account that for a lossless medium \,$\epsilon_{ij}^*=
\epsilon_{ji}$\, and \,$\mu_{ij}^*=\mu_{ji}$,\,  on the basis of
Eqs.~(\ref{eq:5.15}) and (\ref{eq:5.16}) we can obtain the following
expressions
\[
J_{bz}^e\,E_{kz}^* = -\,{1\over i\omega\epsilon_{zz}}\,
(\mbox{\boldmath$\nabla$}_t\times{\bf H}_{kt}^*)\cdot{\bf J}_{bz}^e -
{\epsilon_{xz} E_{kx}^*
+ \epsilon_{yz} E_{ky}^*\over\epsilon_{zz}}\,J_{bz}^e \,=
\]
\begin{equation}
= -\,{1\over i\omega}\,\mbox{\boldmath$\nabla$}_t\cdot \biggl(
{\bf H}_{kt}^*\times{{\bf J}_{bz}^e\over\epsilon_{zz}} \biggr)
\,-\,{1\over i\omega} \biggl( \mbox{\boldmath$\nabla$}_t\times
{{\bf J}_{bz}^e\over\epsilon_{zz}} \biggr) \cdot{\bf H}_{kt}^*
\,-\,{\mbox{\boldmath$\epsilon$}_{tz}\over\epsilon_{zz}}\,J_{bz}^e \,,
\label{eq:5.17}
\end{equation}
and
\[
J_{bz}^m\,H_{kz}^* =\, {1\over i\omega\mu_{zz}}\,
(\mbox{\boldmath$\nabla$}_t\times{\bf E}_{kt}^*)\cdot{\bf J}_{bz}^m -
{\mu_{xz} H_{kx}^*
+ \mu_{yz} H_{ky}^*\over\mu_{zz}}\,J_{bz}^m \,=
\]
\begin{equation}
=\,{1\over i\omega}\,\mbox{\boldmath$\nabla$}_t\cdot \biggl(
{\bf E}_{kt}^*\times{{\bf J}_{bz}^m\over\mu_{zz}} \biggr)
\,+\,{1\over i\omega} \biggl( \mbox{\boldmath$\nabla$}_t\times
{{\bf J}_{bz}^m\over\mu_{zz}} \biggr) \cdot{\bf E}_{kt}^*
\,-\,{\mbox{\boldmath$\mu$}_{tz}\over\mu_{zz}}\,J_{bz}^m \,,
\label{eq:5.18} \\[.3cm]
\end{equation}
where following~\cite{8} we have introduced the auxiliary vectors
\begin{equation}
\mbox{\boldmath$\epsilon$}_{tz} =\,
{\bf x}_0\,\epsilon_{xz}+ {\bf y}_0\,\epsilon_{yz}
\qquad \mbox{and} \qquad
\mbox{\boldmath$\mu$}_{tz} =\,{\bf x}_0\,\mu_{xz}+ {\bf y}_0\,\mu_{yz} \;.
\label{eq:5.19}
\end{equation}

It is easy to see that the terms in Eqs.~(\ref{eq:5.17}) and
(\ref{eq:5.18}) containing the transverse divergence operator
$\mbox{\boldmath$\nabla$}_t\,\cdot$\,, after integrating over
the bulk current area $S_b$, yield the following results
\[
{1\over i\omega} \int_{S_b} \mbox{\boldmath$\nabla$}_t\cdot \biggl(
{\bf E}_{kt}^*\times{{\bf J}_{bz}^m\over\mu_{zz}} \biggr)\,dS =
{1\over i\omega} \oint_{L_b} {\bf n}_b\cdot \biggl(
{\bf E}_{kt}^*\times{{\bf J}_{bz}^m\over\mu_{zz}}
\biggr)\,dl\,\equiv
\]
\begin{equation}
\equiv\, -\int_{L_b}{\bf J}_{s,ef}^e\cdot{\bf E}_k^*\,dl \;,
\label{eq:5.20}    \\[.15cm]
\end{equation}
\[
-\,{1\over i\omega} \int_{S_b} \mbox{\boldmath$\nabla$}_t\cdot \biggl(
{\bf H}_{kt}^*\times{{\bf J}_{bz}^e\over\epsilon_{zz}} \biggr)\,dS =
-\,{1\over i\omega} \oint_{L_b} {\bf n}_b\cdot \biggl(
{\bf H}_{kt}^*\times{{\bf J}_{bz}^e\over\epsilon_{zz}}
\biggr)\,dl\,\equiv
\]
\begin{equation}
\equiv\, -\int_{L_b}{\bf J}_{s,ef}^m\cdot{\bf H}_k^*\,dl \;,
\label{eq:5.21} \\[.2cm]
\end{equation}
where expressions~(\ref{eq:4.32}) for the effective surface currents
have been used.

As is quite evident, the terms~(\ref{eq:5.20}) and (\ref{eq:5.21}),
being inserted in the excitation integral with the bulk currents
by means of equalities~(\ref{eq:5.17}) and (\ref{eq:5.18}), fully
compensate for the contribution from the excitation integral with the
effective surface currents entering into Eq.~(\ref{eq:5.13}) and
(\ref{eq:5.14}). Then the excitation equation~(\ref{eq:5.14}) written for
propagating modes takes the form entirely coincident with that of
Felsen and Marcuvitz~\cite{8} (in~different notation):
\[
{da_k\over dz}+\,\gamma_k\,a_k\,=\,-\,{1\over N_k} \int_{S_b}
({\bf J}_{b,ef}^e\cdot\hat{\bf E}_{kt}^* +
{\bf J}_{b,ef}^m\cdot\hat{\bf H}_{kt}^*)\,dS \,-
\]
\begin{equation}
-\,{1\over N_k} \int_{L_s} ({\bf J}_s^e\cdot\hat{\bf E}_k^* +
{\bf J}_s^m\cdot\hat{\bf H}_k^*)\,dl
\label{eq:5.22} \\[.2cm]
\end{equation}
where following~\cite{8} we have introduced the effective bulk currents
\begin{equation}
{\bf J}_{b,ef}^e\,=\;{\bf J}_{bt}^e \,+\,
{1\over i\omega} \biggl( \mbox{\boldmath$\nabla$}_t\times
{{\bf J}_{bz}^m\over\mu_{zz}} \biggr)\,-\,
{\mbox{\boldmath$\epsilon$}_{tz}\over\epsilon_{zz}}\,J_{bz}^e \;,
\label{eq:5.23}  \\[.15cm]
\end{equation}
\begin{equation}
{\bf J}_{b,ef}^m =\;{\bf J}_{bt}^m \,-\,
{1\over i\omega} \biggl( \mbox{\boldmath$\nabla$}_t\times
{{\bf J}_{bz}^e\over\epsilon_{zz}} \biggr) \,-\,
{\mbox{\boldmath$\mu$}_{tz}\over\mu_{zz}}\,J_{bz}^m \;,
\label{eq:5.24} \\[.2cm]
\end{equation}
with the transverse vectors \,$\mbox{\boldmath$\epsilon$}_{tz}$\, and
\,$\mbox{\boldmath$\mu$}_{tz}$ being defined by formulas~(\ref{eq:5.19}).
The currents~(\ref{eq:5.23}) and (\ref{eq:5.24}) were introduced by
Felsen and Marcuvitz in different designations but of the same structure.

From Eq.~(\ref{eq:5.22}) it follows that the {\it effective bulk
currents\/} ${\bf J}_{b,ef}^e$\, and \,${\bf J}_{b,ef}^m$,
being formed as mixtures of the longitudinal and
transverse components of the {\it actual\/} electric and magnetic
currents, interact only with the transverse eigenfield components
$\hat{\bf E}_{kt}$ and $\hat{\bf H}_{kt}$ of the $k$th mode, but in doing
so take into account the contribution from the {\it effective surface
currents\/} ${\bf J}_{s,ef}^e$\, and \,${\bf J}_{s,ef}^m$ defined by
Eqs.~(\ref{eq:4.32}). It should be mentioned that the contour integral
in Eq.~(\ref{eq:5.22}) allowing for the contribution from the {\it actual
surface currents\/} \,${\bf J}_s^e$\, and \,${\bf J}_s^m$\, is absent
in the appropriate equation of Felsen and Marcuvitz~\cite{8}.   \\[.1cm]
\indent
In view of fundamental importance of the mode excitation equations,
Appendix\,B displays another derivation for the lossless waveguding
structures based on the direct use of Eqs.~(\ref{eq:4.12}) and
(\ref{eq:4.13}) which are an exact consequence of Maxwell's
equations~(\ref{eq:4.4}) and (\ref{eq:4.5}). The general case of lossy
waveguides is studied below on the basis of the reciprocity theorem in
complex-conjugate form.

\subsection{Approach Based on the Reciprocity Theorem}
\label{sec:5B}

\subsubsection{Derivation of the Conjugate Reciprocity Theorem}
\label{sec:5B1}
\indent

The basis of deriving the reciprocity theorem in complex-conjugate
form is constituted by two systems of Maxwell's equations like
Eqs.~(\ref{eq:4.4}) and (\ref{eq:4.5}):
\begin{equation}
\mbox{\boldmath$\nabla$}\times{\bf E}_1\;= -\,
i\omega{\bf B}_1\;-\;{\bf J}_{b1}^m \,,
\quad\qquad
\mbox{\boldmath$\nabla$}\times{\bf E}_2^*\;=\;\;i\omega{\bf B}_2^*\;-\;
{\bf J}_{b2}^{m*},\;
\label{eq:5.25}
\end{equation}
\begin{equation}
\mbox{\boldmath$\nabla$}\times{\bf H}_1\,=\;\;
i\omega{\bf D}_1\;+\;{\bf J}_{b1}^e \,,\;
\quad\qquad
\mbox{\boldmath$\nabla$}\times{\bf H}_2^*\,=-\,i\omega{\bf D}_2^*\;+\;
{\bf J}_{b2}^{e*},
\label{eq:5.26} \\[.2cm]
\end{equation}
written for two different electromagnetic processes (marked with
subscripts~1 and 2) excited by different external currents (bulk and
surface), with the frequency and the constitutive parameters of a
waveguiding medium entering into relations~(\ref{eq:2.8}) and
(\ref{eq:2.9}) assumed to be the same.

Application of the conventional technique to Eqs.~(\ref{eq:5.25}) and
(\ref{eq:5.26}) yields
\[
\mbox{\boldmath$\nabla$}\!\cdot({\bf E}_1\times{\bf H}_2^* +
{\bf E}_2^*\times{\bf H}_1) =
-\,({\bf J}_{b1}^e\cdot{\bf E}_2^* +{\bf J}_{b2}^{e*}\cdot{\bf E}_1) -
({\bf J}_{b1}^m\cdot{\bf H}_2^* +{\bf J}_{b2}^{m*}\cdot{\bf H}_1)\,-
\]
\begin{equation}
-\,i\omega \Bigl[ \,
({\bf D}_1\cdot{\bf E}_2^* -{\bf D}_2^*\cdot{\bf E}_1)\,+\,
({\bf B}_1\cdot{\bf H}_2^* -{\bf B}_2^*\cdot{\bf H}_1)\, \Bigr] \,.
\label{eq:5.27} \\[.2cm]
\end{equation}

After transformation with using the constitutive relations~(\ref{eq:2.8})
and (\ref{eq:2.9}) the last term in the right-hand side of
Eq.~(\ref{eq:5.27}) accepts the following~form
\[
i\omega \Bigl[ \,
({\bf D}_1\cdot{\bf E}_2^* -{\bf D}_2^*\cdot{\bf E}_1)\,+\,
({\bf B}_1\cdot{\bf H}_2^* -{\bf B}_2^*\cdot{\bf H}_1)\, \Bigr] \,=
\]
\[
= i\omega \Bigl[\, (\,\bar{\!\mbox{\boldmath$\epsilon$}}-
\bar{\!\mbox{\boldmath$\epsilon$}}\,^\dagger):
{\bf E}_1{\bf E}_2^* \,+\,
(\,\bar{\!\mbox{\boldmath$\mu$}}-
\bar{\!\mbox{\boldmath$\mu$}}\,^\dagger):
{\bf H}_1{\bf H}_2^* \,+\,
(\,\bar{\!\mbox{\boldmath$\xi$}}-
\bar{\!\mbox{\boldmath$\zeta$}}\,^\dagger):
{\bf H}_1{\bf E}_2^* \,+\,
(\,\bar{\!\mbox{\boldmath$\zeta$}}-
\bar{\!\mbox{\boldmath$\xi$}}\,^\dagger):
{\bf E}_1{\bf H}_2^* \,\Bigr] .                 \\[.2cm]
\]

In accordance with the aforesaid in Sec.\,4.1, the permittivity tensor
\,$\bar{\!\mbox{\boldmath$\epsilon$}}$\, is~regarded here as a sum
$(\,\bar{\!\mbox{\boldmath$\epsilon$}}\,+\;
\bar{\!\mbox{\boldmath$\sigma$}}_c/i\omega)\,$
so that its antihermitian part determines the total tensor of electric
conductivity \,$\bar{\!\mbox{\boldmath$\sigma$}}_e=
\bar{\!\mbox{\boldmath$\sigma$}}_c+\,
\bar{\!\mbox{\boldmath$\sigma$}}_d$\, taking into account
both dielectric (\,$\bar{\!\mbox{\boldmath$\sigma$}}_d$) and conductor
(\,$\bar{\!\mbox{\boldmath$\sigma$}}_c$) losses of a medium.
Magnetic losses (\,$\bar{\!\mbox{\boldmath$\sigma$}}_m$) are taken
into account by the antihermitian part of the permeability tensor
\,$\bar{\!\mbox{\boldmath$\mu$}}$,\, whereas the tensor
\,$\bar{\!\mbox{\boldmath$\sigma$}}_{me} =
i\omega(\,\bar{\!\mbox{\boldmath$\xi$}}-\,
\bar{\!\mbox{\boldmath$\zeta$}}\,^\dagger)$\, reflects the
magneto-electric losses due to bianisotropic properties of a medium.
The use of Eqs.~(\ref{eq:2.21}) -- (\ref{eq:2.23}) converts
relation~(\ref{eq:5.27}) into the differential form of the conjugate
reciprocity theorem
\begin{equation}
\mbox{\boldmath$\nabla$}\cdot{\bf S}_{12}\,+\,q_{12}\,=\,r_{12}^{(b)}
\label{eq:5.28}
\end{equation}
where we have denoted
\begin{equation}
{\bf S}_{12} =\,{\bf E}_1\times{\bf H}_2^* +{\bf E}_2^*\times{\bf H}_1\,,
\label{eq:5.29}    \\[.1cm]
\end{equation}
\begin{equation}
q_{12} =\, 2\,(\,\bar{\!\mbox{\boldmath$\sigma$}}_e:
{\bf E}_1{\bf E}_2^* +
\,\bar{\!\mbox{\boldmath$\sigma$}}_m:{\bf H}_1{\bf H}_2^*) \,+\,
(\,\bar{\!\mbox{\boldmath$\sigma$}}_{me}:{\bf H}_1{\bf E}_2^* +
\,\bar{\!\mbox{\boldmath$\sigma$}}_{me}^{\,\dagger}:
{\bf E}_1{\bf H}_2^*) \,,
\label{eq:5.30} \\[.25cm]
\end{equation}
\begin{equation}
r_{12}^{(b)}\,=\,
-\,({\bf J}_{b1}^e\cdot{\bf E}_2^* +{\bf J}_{b2}^{e*}\cdot{\bf E}_1)\,
-\,({\bf J}_{b1}^m\cdot{\bf H}_2^* +{\bf J}_{b2}^{m*}\cdot{\bf H}_1)\,.
\label{eq:5.31} \\[.2cm]
\end{equation}
Superscript $(b)$ reflects belonging an appropriate quantity
to bulk properties of a system, while the surface properties will be
marked by superscript~$(s)$.      \\[.1cm]
\indent
To obtain the integral form of the reciprocity theorem it is necessary
to integrate Eq.~(\ref{eq:5.28}) over the cross section $S$ of a
waveguiding structure with using the integral relation similar to
Eq.~(\ref{eq:2.25}) which involves the contour integrals taking into
account two physical phenomena:\\
(i) the skin losses expressed by the boundary condition~(\ref{eq:2.26})
with the surface impedance tensor~(\ref{eq:2.27}) given along
a contour $L$\,,\\
(ii) the discontinuity in tangential components of the fields caused both
by the {\it actual surface currents\/} ${\bf J}_s^e$ and ${\bf J}_s^m$
located on a contour $L_s$ with the boundary conditions~(\ref{eq:4.6})
and (\ref{eq:4.7}) and by the {\it effective surface currents\/}
${\bf J}_{s,ef}^e$ and ${\bf J}_{s,ef}^m$ located on a contour $L_b$
with the boundary conditions~(\ref{eq:4.25}) and (\ref{eq:4.26}).

For the sake of brevity it is convenient to write both surface
currents as the overall surface sources
\begin{equation}
{\bf J}_{\scriptscriptstyle\Sigma}^e\,=\,
{\bf J}_s^e\,+\,{\bf J}_{s,ef}^e
\qquad \mbox{and} \qquad
{\bf J}_{\scriptscriptstyle\Sigma}^m\,=\,
{\bf J}_s^m\,+\,{\bf J}_{s,ef}^m
\label{eq:5.32}
\end{equation}
located along the combined contour $L_{\scriptscriptstyle\Sigma}=L_s+L_b$.

Substitution of Eq.~(\ref{eq:5.29}) into the integral relation
(\ref{eq:2.25}) yields
\[
\int_S \mbox{\boldmath$\nabla$}\cdot{\bf S}_{12}\,dS\,=\,
{\partial\over\partial z} \int_S
\Bigl( {\bf E}_1\times{\bf H}_2^* +
{\bf E}_2^*\times{\bf H}_1 \Bigr) \cdot{\bf z}_0\,dS \;-      \\[-0.1cm]
\]
\[
- \int_{L+L_s} \Bigl[ \,
{\bf n}_s^+\!\cdot({\bf E}_1\times{\bf H}_2^* +
{\bf E}_2^*\times{\bf H}_1)^+ +\;
{\bf n}_s^-\!\cdot({\bf E}_1\times{\bf H}_2^* +
{\bf E}_2^*\times{\bf H}_1)^- \Bigr] dl \;+
\]
\[
+ \int_{L_b}  {\bf n}_b\cdot  \Bigl[ \,
({\bf E}_{a1}\times{\bf H}_{b2}^* +
{\bf E}_{b2}^*\times{\bf H}_{a1}) \,+\,
({\bf E}_{b1}\times{\bf H}_{a2}^* +
{\bf E}_{a2}^*\times{\bf H}_{b1}) \Bigr] dl \,=
\]
\[
=\,{\partial\over\partial z} \int_S
\Bigl( {\bf E}_1\times{\bf H}_2^* +
{\bf E}_2^*\times{\bf H}_1 \Bigr) \cdot{\bf z}_0\,dS \,+\,
2 \int_L {\cal R}_s\,({\bf H}_{\tau1}\cdot{\bf H}_{\tau2}^*) \,dl \;+
\]
\[
+ \int_{L_s} \Bigl[ \,
({\bf J}_{s1}^e\cdot{\bf E}_2^* + {\bf J}_{s1}^m\cdot{\bf H}_2^*) \,+\,
({\bf J}_{s2}^{e*}\cdot{\bf E}_1 + {\bf J}_{s2}^{m*}\cdot{\bf H}_1)\,
\Bigr] dl \;+
\]
\[
+ \int_{L_b} \Bigl[ \,
({\bf J}_{s,ef1}^e\cdot{\bf E}_{a2}^* +
{\bf J}_{s,ef1}^m\cdot{\bf H}_{a2}^*) \,+\,
({\bf J}_{s,ef2}^{e*}\cdot{\bf E}_{a1} +
{\bf J}_{s,ef2}^{m*}\cdot{\bf H}_{a1})\, \Bigr] dl \,.      \\[.2cm]
\]

Here we have used:\, (a) the boundary condition~(\ref{eq:2.26}) on
the contour~$L$ with surface impedance~(\ref{eq:2.27}),\, (b) the
boundary conditions~(\ref{eq:4.6}) and~(\ref{eq:4.7}) with the actual
surface currents ${\bf J}_{s1(2)}^e$ and ${\bf J}_{s1(2)}^m$\, given
on the contour~$L_s$,\, (c) the effective surface currents
${\bf J}_{s,ef1(2)}^e = -\,{\bf n}_b\times{\bf H}_{b1(2)}$\, and\,
${\bf J}_{s,ef1(2)}^m = {\bf n}_b\times{\bf E}_{b1(2)}$
defined on the contour $L_b$. Therefore, the line integrals in the
previous formula yield two resulting contributions:\\
(i) from the skin losses on the contour $L$ of a conducting surface
\begin{equation}
q_{12}^\prime =\,2\,{\cal R}_s\,({\bf H}_{\tau1}\cdot{\bf H}_{\tau2}^*)\,,
\label{eq:5.33}
\end{equation}
(ii) from the overall surface currents on the contour
$L_{\scriptscriptstyle\Sigma}= L_s+ L_b$
\begin{equation}
r_{12}^{(s)}\,=\,-\,
({\bf J}_{{\scriptscriptstyle\Sigma}1}^e\cdot{\bf E}_2^* +
{\bf J}_{{\scriptscriptstyle\Sigma}2}^{e*}\cdot{\bf E}_1)\,-\,
({\bf J}_{{\scriptscriptstyle\Sigma}1}^m\cdot{\bf H}_2^* +
{\bf J}_{{\scriptscriptstyle\Sigma}2}^{m*}\cdot{\bf H}_1) \,.
\label{eq:5.34}    \\[.2cm]
\end{equation}

Therefore, the reciprocity theorem in the integral form is given by
the relation
\begin{equation}
{dP_{12}(z)\over dz}\,+\,Q_{12}(z)\,=\,R_{12}(z)
\label{eq:5.35}       \\[.1cm]
\end{equation}
where we have introduced the following integral quantities
(complex-valued) \,(cf.~Eqs.~(\ref{eq:2.29}) and (\ref{eq:2.30}))
\begin{equation}
P_{12}(z)\,\equiv\,
\int_S {\bf S}_{12}({\bf r}_t,z)\cdot{\bf z}_0\,dS\,=\,
\int_S ({\bf E}_1\times{\bf H}_2^* +{\bf E}_2^*\times{\bf H}_1)
\cdot{\bf z}_0\,dS \,,
\label{eq:5.36}       \\[.15cm]
\end{equation}
\[
Q_{12}(z)\,=\,Q_{12}^{(b)}(z) \,+\, Q_{12}^{(s)}(z) \,\equiv\,
\int_S q_{12}({\bf r}_t,z)\,dS\,+
\int_L q_{12}^\prime({\bf r}_t,z)\,dl\,=           \\[.1cm]
\]
\[
=\,2 \int_S (\,\bar{\!\mbox{\boldmath$\sigma$}}_e:
{\bf E}_1{\bf E}_2^*)\,dS\,+\,
2 \int_S (\,\bar{\!\mbox{\boldmath$\sigma$}}_m:
{\bf H}_1{\bf H}_2^*)\,dS\,+                       \\[.1cm]
\]
\begin{equation}
+ \int_S (\,\bar{\!\mbox{\boldmath$\sigma$}}_{me}:{\bf H}_1{\bf E}_2^* +
\,\bar{\!\mbox{\boldmath$\sigma$}}_{me}^{\,\dagger}:
{\bf E}_1{\bf H}_2^*)\,dS\,+\,
2 \int_L {\cal R}_s\,({\bf H}_{\tau1}\cdot{\bf H}_{\tau2}^*)\,dl \,,
\label{eq:5.37} \\[.35cm]
\end{equation}
\[
R_{12}(z)\,=\,R_{12}^{(b)}(z) + R_{12}^{(s)}(z) \,\equiv\,
\int_{S_b} r_{12}^{(b)}({\bf r}_t,z)\,dS\,+
\int_{L_\Sigma} r_{12}^{(s)}({\bf r}_t,z)\,dl\,=     \\[.1cm]
\]
\[
=\,- \int_{S_b} \Bigl[ \,
({\bf J}_{b1}^e\cdot{\bf E}_2^* +{\bf J}_{b1}^m\cdot{\bf H}_2^* )\,+\,
({\bf J}_{b2}^{e*}\cdot{\bf E}_1 +{\bf J}_{b2}^{m*}\cdot{\bf H}_1)\,
\Bigr] \,dS \,-                                      \\[.1cm]
\]
\begin{equation}
- \int_{L_\Sigma} \Bigl[ \,
({\bf J}_{{\scriptscriptstyle\Sigma}1}^e\cdot{\bf E}_2^* +
{\bf J}_{{\scriptscriptstyle\Sigma}1}^m\cdot{\bf H}_2^*)\,+\,
({\bf J}_{{\scriptscriptstyle\Sigma}2}^{e*}\cdot{\bf E}_1 +
{\bf J}_{{\scriptscriptstyle\Sigma}2}^{m*}\cdot{\bf H}_1)\,
\Bigr]\,dl  \,.
\label{eq:5.38} \\[.3cm]
\end{equation}

It is easy to see that with no sources (when $R_{12}= 0$) the second
system (with subscript~2) of Maxwell's equations~(\ref{eq:5.25})
and (\ref{eq:5.26}) describes the same fields as the first (marked by
subscript~1) only with taking complex conjugation. This makes it
possible to replace subscript~2 with 1 and what is more to drop them.
In this case the integral reciprocity theorem~(\ref{eq:5.35})  turns
into the integral Poynting theorem~(\ref{eq:2.28}) in which the real
power flow $P$ and the real power loss (bulk and surface) $Q$ per unit
length of a waveguide are equal to
\begin{equation}  \!\!\!\!\!\!\!\!\!\!
P\,=\,{1\over4} \int_{S} {\bf S}_{11}\cdot{\bf z}_0\,dS \,\equiv\,
\int_S \langle\,{\bf S}\,\rangle\cdot{\bf z}_0\,dS \,,
\label{eq:5.39}    \\[.2cm]
\end{equation}
\begin{equation}    \;\;
Q\,=\,{1\over4} \int_S q_{11}\,dS \,+\,
{1\over4} \int_L q_{11}^\prime\,dl \,\equiv\,
\int_S \langle\,q\,\rangle\,dS \,+\,
\int_L \langle\,q^\prime\,\rangle\,dl \,,
\label{eq:5.40} \\[.25cm]
\end{equation}
where their expressions in terms of fields are given by
Eq.~(\ref{eq:2.29}) and (\ref{eq:2.30}).

\subsubsection{Derivation of the Equations of Mode Excitation}
\label{sec:5B2}
\indent

Inside the source region the reciprocity theorem in the integral form
(\ref{eq:5.35}) is the basis for obtaining the excitation equations.
To this end, the fields marked by subscript~1 (which will be dropped
for the exciting currents) are assumed to be the desired fields excited
by the bulk and surface sources (${\bf J}_{b1}^{e,m}\equiv
{\bf J}_b^{e,m}\ne 0$\, and
\,${\bf J}_{{\scriptscriptstyle\Sigma}1}^{e,m}\equiv
{\bf J}_{\scriptscriptstyle\Sigma}^{e,m}\ne 0$) and represented in the
form of expressions~(\ref{eq:4.8}) and (\ref{eq:4.9}) (with replacing
summation index $k$ by $l$), whereas those marked by subscript~2
are the known fields of the $k$th mode outside the source region
$({\bf J}_{b2}^{e,m}={\bf J}_{{\scriptscriptstyle\Sigma}2}^{e,m}= 0)$
given in the form of Eq.~(\ref{eq:3.5}).

Substitution of Eqs.~(\ref{eq:4.8}) and (\ref{eq:4.9}) into
Eqs.~(\ref{eq:5.36}), (\ref{eq:5.37}), and (\ref{eq:5.38}) yields
the~following expressions:
\[
P_{1k}(z)\,\equiv\,\int_S {\bf S}_{1k}({\bf r}_t,z)\cdot{\bf z}_0\,dS \,=
\int_S ({\bf E}_1\times{\bf H}_k^* +{\bf E}_k^*\times{\bf H}_1)
\cdot{\bf z}_0\,dS\,=
\]
\begin{equation}
=\,\sum_l N_{kl}\,A_l(z)\,{\rm e}^{-\,(\gamma_k^* +\gamma_l) z} ,
\label{eq:5.41}      \\[.1cm]
\end{equation}
\[
Q_{1k}(z)\,\equiv\,\int_S q_{1k}({\bf r}_t,z)\,dS\,+
\int_L q_{1k}^\prime({\bf r}_t,z)\,dl\,=         \\[.1cm]
\]
\begin{equation}
=\,\sum_l M_{kl}\,A_l(z)\,{\rm e}^{-\,(\gamma_k^* +\gamma_l) z} ,
\label{eq:5.42}      \\[.1cm]
\end{equation}
\[
R_{1k}(z)\,=\,R_{1k}^{(b)}(z)\,+\,R_{1k}^{(s)}(z)\,\equiv\,
\int_{S_b} r_{1k}^{(b)}({\bf r}_t,z)\,dS \,+
\int_{L_\Sigma} r_{1k}^{(s)}({\bf r}_t,z)\,dl \,=       \\[.1cm]
\]
\begin{equation}
=\,R_k^{(b)}(z)\,{\rm e}^{-\,\gamma_k^* z} +\,
R_k^{(s)}(z)\,{\rm e}^{-\,\gamma_k^* z}\,\equiv\,
R_k(z)\,{\rm e}^{-\,\gamma_k^* z},
\label{eq:5.43} \\[.2cm]
\end{equation}
where the normalizing and dissipative coefficients $N_{kl}$ and
$M_{kl}$ have the form of Eqs.~(\ref{eq:2.36}) and (\ref{eq:2.37})
and the quantity $R_k(z)= R_k^{(b)}(z)+ R_k^{(s)}(z)$ consists of
two exciting integrals (bulk~and~surface):
\begin{equation}
R_k^{(b)}(z)\,= - \int_{S_b} \Bigl(
{\bf J}_b^e\cdot\hat{\bf E}_k^* +
{\bf J}_b^m\cdot\hat{\bf H}_k^* \Bigr) \,dS \,,
\label{eq:5.44}
\end{equation}
\[
R_k^{(s)}(z)\,= - \int_{L_\Sigma} \Bigl(
{\bf J}_{\scriptscriptstyle\Sigma}^e\cdot\hat{\bf E}_k^* +
{\bf J}_{\scriptscriptstyle\Sigma}^m\cdot\hat{\bf H}_k^* \Bigr)\,dl \,=
\]
\begin{equation}
= - \int_{L_s} \Bigl(
{\bf J}_s^e\cdot\hat{\bf E}_k^* +
{\bf J}_s^m\cdot\hat{\bf H}_k^* \Bigr)\,dl -
\int_{L_b} \Bigl(
{\bf J}_{s,ef}^e\cdot\hat{\bf E}_k^* +
{\bf J}_{s,ef}^m\cdot\hat{\bf H}_k^* \Bigr) \,dS \,.
\label{eq:5.45} \\[.3cm]
\end{equation}

These integrals involve the cross-section eigenfield vectors (marked
with hat) and their dependence on $z$ is due to that of the external
currents ${\bf J}_b^{e,m}(z)$ and
${\bf J}_{\scriptscriptstyle\Sigma}^{e,m}(z)$.

It should be pointed out that Eqs.~(\ref{eq:5.41}) and
(\ref{eq:5.42}) come only from the field contributions of the
mode expansions ${\bf E}_a$ and ${\bf H}_a$ since the orthogonal
complementary fields ${\bf E}_b$ and ${\bf H}_b$, being
proportional to the longitudinal component of external
currents, do not contribute into $P_{1k}$ and cannot
influence the intrinsic losses in a medium related to~$Q_{1k}$.  \\[.1cm]
\indent
Substitution of Eqs.~(\ref{eq:5.41}), (\ref{eq:5.42}), and
(\ref{eq:5.43}) into the integral reciprocity theorem~(\ref{eq:5.35})
(with replacing 2 by $k$) gives a relation
\begin{equation}
\sum_l \biggl\{ N_{kl}\,{dA_l\over dz}\,-\,
\Bigl[ (\gamma_k^* +\gamma_l) N_{kl}- M_{kl} \Bigr] A_l
\biggr\} \,{\rm e}^{-\,\gamma_l z} =\,
R_k\,\equiv R_k^{(b)}\,+\, R_k^{(s)} .
\label{eq:5.46}
\end{equation}

The quasi-orthogonality relation of the general form~(\ref{eq:3.6}) make
the square bracket in Eq.~(\ref{eq:5.46}) vanish so that it reduces to the
desired set of the excitation equations written in the following form:\\
(i) for the excitation amplitudes $A_l(z)$
\begin{equation}
\sum_l N_{kl}\,{dA_l(z)\over dz}\,{\rm e}^{-\,\gamma_l z} =\,
R_k^{(b)}(z)\,+\,R_k^{(s)}(z) \,, \quad\qquad\qquad  k= 1,2,\ldots
\label{eq:5.47}
\end{equation}
(ii) for the mode amplitudes \,$a_l(z)= A_l(z)\,{\rm e}^{-\,\gamma_l z}$
\begin{equation}
\sum_l N_{kl}\, \biggl[ \,
{da_l(z)\over dz}\,+\,\gamma_l a_l(z) \,\biggr] \,=\,
R_k^{(b)}(z)\,+\,R_k^{(s)}(z) \,, \qquad  k= 1,2,\ldots
\label{eq:5.48}
\end{equation}

Discussion of the excitation equations obtained will be put off until the
similar equations for the waveguiding structures with space-dispersive
media are developed in the second part of the paper. \\ [.15cm]
\indent
Up to this point the waveguiding structures under study are assumed to be
closed with a screening metallic boundary, whose spectrum of eigenmodes is
always discrete.

In conclusion, it is pertinent to show features of the excitation
theory peculiar to open waveguiding structures (without losses) in
which an outside homogeneous medium extends to infinity in one or both
transverse directions. As is known \cite{27,29,32}, for the open
waveguides in addition to the discrete part of the spectrum of bound
modes (with the outside medium fields localized near outer
boundaries of the waveguiding layer), there is a continuous part
of the spectrum related to radiation modes (with the fields extending to
infinity in the outside medium). Unlike the eigenfields
${\bf E}_k({\bf r}_t,z)$ and ${\bf H}_k({\bf r}_t,z)$ of discrete modes
marked by the integer-valued subscript $k\!=\! 1,2,\ldots$\, and
expressed by Eqs.~(\ref{eq:3.5}), the fields of a radiation mode
\begin{equation}
{\bf E}({\bf r}_t,z;{\bf k}_t^{\rm o})=
\hat{\bf E}({\bf r}_t;{\bf k}_t^{\rm o})\,
{\rm e}^{-\,i\beta({\bf k}_t^{\rm o})z},   \quad
{\bf H}({\bf r}_t,z;{\bf k}_t^{\rm o})=
\hat{\bf H}({\bf r}_t;{\bf k}_t^{\rm o})\,
{\rm e}^{-\,i\beta({\bf k}_t^{\rm o})z}
\label{eq:5.49}
\end{equation}
are specified by the transverse wave vector \,${\bf k}_t^{\rm o}=
{\bf x}_0\,k_x^{\rm o} + {\bf y}_0\,k_y^{\rm o}$\, of the outside
passive medium.

In this case the modal expansions of the fields
${\bf E}_a({\bf r}_t,z)$ and ${\bf H}_a({\bf r}_t,z)$
inside the source region, besides the series expansion in terms of
discrete modes, involve also the integral expansion in terms of
radiation modes (cf.~Eqs.~(\ref{eq:4.8}) and (\ref{eq:4.9})):
\begin{equation}
{\bf E}_a({\bf r}_t,z) \,=\, \sum_k A_k(z)\,
\hat{\bf E}_k({\bf r}_t)\,{\rm e}^{-\,i\beta_kz} \,+\,
\int A(z;{\bf k}_t^{\rm o})\,
\hat{\bf E}({\bf r}_t;{\bf k}_t^{\rm o})\,
{\rm e}^{-\,i\beta({\bf k}_t^{\rm o})z}\,d{\bf k}_t^{\rm o} \;,
\label{eq:5.50}
\end{equation}
\begin{equation}
{\bf H}_a({\bf r}_t,z) \,=\, \sum_k A_k(z)\,
\hat{\bf H}_k({\bf r}_t)\,{\rm e}^{-\,i\beta_kz} \,+\,
\int A(z;{\bf k}_t^{\rm o})\,
\hat{\bf H}({\bf r}_t;{\bf k}_t^{\rm o})\,
{\rm e}^{-\,i\beta({\bf k}_t^{\rm o})z}\,d{\bf k}_t^{\rm o} \;,
\label{eq:5.51} \\[.2cm]
\end{equation}
where integrating over $k_x^{\rm o}$ and $k_y^{\rm o}$ is taken
along the real axes from $-\infty$~to~$\infty$.

The orthonormalization relation for radiation modes can be written
by analogy with relation~(\ref{eq:3.17}) for discrete modes in
the following form
\[
N({\bf k}_t^{\rm o},\,{\bf k}_t^{{\rm o}\prime}) \equiv
\int_S \Bigl[ \,
(\hat{\bf E}^*({\bf r}_t;{\bf k}_t^{\rm o})\times
\hat{\bf H}({\bf r}_t;{\bf k}_t^{{\rm o}\prime}) +
\hat{\bf E}({\bf r}_t;{\bf k}_t^{{\rm o}\prime})\times
\hat{\bf H}^*({\bf r}_t;{\bf k}_t^{\rm o})\, \Bigr]
\!\cdot{\bf z}_0\,dS =
\]
\begin{equation}
=\,\delta({\bf k}_t^{\rm o} -
{\bf k}_t^{{\rm o}\prime})\,N({\bf k}_t^{\rm o})
\label{eq:5.52}    \\[.1cm]
\end{equation}
where the Dirac delta function $\delta({\bf k}_t^{\rm o} -
{\bf k}_t^{{\rm o}\prime}) = \delta(k_x^{\rm o} - k_x^{{\rm o}\prime})\,
\delta(k_y^{\rm o} - k_y^{{\rm o}\prime})$ replaces the Kronecker delta
function $\delta_{kl}$.
It should be mentioned that since $\delta({\bf k}_t^{\rm o} -
{\bf k}_t^{{\rm o}\prime})$ has dimensions of (length)$^2$, the
dimensionality of the norm $N({\bf k}_t^{\rm o})$ and the excitation
amplitude $A(z;{\bf k}_t^{\rm o})$ for the radiation modes is equal to
\,watts/m$^2$\, and \,m$^2$,\, respectively, as distinct from the
bounded modes for which the similar quantities are taken in watts and
as dimensionless.

The equation for the excitation amplitude $A(z;{\bf k}_t^{\rm o})$ of the
radiation mode has the form similar to Eq.~(\ref{eq:5.13}):
\[
{dA({\bf k}_t^{\rm o})\over dz} \,=\,-\,
{1\over N({\bf k}_t^{\rm o})} \int_{S_b}
\Bigl( {\bf J}_b^e\cdot{\bf E}^*({\bf k}_t^{\rm o}) \,+\,
{\bf J}_b^m\cdot{\bf H}^*({\bf k}_t^{\rm o}) \Bigr) \,dS \,-
\]
\begin{equation}
-\,{1\over N({\bf k}_t^{\rm o})} \int_{L_\Sigma}
\Bigl( {\bf J}_{\scriptscriptstyle\Sigma}^e\cdot
{\bf E}^*({\bf k}_t^{\rm o}) \,+\,
{\bf J}_{\scriptscriptstyle\Sigma}^m\cdot
{\bf H}^*({\bf k}_t^{\rm o}) \Bigr) \,dl
\label{eq:5.53} \\[.2cm]
\end{equation}
where the coordinate variables are dropped for simplicity.

The orthogonal complementary fields ${\bf E}_b= {\bf z}_0 E_b$ and
${\bf H}_b= {\bf z}_0 H_b$ obtained in the form of Eqs.~(\ref{eq:4.21})
and (\ref{eq:4.22}) remain valid for open waveguides.

\section{CONCLUSION}
\label{sec:6}

We have shown a unified treatment of the electrodynamic theory of the
guided wave excitation by external sources applied to any waveguiding
structure involving the complex media with bianisotropic properties.
Allowing for losses in such media has reduced to the power loss density
in Poynting's theorem due to the {\it magneto-electric conductivity\/}
$\bar{\!\mbox{\boldmath$\sigma$}}_{me}$ defined by Eq.~(\ref{eq:2.23}),
in addition to the usual electric and magnetic conductivities.

Application of the desired field expansions in terms of eigenmode fields
gives the {\it self-power\/} and {\it cross-power\/} quantities
(flows and losses) transmitted and dissipated by the eigenmodes of a
lossy waveguide, as well as the time-average energy density stored by
the propagating modes in a lossless waveguide  which involves the
additional contributions from bianisotropic properties of a medium.

The basis of developing the excitation theory for lossy waveguides is
the novel relation~(\ref{eq:3.6}) called the {\it quasi-orthogonality
relation\/} whose general form is always true including the propa\-gating
(active) and nonpropagating (reactive) modes in lossless waveguides
considered as a special case. Among the external sources exciting the
waveguiding structure we have included the bulk sources (currents, fields,
and medium perturbations) and the actual surface currents. Inside the
source region the modal expansions~(\ref{eq:2.31}) and (\ref{eq:2.32})
have proved to be incomplete and must be supplemented with the
{\it ortho\-gonal complementary fields\/}~(\ref{eq:4.21}) and
(\ref{eq:4.22}), as it is done by Eqs.~(\ref{eq:4.8}) and
(\ref{eq:4.9}). Generally these complementary fields generate the
{\it effective surface currents\/}~(\ref{eq:4.27}) and (\ref{eq:4.28}).
So in the most general case the external source region contains the
bulk currents \,${\bf J}_b^{e,m}$, the actual surface currents
\,${\bf J}_s^{e,m}$, and the effective surface currents
\,${\bf J}_{s,ef}^{e,m}$ brought about by the longitudinal components of
the bulk currents.

The equations of mode excitation in the form of~(\ref{eq:5.47}) or
(\ref{eq:5.48}) have been derived by using three approaches based on:\,
(i) the direct derivation from Maxwell's equations (see Appendix\,B),\,
(ii) the electrodynamic analogy with the mathematical method of
variation of constants (see Sec.\,5.1),\, (iii) the reciprocity theorem
in the complex-conjugate form (see Sec.\,5.2).

All the results obtained are valid for the time-dispersive media
specified by macroscopically-local and frequency-dependent parameters. An
extension of the theory to space-dispersive media which require for their
description the special equations of motion with regard for nonlocal
effects will be examined in the second part of the paper where the
orthogonal complementary fields are explained as a part of the
contribution from the potential fields of external sources.

\vspace{2\baselineskip}
\noindent
{\Large{\bf APPENDIX}}
\appendix
\renewcommand{\thesection}{\Alph{section}.}
\renewcommand{\thesubsection}{\Alph{section}.\arabic{subsection}}
\renewcommand{\theequation}{\Alph{section}.\arabic{equation}}

\setcounter{equation}{0}
\section{BASIC RELATIONS OF FUNCTIONAL ANALYSIS
                      AND THEIR ELECTRODYNAMIC ANALOGS}
\label{app:A}

\subsection{Mathematical formulation (in notation of [42])}
\label{app:A1}
\indent

Unlike~\cite{45}, we shall examine the general case of
{\it nonorthogonal\/} base functions which gives the orthogonal basis as
a special case.

Consider a countable set of complex functions $\psi_1(x), \psi_2(x),
\ldots$ quadratically integ\-rable in the sense of Lebesgue on a given
set $S$ of points $(x)$. The class $L_2(S)$ of such functions (regarded
as vectors) constitutes an infinite-dimensional {\it unitary\/} functional
(vector) space if, in addition to two binary operations of the vector sum
$\psi_k(x)+\psi_l(x)$ and the product $a_k\psi_k(x)$ by a complex scalar
$a_k$, one defines the inner product of $\psi_k(x)$ and $\psi_l(x)$ as
\begin{equation}
(\psi_k, \psi_l)= \int_S \psi_k^*(x)\gamma(x)\psi_l(x)\,dx
\label{eq:A1}
\end{equation}
where the weighting function $\gamma(x)$ is a given real nonnegative
function quadra\-tically integrable on $S$, in particular, may be
$\gamma(x)\equiv 1$.

If Gram's determinant $\det[(\psi_k, \psi_l)]$ built up on the inner
products of the form~(\ref{eq:A1}) differs from zero, the functions
$\psi_k(x),\;k\!=\!1,2,\ldots$\, are linearly independent in $L_2$ and
can be chosen as a basis of the unitary functional space, with their
mutual orthogonality not being necessarily required in ge\-neral. The
given set of functions $\psi_k(x)$ spans a linear manifold comprising
all linear combinations of $\psi_1(x), \psi_2(x), \ldots$ .

Let us compose a partial sum of the $n$th order
\begin{equation}
s_n(x)= \sum_{k=1}^n a_k^{(n)}\psi_k(x)
\label{eq:A2}
\end{equation}
with scalar coefficients $a_k^{(n)}$ not yet defined. Given a function
$\psi(x)$ fully belonging to the linear manifold spanned by
$\psi_1(x), \psi_2(x),\ldots$ , these coefficients can be found from
the requirement that the weighted mean-square difference
\begin{equation}
D_n= \int_S \gamma(x)|s_n(x)- \psi(x)|^2\,dx
\label{eq:A3}
\end{equation}
between $s_n(x)$ and $\psi(x)$ would be minimum. With the help of
Eq.~(\ref{eq:A2}) the quantity $D_n$ can be rewritten in the following
form
\begin{equation}
D_n= \int_S \biggl[\;\sum_{k=1}^n a_k^{(n)*}\psi_k^*(x)-
\psi^*(x)\biggr] \gamma(x) \biggl[\;\sum_{l=1}^n a_l^{(n)}\psi_l(x)-
\psi(x)\biggr]\,dx \;.
\label{eq:A4}
\end{equation}
Then the conditions of its minimality with respect to the set of
coefficients $a_k^{(n)}$ are written~as
\begin{equation}
{\partial D_n\over\partial a_k^{(n)*}}= \int_S \psi_k^*(x)\gamma(x)
\biggl[\;\sum_{l=1}^n a_l^{(n)}\psi_l(x)- \psi(x)\biggr]\,dx= 0 \;,
\label{eq:A5}
\end{equation}
\begin{equation}
{\partial^2 D_n\over\partial a_k^{(n)*}\partial a_k^{(n)}}=
\int_S \psi_k^*(x)\gamma(x)\psi_k(x)\,dx >0 \;.
\label{eq:A6}   \\[.2cm]
\end{equation}

Eq.~(\ref{eq:A6}) complies with the requirement of quadratic
integrability initially imposed on the base functions $\psi_k(x)$,
while the condition~(\ref{eq:A5}) yields the following system of
equations to find~$a_k^{(n)}$:
\begin{equation}
\sum_{l=1}^n N_{kl} a_l^{(n)}= R_k \;, \qquad k= 1,\,2,\ldots
\label{eq:A7}
\end{equation}
where we have denoted
\begin{equation}
N_{kl}=\int_S\psi_k^*(x)\gamma(x)\psi_l(x)\,dx\equiv(\psi_k, \psi_l)\;,
\label{eq:A8}
\end{equation}
\begin{equation}
R_k= \int_S \psi_k^*(x)\gamma(x)\psi(x)\,dx\equiv (\psi_k, \psi) \;.
\label{eq:A9}   \\[.2cm]
\end{equation}

Metric convergence in $L_2$ is defined as {\it convergence in mean\/}
(with index 2) of the sequence of partial sums $s_n(x)$
(with coefficients $a_k^{(n)}$ from Eqs.~(\ref{eq:A7})) to the function
$\psi(x)$,\, i.\,e.,\, $s_n(x)\buildrel\rm{mean}\over\longrightarrow
\psi(x)$  as $n\!\to\!\infty$, {} which occurs if and only if
\begin{equation}
D_n \equiv \int_S \gamma(x)|s_n(x)- \psi(x)|^2\,dx \to 0
\quad \mbox{as} \quad  n\to\infty \;.
\label{eq:A10}
\end{equation}

Using Eqs.~(\ref{eq:A7}) through (\ref{eq:A9}) and the equality
$N_{kl}= N_{lk}^*$ allows Eq. (\ref{eq:A4}) to take the following form
\[
D_n= \sum_{k=1}^n\sum_{l=1}^n N_{kl} a_k^{(n)*} a_l^{(n)} -
\sum_{k=1}^n R_k a_k^{(n)*} -
\sum_{l=1}^n R_l^*a_l^{(n)}+\int_S\gamma(x)|\psi(x)|^2\,dx =    \\[-0.1cm]
\]
\[
= - \sum_{k=1}^n\sum_{l=1}^n N_{kl}a_k^{(n)*} a_l^{(n)} +
\int_S \gamma(x)|\psi(x)|^2\,dx \;.
\]

From here for limiting case~(\ref{eq:A10}), when
$a_k^{(n)}(z)\!\to\!a_k(z)$ as $n\!\to\!\infty$, it~follows that
\begin{equation}
(\psi, \psi) \equiv \int_S \psi^*(x)\gamma(x)\psi(x)\,dx =
\sum_{k=1}^\infty \sum_{l=1}^\infty N_{kl}a_k^* a_l \;.
\label{eq:A11}
\end{equation}

Relation (\ref{eq:A11}) is realizable only for functions $\psi(x)$
quadratically integrable on~$S$ (with the weighting function
$\gamma(x)$),\, i.\,e., for which there exists an integral
on the left. This relation expresses  completeness of the set
of the base functions $\psi_k(x)$ (also quadratically integrable on~$S$)
inside the class of functions $\psi(x)$.\, The completeness
property establishes the space~$L_2$ as the Hilbert space for
which the series expansion
\begin{equation}
\psi(x) \buildrel\rm{mean}\over = \sum_{k=1}^\infty a_k\psi_k(x)
\label{eq:A12}
\end{equation}
interpreted in the sense of convergence in mean given by formula
(\ref{eq:A10}) is valid. Uniqueness of this expansion arises from the
following reasoning.

By conradiction,\, let two different series expansions\,
$\sum_k a_k^\prime\psi_k(x)$\, and\,
$\sum_k a_k^{\prime\prime}\psi_k(x)$ correspond to
the same function $\psi(x)$ in the sense of convergence
in mean. To determine the expansion coefficients $a_k^\prime$ and
$a_k^{\prime\prime}$ there are two  systems of form~(\ref{eq:A7})
with the same right-hand sides $R_k$. When resulted from them, the
difference system of equations $\sum_l N_{kl}(a_l^\prime -
a_l^{\prime\prime})= 0,\;\,k=1,2,\ldots$\, gives
\,$a_l^\prime\equiv a_l^{\prime\prime}$\, by virtue of
\,$N_{kl}\ne 0$,\, i.\,e., the initial series expansions coincide.
If on the contrary one assumes that the same series expansion
$\sum_k a_k\psi_k(x)$ corresponds to two different functions
$\psi^\prime(x)$ and $\psi^{\prime\prime}(x)$, then the difference
function $\psi^-(x)= \psi^\prime(x)- \psi^{\prime\prime}(x)$ has the
expansion coefficients identically equal to zero. So the right-hand
side of the completeness relation~(\ref{eq:A11}) vanishes, which
necessarily provides $\psi^-(x)\equiv 0$,\, i.\,e., the initial
functions coincide.

The completeness relation~(\ref{eq:A11}) is a generalization of the
conventional Parseval identity (see Eq.~(\ref{eq:A16})) to the case of
nonorthogonal bases. All the aforestated convince us that linear
independence and completeness of the set of base functions $\psi_k(x)$
are fundamental properties of the basis, whereas their mutual
orthogonality is not obligatory requirement and merely facili\-tates the
problem of finding the expansion coefficients $a_k$. Indeed, for the
orthogonal basis
\begin{equation}
(\psi_k, \psi_l) \equiv \int_S \psi_k^*(x)\gamma(x)\psi_l(x)\,dx= 0
\quad  \mbox{for}  \quad   k\ne l
\label{eq:A13}
\end{equation}
so that
\begin{equation}
N_{kl}= N_k\delta_{kl}    \quad \mbox{with} \quad
N_k= \int_S \psi_k^*(x)\gamma(x)\psi_k(x)\,dx \equiv \|\psi_k\|^2
\label{eq:A14}
\end{equation}
where $\|\psi_k\|= \sqrt{(\psi_k, \psi_k)} \equiv \sqrt{N_k}$ is
conventionally called the norm of a function $\psi_k(x)$~\cite{45}.
In addition, we extend this term to quantities $N_{kl}$ recognizing
the~{\it self norm\/} $N_k\equiv N_{kk}$ for $l=k$ and
the {\it cross norm\/} $N_{kl}$  for $l\ne k$.

Hence, in the special case of the orthogonal basis satisfying
Eq.~(\ref{eq:A14}): \\
(i) the system of coupled equations (\ref{eq:A7}) falls apart into
separate equations yielding
\begin{equation}
a_k= {R_k\over N_k}\equiv{(\psi_k, \psi)\over (\psi_k, \psi_k)}=
{1\over N_k} \int_S \psi_k^*(x)\gamma(x)\psi(x)\,dx \;,
\label{eq:A15}  \\[.2cm]
\end{equation}
(ii) the general completeness relation (\ref{eq:A11}) gives
the conventional Parseval identity
\begin{equation}
(\psi, \psi) \equiv \int_S \psi^*(x)\gamma(x)\psi(x)\,dx=
\sum_{k=1}^\infty N_k |a_k|^2 \;.
\label{eq:A16}  \\[.2cm]
\end{equation}

It should be remembered that the use of the known Gram-Schmidt
orthogo\-nalization process~\cite{45}, in principle, allows one to
construct the orthonormal basis.

The above completeness property of a basis expressed by relation
(\ref{eq:A11}) or (\ref{eq:A16}) concerns only such functions $\psi(x)$
that fully belong to the linear manifold spanned by the functions
$\psi_1(x), \psi_2(x),\ldots$ . However, for the most general functions
$f(x)$ this~is~not the case.

Any given function $f(x)$ quadratically integrable on $S$ (with the
weighting function $\gamma(x)$, in general) can formally
be associated with the function $\psi(x)$ represented by series
(\ref{eq:A12}) if one assumes that its coefficients $a_k$ satisfying
Eqs.~(\ref{eq:A7}) through (\ref{eq:A9}) are due to $f(x)$ and not to
$\psi(x)$,\, i.\,e., the quantities $R_k(x)$ contain $f(x)$ in place of
$\psi(x)$ under the integral sign of Eq.~(\ref{eq:A9}). Let us prove
that the difference $c\,(x)= f(x)- \psi(x)$ is orthogonal to every
base function $\psi_k(x)$ in the sense of relation~(\ref{eq:A13}):
\[
(\psi_k, c) \equiv (\psi_k, f-\psi)= (\psi_k, f)- (\psi_k, \psi) =
\]
\begin{equation}
= (\psi_k, f)- \sum_l (\psi_k, \psi_l) a_l= R_k - \sum_l N_{kl}a_l= 0
\label{eq:A17}
\end{equation}
where the relations $N_{kl}= (\psi_k, \psi_l)$ and $R_k= (\psi_k, f)$
have been used.

Thus, any arbitrary function $f(x)$ not belonging fully to the Hilbert
space (spanned, for instance, by eighenfunctions of a boundary-value
problem) can be represented in the following form
\begin{equation}
f(x)= \psi(x) + c\,(x) = \sum_k a_k\psi_k(x) + c\,(x) \;.
\label{eq:A18}
\end{equation}

Here the function $\psi(x)$ written as a series expansion in terms of
base functions (convergent in mean) and considered as tangential
to the given Hilbert space is called the {\it projection\/}
of $f(x)$ on this space, while $c\,(x)$ is a function orthogonal to
the Hilbert space and reffered to as the {\it orthogonal complement\/}
because $(\psi_k, c)= 0$. For such a function $f(x)$ instead of the
generalized Parseval identity~(\ref{eq:A11}) there exists the
generalized Bessel inequality
\begin{equation}
(f, f) \equiv \|\psi+ c\|^2 \equiv \int_S f^*(x)\gamma(x) f(x)\,dx
\ge \sum_{k=1}^\infty \sum_{l=1}^\infty N_{kl} a_k^* a_l
\;\; \mbox{or} \;\; \ge \sum_{k=1}^\infty N_k |a_k|^2
\label{eq:A19}
\end{equation}
where the last single sum corresponds to the orthogonal basis.

\subsection{Electrodynamic treatment (in notation of [8])}
\label{app:A2}
\indent

Let us consider a relevant aspect of the electrodynamic modal theory
on the basis of analogy with the foregoing mathematical relations.

Given an infinite set of eigenfunctions of a boundary-value problem
defined on the cross section $S$ of a waveguiding structure with the
transverse radius vector ${\bf r}_t$ and the longitudinal axis $z$,
any eigenfunction ${\bf\Psi}_k({\bf r}_t)$ and its adjoint (hermitian
conjugate) ${\bf\Psi}_k^\dagger({\bf r}_t)$ are denoted in the
two-vector notation as
\begin{equation}
{\bf\Psi}_k({\bf r}_t)= {\hat{\bf E}_k({\bf r}_t) \choose
\hat{\bf H}_k({\bf r}_t)}  \quad \mbox{and} \quad
{\bf\Psi}_k^\dagger({\bf r}_t)= \Bigl( \hat{\bf E}_k^*({\bf r}_t)\;\,
\hat{\bf H}_k^*({\bf r}_t) \Bigr)
\label{eq:A20}  \\[.2cm]
\end{equation}
where the hat over field vectors means the absence of their
dependence on $z$.

By analogy with Eq.~(\ref{eq:A1}), the inner product of two
eigenfunctions ${\bf\Psi}_k({\bf r}_t)$ and
${\bf\Psi}_l({\bf r}_t)$ can be defined in the following form
\begin{equation}
({\bf\Psi}_k, {\bf\Psi}_l)= \int_S {\bf\Psi}_k^\dagger({\bf r}_t)
\cdot\bar{\bf\Gamma}\cdot{\bf\Psi}_l({\bf r}_t)\;dS
\label{eq:A21}  \\[.2cm]
\end{equation}
with the weighting function given in the form of a special dyadic
\begin{equation}
\bar{\bf\Gamma} =
\pmatrix{ 0& -{\bf z}_0\times \bar{\bf I}\cr
              {\bf z}_0\times \bar{\bf I}&0\cr}
\label{eq:A22}
\end{equation}
where ${\bf z}_0$ is the  unit vector of the axis $z$ and
$\bar{\bf I}$ is the unit dyadic such that
$({\bf z}_0\times\bar{\bf I})\cdot{\bf a}=
-\,{\bf a}\cdot(\bar{\bf I}\times{\bf z}_0)= {\bf z}_0\times{\bf a}$
\,for any vector $\bf a$. The weighting dyadic $\bar{\bf\Gamma}$ is
constructed so as to make the double scalar product
${\bf\Psi}_k^\dagger\cdot\bar{\bf\Gamma}\cdot{\bf\Psi}_l =
{\bf\Psi}_k^\dagger\cdot \Bigl( \bar{\bf\Gamma}\cdot{\bf\Psi}_l \Bigr) =
\Bigl( {\bf\Psi}_k^\dagger\cdot\bar{\bf\Gamma} \Bigr) \cdot{\bf\Psi}_l
\equiv \bar{\bf\Gamma}:{\bf\Psi}_l{\bf\Psi}_k^\dagger$\,
under the integral sign of Eq.~(\ref{eq:A21}) be equal to\,
$(\hat{\bf E}_k^*\times\hat{\bf H}_l\,+\,
\hat{\bf E}_l\times\hat{\bf H}_k^*)\cdot{\bf z}_0$.

Thus, the {\it cross norm} $N_{kl}$\, for the $k$th\, and \,$l$th
modes and the {\it self norm} $N_k\equiv N_{kk}$~for the $k$th mode,
according to Eqs.~(\ref{eq:A8}) and (\ref{eq:A21}), can be
represented as
\[
N_{kl} \equiv ({\bf\Psi}_k, {\bf\Psi}_l)= \int_S
{\bf\Psi}_k^\dagger \cdot \bar{\bf\Gamma}
\cdot{\bf\Psi}_l \;dS =
\]
\begin{equation}
=\int_S (\hat{\bf E}_k^*\times\hat{\bf H}_l+\hat{\bf E}_l\times
\hat{\bf H}_k^*) \cdot {\bf z}_0\,dS
\label{eq:A23}
\end{equation}
and
\[
N_k \equiv ({\bf\Psi}_k, {\bf\Psi}_k)= \int_S {\bf\Psi}_k^\dagger \cdot
\bar{\bf\Gamma} \cdot {\bf\Psi}_k \;dS =
\]
\begin{equation}
= 2\,{\rm Re}\int_S(\hat{\bf E}_k^*\times\hat{\bf H}_k)\cdot{\bf z}_0\,dS\,.
\label{eq:A24}
\end{equation}

By analogy with the series expansion~(\ref{eq:A18}), an arbitrary
function ${\bf F}({\bf r}_t, z)$ quadratically integrable on $S$ can be
represented as a sum of the modal expansion ${\bf\Psi}({\bf r}_t, z)$
in terms of eigenfunctions ${\bf\Psi}_k({\bf r}_t)$ (the
{\it projection} of ${\bf F}({\bf r}_t, z)$ tangent to Hilbert
space and convergent in mean) and the {\it orthogonal complement}
${\bf C}({\bf r}_t, z)$:
\begin{equation}
{\bf F}({\bf r}_t, z)= {\bf\Psi}({\bf r}_t, z)+{\bf C}({\bf r}_t, z)=
\sum_k a_k(z)\,{\bf\Psi}_k({\bf r}_t) + {\bf C}({\bf r}_t, z)
\label{eq:A25}
\end{equation}
where we have denoted
\[
{\bf F}({\bf r}_t, z)={{\bf E}({\bf r}_t, z) \choose
{\bf H}({\bf r}_t, z)}\;,  \quad
{\bf\Psi}({\bf r}_t, z)={{\bf E}_a({\bf r}_t, z) \choose
{\bf H}_a({\bf r}_t, z)}\;, \quad
{\bf C}({\bf r}_t, z)=
{{\bf E}_b({\bf r}_t, z) \choose {\bf H}_b({\bf r}_t, z)}
\]
and by analogy with Eq.~(\ref{eq:A17}) the orthogonal complement
${\bf C}$ satisfy the relation
\begin{equation}
({\bf\Psi}_k, {\bf C})\equiv \int_S {\bf\Psi}_k^\dagger \cdot
\bar{\bf\Gamma} \cdot {\bf C} \;dS =
\int_S (\hat{\bf E}_k^*\times{\bf H}_b+ {\bf E}_b\times
\hat{\bf H}_k^*) \cdot {\bf z}_0\,dS \,=\,0 \;.
\label{eq:A26}  \\[.2cm]
\end{equation}

Eqs.~(\ref{eq:A25}) and (\ref{eq:A26}) allow the electromagnetic fields
to be represented in the following form
\begin{equation}
{\bf E}({\bf r}_t, z)\,=\,{\bf E}_a({\bf r}_t, z)\,+\,
{\bf E}_b({\bf r}_t, z) =
\sum_k a_k(z)\,\hat{\bf E}_k({\bf r}_t)+ {\bf E}_b({\bf r}_t, z)\;,
\label{eq:A27}      \\[-0.1cm]
\end{equation}
\begin{equation}
{\bf H}({\bf r}_t, z) = {\bf H}_a({\bf r}_t, z) +
{\bf H}_b({\bf r}_t, z) =
\sum_k a_k(z)\,\hat{\bf H}_k({\bf r}_t)+ {\bf H}_b({\bf r}_t, z)\;,
\label{eq:A28}
\end{equation}
where  the orthogonal complementary fields ${\bf E}_b({\bf r}_t, z)$
and ${\bf H}_b({\bf r}_t, z)$ as well as the mode amplitudes $a_k(z)$
of the modal expansions
\begin{equation}
{\bf E}_a({\bf r}_t, z)= \sum_k a_k(z)\,\hat{\bf E}_k({\bf r}_t)
\qquad \mbox{and} \qquad
{\bf H}_a({\bf r}_t, z)= \sum_k a_k(z)\,\hat{\bf H}_k({\bf r}_t)
\label{eq:A29}
\end{equation}
should be determined. The  amplitudes  $a_k(z)$, in principle, can
be found from the equations similar to Eqs.~(\ref{eq:A7}) for the
nonorthogonal basis or to Eq.~(\ref{eq:A15}) for the orthogonal basis,
with $R_k$ being given as follows
\[
R_k \equiv ({\bf\Psi}_k, {\bf\Psi})= ({\bf\Psi}_k, {\bf F})=
\int_S {\bf\Psi}_k^\dagger\cdot\bar{\bf\Gamma} \cdot {\bf F}\;dS=
\]
\begin{equation}
= \int_S (\hat{\bf E}_k^*\times{\bf H}+ {\bf E}\times\hat{\bf H}_k^*)
\cdot{\bf z}_0\,dS \,,
\label{eq:A30} \\[.2cm]
\end{equation}
in particular, by analogy with Eq.~(\ref{eq:A15})
\begin{equation}
a_k= {R_k\over N_k}\equiv
{({\bf\Psi}_k, {\bf F})\over({\bf\Psi}_k, {\bf\Psi}_k)} = {1\over N_k}
\int_S (\hat{\bf E}_k^*\times{\bf H}+ {\bf E}\times\hat{\bf H}_k^*)
\cdot{\bf z}_0\,dS \;.
\label{eq:A31}
\end{equation}

It is of great importance in electrodynamic applications that such
a procedure of determining the mode amplitude $a_k(z)$ based
on Eqs.~(\ref{eq:A7}) or (\ref{eq:A15}) allows us instead of
the series expansion ${\bf\Psi}({\bf r}_t, z)$ in terms of eigenmodes
to apply its finite sum of the $n$th order
\begin{equation}
{\bf S}_n({\bf r}_t, z)= \sum_{k=1}^n a_k^{(n)}(z)
{\bf\Psi}_k({\bf r}_t)
\label{eq:A32}
\end{equation}
like Eq.~(\ref{eq:A2}), which yields the least mean-square error
\begin{equation}
D_n= \int_S \Bigl[ {\bf S}_n^\dagger({\bf r}_t, z)-
{\bf\Psi}^\dagger({\bf r}_t, z) \biggr] \cdot
\bar{\bf\Gamma} \cdot \biggl[ {\bf S}_n({\bf r}_t, z)-
{\bf\Psi}({\bf r}_t, z) \Bigr]\,dS \,,
\label{eq:A33}
\end{equation}
analogously to Eq.~(\ref{eq:A4}).

The above general reasoning concerning the convergence in mean,
completeness, and orthogonality properties of base functions can be
extended to the electrodynamic basis of eigenfunctions so that, in
particular, the gene\-ralized Parseval identity~(\ref{eq:A11}) and
Bessel inequality~(\ref{eq:A19}) take the following form
\[
({\bf\Psi}, {\bf\Psi}) \equiv \|{\bf\Psi}\|^2 =
\int_S {\bf\Psi}^\dagger \cdot \bar{\bf\Gamma}
\cdot {\bf\Psi}\;dS=
\]
\begin{equation}
= \int_S ({\bf E}_a^*\times{\bf H}_a +
{\bf E}_a\times{\bf H}_a^*)\cdot{\bf z}_0\,dS=
\!\sum_{k=1}^\infty \sum_{l=1}^\infty N_{kl}a_k^*a_l
\;\;\; \mbox{or} \;\;=\!\sum_{k=1}^\infty N_k |a_k|^2
\label{eq:A34}
\end{equation}
and
\[
({\bf F}, {\bf F}) \equiv \|{\bf\Psi}+ {\bf C}\|^2 =
\int_S {\bf F}^\dagger \cdot \bar{\bf\Gamma}
\cdot{\bf F}\;dS=
\]
\begin{equation}
= \int_S ({\bf E}^*\times{\bf H} +
{\bf E} \times{\bf H}^*) \cdot{\bf z}_0\,dS \ge
\sum_{k=1}^\infty \sum_{l=1}^\infty N_{kl}a_k^*a_l
\quad \mbox{or} \quad \ge\sum_{k=1}^\infty N_k |a_k|^2
\label{eq:A35} \\[.2cm]
\end{equation}
where the last single sums correspond to the orthogonal basis.

As noted above, the orthogonality property of a basis is not mandatory
but its existence facilitates the determination of the expansion
coefficients $a_k(z)$. Such a property is inherent in lossless physical
systems, whereas losses destroy the "pure" orthogonality and convert
it into the so-called {\it quasi-orthogonality\/} (see Sec.\,3.1).

\setcounter{equation}{0}
\section{DIRECT DERIVATION OF THE EQUATIONS OF MODE
                        EXCITATION FROM MAXWELL'S EQUATIONS}
\label{app:B}

Starting point to derive the equation of mode excitation is
formulas~(\ref{eq:4.12}) and (\ref{eq:4.13}) which are a result of
transforming Maxwell's equations~(\ref{eq:4.4}) and (\ref{eq:4.5})
inside the source region. Let us rewrite Eqs.~(\ref{eq:4.12}) and
(\ref{eq:4.13}) for transverse components:
\begin{equation}
\sum_l {dA_l\over dz}\,({\bf z}_0\times{\bf E}_l) =
-\,\mbox{\boldmath$\nabla$}\times{\bf E}_b - i\omega\mu_0{\bf M}_{bt} -
{\bf J}_{bt}^m \;,
\label{eq:B1}
\end{equation}
\begin{equation}
\sum_l {dA_l\over dz}\,({\bf z}_0\times{\bf H}_l) =
-\,\mbox{\boldmath$\nabla$}\times{\bf H}_b\;+\;i\omega{\bf P}_{bt}\;+\;
{\bf J}_{bt}^e \;.
\label{eq:B2} \\[.2cm]
\end{equation}

Here, in accordance with Eq.~(\ref{eq:2.2}), the orthogonal complements
for the polarization ${\bf P}_b$ and magnetization ${\bf M}_b$ are
defined as
\begin{equation}
{\bf P}_b= {\bf D}_b- \epsilon_0{\bf E}_b  \qquad \mbox{and} \qquad
\mu_0{\bf M}_b= {\bf B}_b- \mu_0{\bf H}_b
\label{eq:B3}
\end{equation}
so that, as follows from Eqs.~(\ref{eq:4.14}), (\ref{eq:4.15}),
(\ref{eq:4.17}), and (\ref{eq:4.18}), their longitudinal components
contribute to the complementary fields:
\begin{equation}
{\bf E}_b\,\equiv\,{\bf z}_0\,E_b \;=\,-\;{1\over i\omega\epsilon_0}\,
(\,{\bf J}_{bz}^e \;+\; i\omega{\bf P}_{bz}) \,,
\label{eq:B4}           \\[-0.1cm]
\end{equation}
\begin{equation}
{\bf H}_b\,\equiv\,{\bf z}_0\,H_b = -\,{1\over i\omega\mu_0}
({\bf J}_{bz}^m + i\omega\mu_0{\bf M}_{bz}) \,.
\label{eq:B5} \\[.2cm]
\end{equation}

If we scalar-multiply Eqs.~(\ref{eq:B1}) and (\ref{eq:B2}) by
${\bf H}_k^*$ and $-{\bf E}_k^*$, respectively, and add the results,
then after integrating over $S$ we obtain
\[
\sum_l {dA_l\over dz} \int_S ({\bf E}_k^*\times{\bf H}_l+
{\bf E}_l\times{\bf H}_k^*)\cdot{\bf z}_0\,dS =
\int_S ({\bf E}_k^*\cdot\mbox{\boldmath$\nabla$}\times{\bf H}_b -
{\bf H}_k^*\cdot\mbox{\boldmath$\nabla$}\times{\bf E}_b)\,dS \,-
\]
\begin{equation}
- \int_S ({\bf J}_{bt}^e\cdot{\bf E}_{kt}^* +
{\bf J}_{bt}^m\cdot{\bf H}_{kt}^*)\,dS
-\,i\omega \int_S ({\bf P}_{bt}\cdot{\bf E}_{kt}^* +
\mu_0{\bf M}_{bt}\cdot{\bf H}_{kt}^*)\,dS \,.
\label{eq:B6} \\[.2cm]
\end{equation}

Now it is necessary to transform the first integral in the right-hand
side of Eq.~(\ref{eq:B6}). The terms of its integrand can be rearranged
as follows
\[
{\bf E}_k^*\cdot\mbox{\boldmath$\nabla$}\times{\bf H}_b \,=\,
{\bf z}_0\cdot({\bf E}_k^*\times\mbox{\boldmath$\nabla$} H_b) \,=\,
{\bf H}_b\cdot\mbox{\boldmath$\nabla$}\times{\bf E}_k^* \,-\,
{\bf z}_0\cdot\mbox{\boldmath$\nabla$}\times({\bf E}_k^*\,H_b)\,= \\[-0.1cm]
\]
\[
=\,i\omega\mu_0\,({\bf H}_k^* + {\bf M}_k^*)\cdot{\bf H}_b\,-\,
{\bf z}_0\cdot\mbox{\boldmath$\nabla$}\times({\bf E}_k^*\,H_b) \;, \\[.2cm]
\]
\[
{\bf H}_k^*\cdot\mbox{\boldmath$\nabla$}\times{\bf E}_b \,=\,
{\bf z}_0\cdot({\bf H}_k^*\times\mbox{\boldmath$\nabla$} E_b) \,=\,
{\bf E}_b\cdot\mbox{\boldmath$\nabla$}\times{\bf H}_k^* \,-\,
{\bf z}_0\cdot\mbox{\boldmath$\nabla$}\times({\bf H}_k^*\,E_b)\,=  \\[.1cm]
\]
\[
=\,-\,i\omega\,(\epsilon_0{\bf E}_k^*+{\bf P}_k^*)\cdot{\bf E}_b\,-\,
{\bf z}_0\cdot\mbox{\boldmath$\nabla$}\times({\bf H}_k^*\,E_b) \;,  \\[.2cm]
\]
where in the last equalities for the $k$th mode we have used
Eqs.~(\ref{eq:5.9}) and (\ref{eq:5.10}).
Then the first integral in Eq.~(\ref{eq:B6})  turns into the sum of
three integrals:
\[
\int_{S_b} ({\bf E}_k^*\cdot\mbox{\boldmath$\nabla$}\times{\bf H}_b -
{\bf H}_k^*\cdot\mbox{\boldmath$\nabla$}\times{\bf E}_b)\,dS =
\int_{S_b} \mbox{\boldmath$\nabla$}\times({\bf H}_k^*\,E_b -
{\bf E}_k^*\,H_b)\cdot{\bf z}_0\,dS \,+
\]
\begin{equation}
+\,i\omega \int_{S_b} ({\bf E}_k^*\cdot\epsilon_0{\bf E}_b +
{\bf H}_k^*\cdot\mu_0{\bf H}_b)\,dS \,+\,
i\omega \int_{S_b} ({\bf P}_k^*\cdot{\bf E}_b +
\mu_0{\bf M}_k^*\cdot{\bf H}_b)\,dS \;.
\label{eq:B7}  \\[.2cm]
\end{equation}

The first integral in the right-hand side of Eq.~(\ref{eq:B7}) is
transformed by using the Stokes theorem~\cite{45} into the following form
\[
\int_{S_b} \mbox{\boldmath$\nabla$}\times({\bf H}_k^*\,E_b -
{\bf E}_k^*\,H_b)\cdot{\bf z}_0\,dS \,=\,
\oint_{L_b} ({\bf H}_k^*\,E_b-{\bf E}_k^*\,H_b)\cdot
\mbox{\boldmath$\tau$}\,dl\,=
\]
\begin{equation}
=\,\oint_{L_b} \Bigl[ \,({\bf n}_b\times{\bf H}_b)\cdot{\bf E}_k^* -
({\bf n}_b\times{\bf E}_b)\cdot{\bf H}_k^*\, \Bigr] \,dl
\label{eq:B8} \\[.2cm]
\end{equation}
where ${\bf n}_b$ and ${\mbox{\boldmath$\tau$}}=
{\bf z}_0\times{\bf n}_b$ are the
unit vectors, respectively, normal (outward) and tangential to the
contour $L_b$ bounding the bulk current area $S_b$.

The second integral in the right-hand side of Eq.~(\ref{eq:B7}) is
rearranged by using Eqs.~(\ref{eq:B4}) and (\ref{eq:B5}) to the
following form
\[
i\omega \int_{S_b} ({\bf E}_k^*\cdot\epsilon_0{\bf E}_b +
{\bf H}_k^*\cdot\mu_0{\bf H}_b)\,dS \,=
\]
\begin{equation}
=\,- \int_{S_b} (J_{bz}^e\,E_{kz}^* + J_{bz}^m\,H_{kz}^*)\,dS \,-\,
i\omega \int_{S_b} (P_{bz}\,E_{kz}^* + \mu_0 M_{bz}\,H_{kz}^*)\,dS \,.
\label{eq:B9}  \\[.2cm]
\end{equation}

After inserting Eqs.~(\ref{eq:B8}) and (\ref{eq:B9}) into
Eq.~(\ref{eq:B7}) we obtain
\[
\int_{S_b} \Bigl( {\bf E}_k^*\cdot\mbox{\boldmath$\nabla$}\times{\bf H}_b -
{\bf H}_k^*\cdot\mbox{\boldmath$\nabla$}\times{\bf E}_b \Bigr)\,dS =
\oint_{L_b} \Bigl[ \,({\bf n}_b\times{\bf H}_b)\cdot{\bf E}_k^* -
({\bf n}_b\times{\bf E}_b)\cdot{\bf H}_k^*\, \Bigr] dl -  \\[-0.1cm]
\]
\[
- \int_{S_b} (J_{bz}^e\,E_{kz}^* + J_{bz}^m\,H_{kz}^*)\,dS \,-\,
i\omega \int_{S_b} (P_{bz}\,E_{kz}^* +
\mu_0 M_{bz}\,H_{kz}^*)\,dS\,+          \\[.1cm]
\]
\begin{equation}
+\,i\omega \int_{S_b} ({\bf P}_k^*\cdot{\bf E}_b +
\mu_0{\bf M}_k^*\cdot{\bf H}_b)\,dS \;.
\label{eq:B10} \\[.2cm]
\end{equation}

The first integral in the right-hand side of Eq.~(\ref{eq:B10})
involves the effective surface currents ${\bf J}_{s,ef}^e\!=\!
-{\bf n}_b\times{\bf H}_b$ and ${\bf J}_{s,ef}^m\!=\! {\bf n}_b
\times{\bf E}_b$ defined by Eqs.~(\ref{eq:4.27}) and (\ref{eq:4.28}).
With allowing for this and employing the expression for the
normalizing~coefficient
\[
N_{kl}\,=\, \int_S ({\bf E}_k^*\times{\bf H}_l+
{\bf E}_l\times{\bf H}_k^*)\cdot{\bf z}_0\,dS \;,   \\[.2cm]
\]
the substitution of Eq.~(\ref{eq:B10}) into Eq.~(\ref{eq:B6}) yields
\[
\sum_l N_{kl}\,{dA_l\over dz} = - \int_{S_b}
({\bf J}_b^e\cdot{\bf E}_k^* +
{\bf J}_b^m\cdot{\bf H}_k^*)\,dS - \int_{L_b}
({\bf J}_{s,ef}^e\cdot{\bf E}_k^* +
{\bf J}_{s,ef}^m\cdot{\bf H}_k^*)\,dl \,-
\]
\begin{equation}
-\,i\omega \int_{S_b} \Bigl[\,
({\bf P}_b\cdot{\bf E}_k^* - {\bf P}_k^*\cdot{\bf E}_b) +
(\mu_0{\bf M}_b\cdot{\bf H}_k^* -
\mu_0{\bf M}_k^*\cdot{\bf H}_b)\, \Bigr]\,dS \;.
\label{eq:B11} \\[.2cm]
\end{equation}

The last integral in the right-hand side of Eq.~(\ref{eq:B11})
vanishes because of
\[
\Bigl[ \,({\bf P}_b\cdot{\bf E}_k^* -
{\bf P}_k^*\cdot{\bf E}_b) +
(\mu_0{\bf M}_b\cdot{\bf H}_k^* -
\mu_0{\bf M}_k^*\cdot{\bf H}_b)\, \Bigr]\,=     \\[-0.1cm]
\]
\[
=\,\Bigl[ \,({\bf D}_b\cdot{\bf E}_k^* - {\bf D}_k^*\cdot{\bf E}_b) +
({\bf B}_b\cdot{\bf H}_k^* -{\bf B}_k^*\cdot
{\bf H}_b)\, \Bigr]\,=                \\[.1cm]
\]
\[
=\! \Bigl[ (\,\bar{\!\mbox{\boldmath$\epsilon$}}-
\bar{\!\mbox{\boldmath$\epsilon$}}\,^\dagger):
{\bf E}_b{\bf E}_k^* \,+\,
(\,\bar{\!\mbox{\boldmath$\mu$}}-
\bar{\!\mbox{\boldmath$\mu$}}\,^\dagger):
{\bf H}_b{\bf H}_k^* \,+\,
(\,\bar{\!\mbox{\boldmath$\xi$}}-
\bar{\!\mbox{\boldmath$\zeta$}}\,^\dagger):
{\bf H}_b{\bf E}_k^* \,+\,
(\,\bar{\!\mbox{\boldmath$\zeta$}}-
\bar{\!\mbox{\boldmath$\xi$}}\,^\dagger):
{\bf E}_b{\bf H}_k^* \Bigr] \!= 0                \\[.2cm]
\]
where the constitutive relations~(\ref{eq:2.8}), (\ref{eq:2.9}), and
(\ref{eq:2.13}) have been used for lossless bianisotropic media. \\[.15cm]
\indent
For the most general case of the reactive $k$th mode from the
orthonormalization relation~(\ref{eq:3.23}) we have $N_{kl} =
N_k\delta_{{\tilde k}l}$. Then formula~(\ref{eq:B11}) finally gives
the excitation equation for the $\tilde k$th mode:
\[
{dA_{\tilde k}\over dz} = -\,{1\over N_k} \int_{S_b}
({\bf J}_b^e\cdot{\bf E}_k^* \,+\,
{\bf J}_b^m\cdot{\bf H}_k^*)\,dS \,-        \\[.1cm]
\]
\begin{equation}
- {1\over N_k} \int_{L_b} ({\bf J}_{s,ef}^e\cdot{\bf E}_k^* \,+\,
{\bf J}_{s,ef}^m\cdot{\bf H}_k^*)\,dl \,.
\label{eq:B12} \\[.2cm]
\end{equation}

This formula is in agreement with the similar equation (\ref{eq:5.13})
obtained by another method, not counting the absence of the actual
surface currents which can be considered as enclosed implicitly into
the bulk currents.

\vspace{2\baselineskip}

\end{document}